\documentclass[Afour,sagev,times]{sagej}

\usepackage{moreverb,url}
\usepackage[colorlinks,bookmarksopen,bookmarksnumbered,citecolor=red,urlcolor=red]{hyperref}
\newcommand\BibTeX{{\rmfamily B\kern-.05em \textsc{i\kern-.025em b}\kern-.08em
T\kern-.1667em\lower.7ex\hbox{E}\kern-.125emX}}
\def\volumeyear{2016}

\usepackage[utf8]{inputenc}
\usepackage{subfigure}
\usepackage{amsmath}
\usepackage{algorithm}
\usepackage{algorithmic}
\usepackage{multirow}
\graphicspath{{fig/}}

\begin{document}

\runninghead{Risco-Martín et al.}

\title{Reconsidering Performance of DEVS Modeling and Simulation Environments Using the DEVStone Benchmark}

\author{José L. Risco-Martín\affilnum{1} and Saurabh Mittal\affilnum{2} and Juan Carlos Fabero Jiménez\affilnum{1} and Marina Zapater\affilnum{1} and Román Hermida Correa\affilnum{1}}

\affiliation{\affilnum{1}Complutense University of Madrid, Spain\\
\affilnum{2}Dunip Technologies, LLC, USA}

\corrauth{José L. Risco-Martín, Department of Computer Architecture and Automation,
Complutense University of Madrid,
C/Prof. José García Santesmases 9,
28040 Madrid,
Spain.\\
Tel. +34-91-3947603\\
Fax. +34-91-3947527}

\email{jlrisco@ucm.es}

\begin{abstract}
The Discrete Event System Specification formalism (DEVS), which supports hierarchical and modular model composition, has been widely used to understand, analyze and develop a variety of systems. DEVS has been implemented in various languages and platforms over the years. The DEVStone benchmark was conceived to generate a set of models with varied structure and behavior, and to automate the evaluation of the performance of DEVS-based simulators. However, DEVStone is still in a preliminar phase and more model analysis is required. In this paper, we revisit DEVStone introducing new equations to compute the number of events triggered. We also introduce a new benchmark, called HOmem, designed as an alternative version of HOmod, with similar CPU and memory requirements, but with an easier implementation and analytically more manageable. Finally, we compare both the performance and memory footprint of five different DEVS simulators in two different hardware platforms.
\end{abstract}

\keywords{DEVS, DEVStone, Synthetic Benchmarks, Performance, Memory Usage}

\maketitle

\section{Introduction}\label{sec:Introduction}
In the last four decades, various Modeling and Simulation (M\&S) methodologies have provided excellent approaches to solve problems. Among them, one of the M\&S techniques that has gained popularity is the Discrete Event System Specification (DEVS): a sound, formal definition, based on generic dynamic systems theoretical concepts, which supports efficient event based simulation, verification and validation~\cite{Zeigler1976}~\cite{Zeigler2000}.

DEVS divides the system into basic models, called atomic models, and composite models called coupled models. Atomic models define the behavior of the system, whereas coupled models specify the structure. We can distinguish between classic DEVS and Parallel DEVS (PDEVS)~\cite{Zeigler2000}. In classic DEVS, when two or more models are scheduled for state transitions at the same time, one of the models is chosen according to a \emph{select} function provided in the coupled model specification. PDEVS is an extension of classic DEVS to allow all imminent components to be activated and to send their output to other components. Removing the \emph{select} function and adding a new \emph{confluent} transition function, PDEVS introduces the possibility to manage simultaneous events in a natural manner.

DEVS has been successfully used for modeling a wide range of application domains. For example, it has been used in urban traffic analysis~\cite{Wainer2007}, logistics and supply chain~\cite{Byon2011}, computer architecture~\cite{Wainer2001},~\cite{Moreno2013}, embedded system designs~\cite{Moallemi2011}, unmanned aerial vehicles~\cite{Moreno2009}, decision support systems~\cite{Perez2010}, etc. Because of the ease of model definition, model composition, reuse, and hierarchical coupling, DEVS has always been successfully applied in such a variety of applications.

In contrast to time-stepped discrete time simulation, DEVS advances time through the concept of minimum time to next event, thereby advancing time asynchronously and achieving significant speedup over the former method ~\cite{Zeigler2000}. As a result, the DEVS formalism has been implemented in major object-oriented programming languages, like Lisp, Scheme, C++, Java, Python, SmallTalk, leading to many DEVS simulation engines across the globe, like DEVSJAVA~\cite{DEVSJAVA}, DEVS-Suite~\cite{DEVSuite}, COSMOS~\cite{CoSMoS}, CD++~\cite{CD++}, PyPDEVS~\cite{Tendeloo2014}, aDEVS~\cite{aDEVS}, JAMES-II~\cite{JAMES-II}, DEVSim++~\cite{Kim2011}, xDEVS~\cite{DUNIP}, to name but a few.

This variety of simulation engines has generated an extensive study in DEVS performance, commonly focused on particular application domains. However, after several years of research, a final version of a discrete event simulation benchmark was published, named DEVStone~\cite{Glinsky2005},~\cite{Gutierrez-Alcaraz2008},~\cite{Wainer2011}. DEVStone can be used to automatically generate a vast variety of models with different shapes and sizes. These models can then be simulated to test different features with respect to the corresponding simulator. 

These benchmarks incorporate several benefits but some of them suffer from shortcomings in their mathematical descriptions, like the formal computation of the total number of events triggered. In this paper, we reconsider these benchmarks. Firstly, we include the computation of the total number of events triggered inside each benchmark. Secondly, we fix some equations that in~\cite{Wainer2011} did not give the exact number of transitions, concretely equations (2), (3) and (4) in \cite{Wainer2011}. It is worthwhile to mention that these errors have not affected the reliability of previous paper results, since these models have been always used to compare wall-clock execution times. Finally, we define an additional benchmark, called HOmem, that demands the same computational effort than HOmod, the more complex model in DEVStone, but is analytically more manageable, as is demonstrated in the research work.

The remainder of this paper is organized as follows: we show a brief description of the DEVS formalism and introduce several DEVS simulation engines in Section~\ref{sec:DevsIntro}. The DEVStone benchmark is revisited in Section~\ref{sec:DevStone}, including all the contributions to the benchmark performed in this work. In Section~\ref{sec:ExpMethodology} we describe our experimental infrastructure and methodology. In Section~\ref{sec:Results} we present experimental results, including a comparison of up to five simulators and more than 1400 DEVStone models. Finally, we present conclusions in  Section~\ref{sec:Conclusions}.

\section{DEVS: Formalism and Simulation Engines}\label{sec:DevsIntro}

\subsection{The Discrete Event System Specification}
DEVS is a general formalism for discrete event system modeling based on set theory~\cite{Zeigler2000}. The DEVS formalism provides the framework for information modeling which gives several advantages to analyze and design complex systems: completeness, verifiability, extensibility, and maintainability. Once a system is described in terms of the DEVS theory, it can be easily implemented using an existing computational library. As stated in Section \ref{sec:Introduction}, the parallel DEVS (PDEVS) approach was introduced, after 15 years, as a revision of Classic DEVS. Currently, PDEVS is the prevalent DEVS, implemented in many libraries. In the following, unless it is explicitly noted, the use of DEVS implies PDEVS.

DEVS enables the representation of a system by three sets and five functions: input set $(X)$, output set $(Y)$, state set $(S)$, external transition function ($\delta_{\rm ext}$), internal transition function ($\delta_{\rm int}$), confluent function ($\delta_{\rm con}$), output function ($\lambda$), and time advance function ($ta$). 

DEVS models are of two types: atomic and coupled. Atomic models are directly expressed in the DEVS formalism specified above. Atomic DEVS processes input events based on their model's current state and condition, generates output events and transition to the next state. The coupled model is the aggregation/composition of two or more atomic and coupled models connected by explicit couplings. Particularly, an atomic model is defined by the following equation:

\begin{equation}
A=\langle X, Y, S, \delta_{\rm ext},  \delta_{\rm int}, \delta_{\rm con}, \lambda, ta\rangle
\label{eq:DevsAtomicModel}
\end{equation}

\noindent where:

\begin{itemize}

\item $X$ is the set of inputs described in terms of pairs port-value: $\left\{ p \in IPorts,v \in X_p \right\} $.

\item $Y$ is the set of outputs, also described in terms of pairs port-value: $\left\{ p \in OPorts,v \in Y_p \right\} $.

\item $S$ is the set of sequential states.

\item $\delta_\mathrm{ext}: Q\times X^b \rightarrow S$ is the external transition function. It is automatically executed when an external event arrives to one of the input ports, changing the current state if needed.
\begin{itemize}
\item $Q={(s,e) s \in S, 0 \leq e \leq ta(s)}$ is the total state set, where $e$ is the time elapsed since the last transition.
\item $X^b$ is the set of bags over elements in $X$.
\end{itemize}

\item $\delta_\mathrm{int}: S \rightarrow S$ is the internal transition function. It is executed right after the output ($\lambda$) function and is used to change the state $S$.

\item $\delta_\mathrm{con}: Q\times X^b \rightarrow S$ is the confluent function, subject to $\delta_\mathrm{con}(s,ta(s),\emptyset)=\delta_\mathrm{int}(s)$. This transition decides the next state in cases of collision between external and internal events, i.e., an external event is received and elapsed time equals time-advance. Typically, $\delta_\mathrm{con}(s,ta(s),x) = \delta_\mathrm{ext}(\delta_\mathrm{int}(s),0,x)$.

\item $\lambda: S \rightarrow Y^b$ is the output function. $Y^b$ is the set of bags over elements in $Y$. When the time elapsed since the last output function is equal to $ta(s)$, then $\lambda$ is automatically executed.

\item $ta(s): S \rightarrow \Re_0^+ \cup \infty$ is the time advance function.
\end{itemize}

The formal definition of a coupled model is described as:
\begin{equation}
M = \langle X, Y, C, EIC, EOC, IC \rangle
\end{equation}

\noindent where:
\begin{itemize}
\item $X$ is the set of inputs described in terms of pairs port-value: $\left\{ p \in IPorts,v \in X_p \right\} $.
\item $Y$ is the set of outputs, also described in terms of pairs port-value: $\left\{ p \in OPorts,v \in Y_p \right\} $.
\item $C$ is a set of DEVS component models (atomic or coupled). Note that $C$ makes this definition recursive.
\item $EIC$ is the external input coupling relation, from external inputs of $M$ to component inputs of $C$.
\item $EOC$ is the external output coupling relation, from component outputs of $C$ to external outputs of $M$.
\item $IC$ is the internal coupling relation, from component outputs of $c_i \in C$ to component outputs of $c_j \in C$, provided that $i \neq j$.
\end{itemize}
Given the recursive definition of $M$, a coupled model can itself be a part of a component in a larger coupled model system giving rise to a hierarchical DEVS model construction.

\subsection{DEVS Simulation Engines}
In the last decade, many DEVS M\&S engines have come into existence. All of them offer a programmer-friendly Application Programming Interface (API) to define new models using a high level language. However, only a few of them provide a user-friendly Graphical User Interface (GUI) for model specification. In the following, we describe some of the most referenced DEVS M\&S simulation frameworks:

\subsubsection{DEVSJAVA:}
DEVSJAVA has been developed by Bernard P. Zeigler (University of Arizona, U.S.A.) and Hessam Sarjoughian (Arizona State University, U.S.A.)~\cite{DEVSJAVA}. It is written in Java and supports virtual time, real time, and sequential and parallel execution. The definition of new models is performed through an API. Several M\&S tools have been defined around DEVSJAVA (GUIs for results visualization, GUIs for models definition, etc.), as DEVSJAVA is one of the primary DEVS M\&S reference simulators in the community.

\subsubsection{DEVS-Suite and COSMOS:}
DEVS-Suite is a simulator built based on the Parallel DEVS formalism. This software provides a library of examples proving some experimental concepts. It also incorporates simulation visualization techniques consisting of displaying static structure of models, animation of models, and run-time viewing of time-based trajectories~\cite{DEVSuite}. CoSMoS (Component-Based System Modeling and Simulation) is a framework aimed at integrated visual model development, model configuration and automatic data collection simulation~\cite{CoSMoS}. The CoSMoS environment supports component-based modeling with direct support for DEVS formalism and XML Schema. DEVS-Suite's core is largely DEVSJAVA. It is bundled within the CoSMoS distribution and thus enables both modeling and simulation of Parallel DEVS models.

\subsubsection{CD++:}
CD++ has been developed by Gabriel Wainer and his students (Carleton University, Canada; Universidad de Buenos Aires, Argentina). Written in C++, it allows the definition of DEVS and Cell-DEVS models graphically. These models are also defined using an API. CD++ supports virtual and real time, as well as sequential, parallel and distributed simulations~\cite{CD++}.

\subsubsection{PyPDEVS:}
PythonPDEVS (a.k.a. PyPDEVS) implements both Classic and Parallel DEVS in the Python language, with a matching simulator~\cite{Tendeloo2014}. Models are defined through the provided API, allowing the execution of virtual time or real time simulations. The latest release of PyPDEVS is focused on improving the performance, mainly because Python is an interpreted language. To this end, several schedulers have been defined, obtaining good performance metrics.

\subsubsection{aDEVS:}
ADEVS (A Discrete EVent System simulator) is a C++ library for constructing discrete event simulations based on the Parallel DEVS and Dynamic DEVS (dynDEVS) formalisms~\cite{aDEVS}. Developed by Jim Nutaro, it allows the implementation of both sequential and parallel simulations using the provided C++ API. This tool has usually displayed the best performance.

\subsubsection{JAMES-II:}
Developed at the University of Rostock, the Java-based Multipurpose Environment for Simulation II (JAMES II) provides support for many different formalisms, including various variants of DEVS. Besides an API to define models, this framework also provides a GUI to configure experiments and check simulation results. This simulation engine supports sequential and parallel execution~\cite{JAMES-II}.

\subsubsection{DEVSim++:}
Developed by Tag Gon Kim and his group at Korea Advanced Institute of Technology (KAIST)~\cite{Kim2011}, this is a C++ based engine and used extensively for large simulations focusing on wargaming and simulation interoperability.

\subsubsection{xDEVS:}
xDEVS engine is Java-based and is released under the GNU Public License (GPL). This facilitates the rapid development of new components and extensions, and wide adoption of the core engine. xDEVS provides the user with a set of base classes that can be used to develop new DEVS models, or to develop new DEVS simulation engines. It is based on the fundamental separation of model and the underlying corresponding simulator~\cite{Zeigler2000} and rightly so, provides, the modeling Application Program Interface (API) and the simulation API~\cite{DUNIP}. It is made available as a standalone executable jar and as an Eclipse plugin.

\subsubsection{Others:}
In addition to the above DEVS implementations used widely, there are others with selective adoption such as GALATEA~\cite{Davila2000} for Multi-Agent Systems (MAS), SimStudio~\cite{Traore2010}, PowerDEVS~\cite{PowerDEVS} for hybrid systems, MS4Me based on DEVSJAVA~\cite{MS4Me} and last but not the least, Virtual Laboratory Environment (VLE)~\cite{VLE}, that based on C++, is a multiparadigm environment based on several DEVS extensions. 

We have selected five well-known DEVS simulation frameworks distributed among three implementation languages and compared their performance against a revisited DEVStone benchmark. The current diversity on the programming languages used is concentrated on C++, JAVA and Python. As a consequence, we have selected two JAVA-based simulators (DEVSJAVA and xDEVS), two C++-based simulators (aDEVS and CD++) and PyPDEVS as the Python-based simulation engine. In the following, we describe the revisited DEVStone benchmark.

\section{DEVStone}\label{sec:DevStone}
DEVStone is a synthetic benchmark~\cite{Wainer2011} that has been used in recent years to evaluate the performance of different DEVS simulators~\cite{Wainer2011}~\cite{VanTendeloo2014}~\cite{Vicino2015}. DEVStone can be used to automatically generate a vast variety of models with different shapes and sizes. These models can then be simulated to test different features with respect to the corresponding simulator. A DEVStone benchmark is defined with five parameters:

\begin{description}
\item[Type:] Different structure and interconnection schemes between the components in the model.
\item[Width] This parameter is based the number of components in each intermediate coupled model.
\item[Depth] The number of levels in the model hierarchy.
\item[Internal transition time:] The execution time spent by each internal transition function.
\item[External transition time:] The execution time spent by each external transition function.
\end{description}

According to the DEVStone specifications, both the internal and external transition function times are spent executing Dhrystones~\cite{Weicker1984} to keep the CPU busy.

In~\cite{Wainer2011}, four different DEVStone benchmarks were presented (named LI, HI, HO and HOmod), deriving different equations to compute the number of external and internal transition functions. 

In the following a formal definition of the DEVStone atomic model is introduced. Next, all the five benchmarks considered in this work (LI, HI, HO, HOmod and the newly introduced HOmem) are presented. To simplify the computation of the total number of events triggered, it is assumed that: (1) the execution time spent by the external or internal transition function is equal to 0 seconds, i.e. the transition is instantaneous in a computational sense, and (2) all the events injected to the DEVStone benchmarks are separated in time by more than 0 seconds.

\subsection{DEVStone atomic model}
The atomic model of DEVStone can be defined as shown in Algorithm~\ref{alg:atomic}. 

\begin{algorithm}[ht]
\caption{DEVStone atomic model
\label{alg:atomic}}
\begin{algorithmic}

\REQUIRE{\texttt{NUM\_DELT\_INTS}, \texttt{NUM\_DELT\_EXTS} and \texttt{NUM\_OF\_EVENTS} are global variables, and store the total number of internal transition functions, external transition functions and events triggered inside the whole model. $\Delta_{\rm int}$  and $\Delta_{\rm ext}$ are the delays introduced in the internal and external transition functions, respectively.}

\vspace{0.25cm}
\textbf{function} [list,\emph{phase},$\sigma$] = init()
\STATE list = [] \COMMENT{list is part of the state, and stores all the events received by this atomic model}
\STATE $\sigma = \infty$

\vspace{0.25cm}
\textbf{function} [list,\emph{phase},$\sigma$] = $\delta_{\rm int}$(list,\emph{phase},$\sigma$)
\STATE \texttt{NUM\_DELT\_INTS} = \texttt{NUM\_DELT\_INTS} + 1
\STATE Dhrystone($\Delta_{\rm int}$)
\STATE list = []
\STATE $\sigma = \infty$

\vspace{0.25cm}
\textbf{function} [list,\emph{phase},$\sigma$] = $\delta_{\rm ext}$(list,\emph{phase},$\sigma$,$e$,$X^b$)
\STATE \texttt{NUM\_DELT\_EXTS} = \texttt{NUM\_DELT\_EXTS} + 1
\STATE Dhrystone($\Delta_{\rm ext}$)
\STATE values = $X^b(in)$ \COMMENT{$X^b(in)$ is a list containing all the events waiting in the ``in'' input port}
\STATE \texttt{NUM\_OF\_EVENTS} = \texttt{NUM\_OF\_EVENTS} + values.size()
\STATE list = [list;values] \COMMENT{Concatenate both lists}
\STATE \emph{phase = ``active''}
\STATE $\sigma = 0$

\vspace{0.25cm}
\textbf{function} [list,\emph{phase},$\sigma$] = $\delta_{\rm con}$(list,\emph{phase},$\sigma$,$ta(s)$,$X^b$)
\STATE $\delta_{\rm ext}$($\delta_{\rm int}$(list,\emph{phase},$\sigma$),0,$X^b$)

\vspace{0.25cm}
\textbf{function} $\lambda ()$
 \STATE send(``out'', list) \COMMENT{sends the whole list by the ``out'' output port}

\vspace{0.25cm}
\textbf{function} $\sigma$ = $ta$(list,\emph{phase},$\sigma$)
 \STATE $\sigma = \sigma$

\end{algorithmic}
\end{algorithm}

\subsection{LI (Low level of Interconnections) models}

\begin{figure}[ht]
\centering
\subfigure[DEVStone LI regular coupled component]{
\includegraphics[width=0.95\columnwidth]{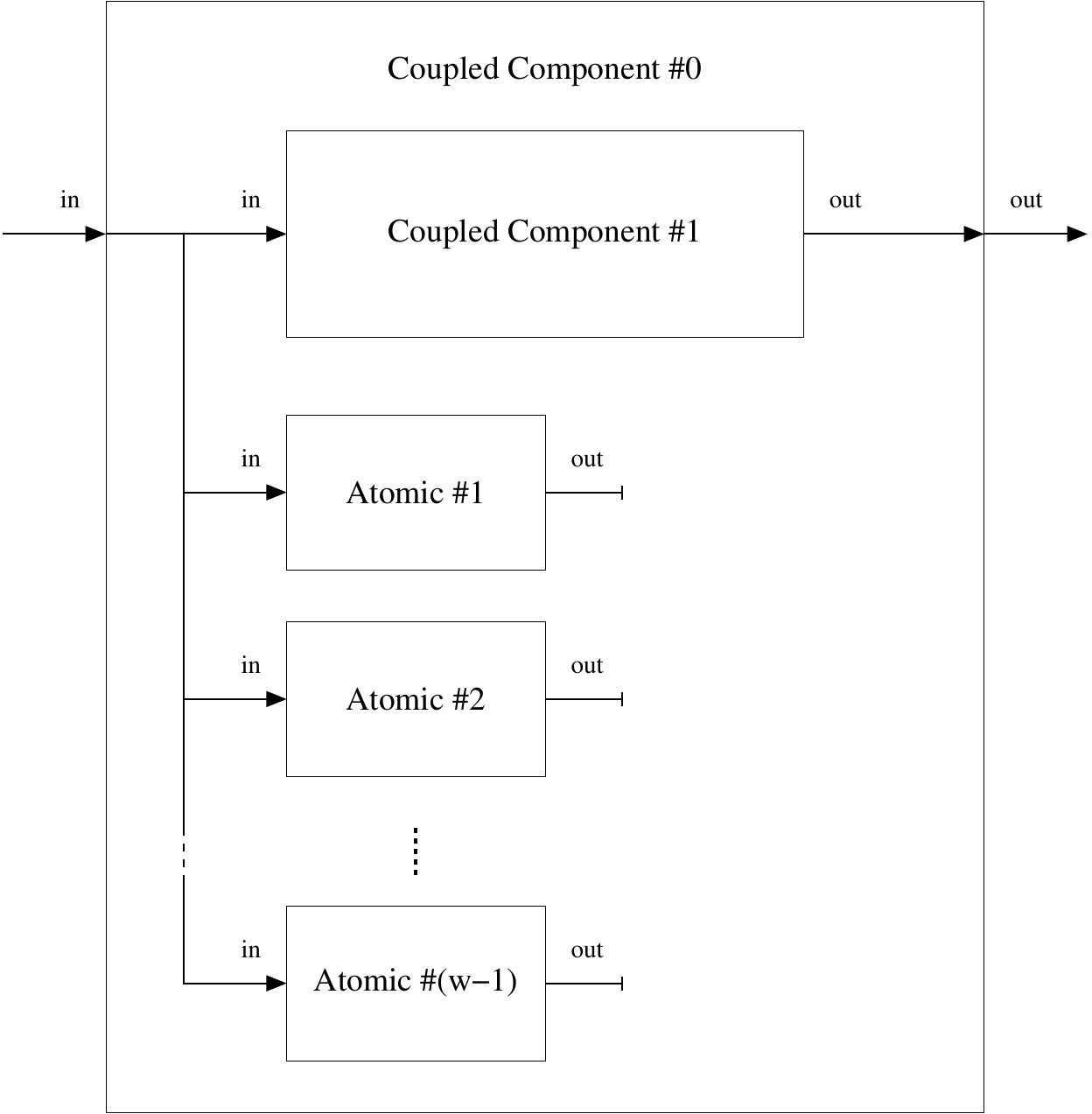}
\label{fig:LiRegularModel}
}
\subfigure[DEVStone LI deepest coupled component]{
\includegraphics[width=0.65\columnwidth]{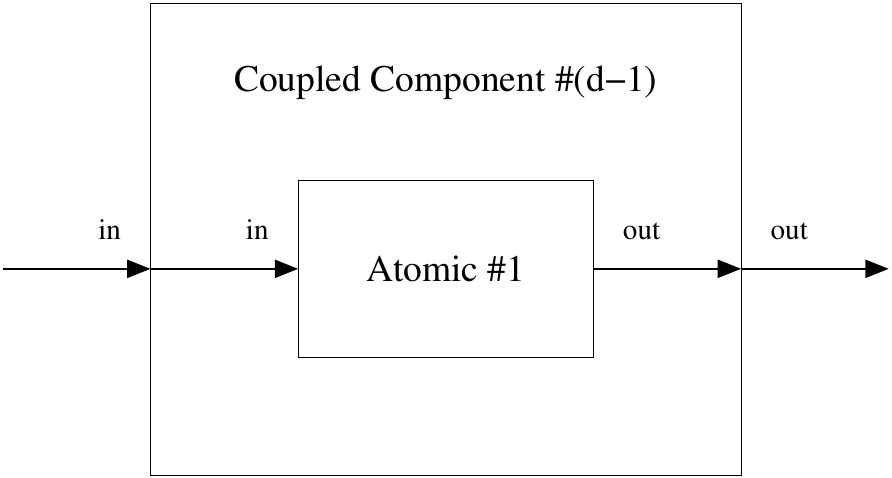}
\label{fig:LiDeepestModel}
}
\caption{DEVStone LI components}
\label{fig:LiModels}
\end{figure}

Figure~\ref{fig:LiModels} shows the general structure of an LI model. With $d$ layers (depth), the first $d-1$ (with $d \ge 1$) layers have the structure of Figure \ref{fig:LiRegularModel}. All these layers have one coupled model and $w-1$ (with $w \ge 2$) atomic models (where $w$ is the width). On the other hand, the $d$-th layer has the structure given in Figure \ref{fig:LiDeepestModel}, just with one atomic model. The arrows in the Figure represent the connection between the input and output ports in the whole model.

As stated above, two metrics are measured: execution time and memory footprint (also known as memory high-water mark). Obviously, these two metrics depend on the number of atomic models, number of internal transitions, number of external transitions, and the total number of events internally generated. Additionally, memory footprint depends on the concurrency of the model, that is, the number of pending events simultaneously waiting at the input ports.

Since the model structure is known, and the simplification $\Delta_\mathrm{int}=\Delta_\mathrm{ext}=0$ is made, the theoretical execution time and the total number of events generated can be easily computed.

Firstly, considering the model's $d-1$ levels with $w-1$ atomic models and 1 level with 1 atomic model, the total number of atomic models is:

\begin{equation}
\#{\rm Atomic} = (w-1) \cdot (d-1) + 1
\end{equation}

Secondly, LI models produce one external transition, output event and internal transition for each atomic model and external events injected. Thus, in LI models, the number of transitions and events generated is equal to the number of atomic models multiplied by the total number of external events injected $N$:

\begin{eqnarray}
\#\delta_{\rm int} & = & N \cdot \left( (w-1) \cdot (d-1) + 1 \right) \\
\#\delta_{\rm ext} & = & N \cdot \left( (w-1) \cdot (d-1) + 1 \right) \\
\#{\rm Events} & = & N \cdot \left( (w-1) \cdot (d-1) + 1 \right) 
\end{eqnarray}

In the following DEVStone benchmarks, we derive the equations for the number of transition functions and events internally generated given a single external event injected, i.e., for this benchmark:

\begin{eqnarray}
\#\delta_{\rm int} & = & (w-1) \cdot (d-1) + 1 \\
\#\delta_{\rm ext} & = & (w-1) \cdot (d-1) + 1 \\
\#{\rm Events} & = & (w-1) \cdot (d-1) + 1
\end{eqnarray}

\subsection{HI (High Input couplings) models}

\begin{figure}[ht]
\centering
\subfigure[DEVStone HI regular coupled component]{
\includegraphics[width=0.95\columnwidth]{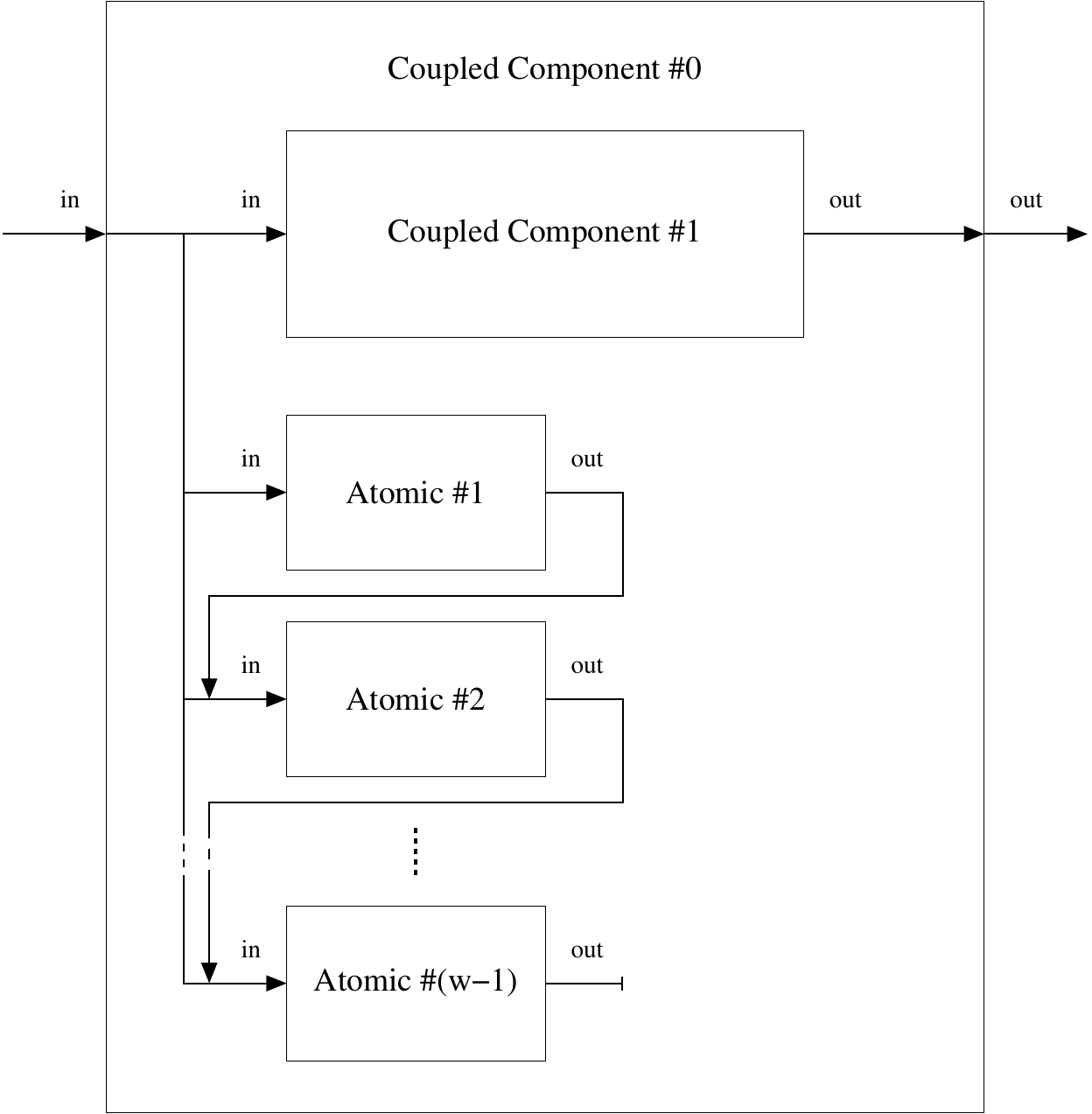}
\label{fig:HiRegularModel}
}
\subfigure[DEVStone HI deepest coupled component]{
\includegraphics[width=0.65\columnwidth]{atomic_LI.pdf}
\label{fig:HiDeepestModel}
}
\caption{DEVStone HI components}
\label{fig:HiModels}
\end{figure}

Figure~\ref{fig:HiModels} shows the general structure of a HI model. It is equal to the LI model, but where the output port of an atomic component $i$ is connected to the input port of the next atomic component $i+1$, as seen in Figure~\ref{fig:HiRegularModel}.

Therefore, the number of atomic models is equal to the LI model. However, the number of transition functions and events generated are quite different, because for each external input, the set of $w-1$ atomic models acts as a shift register, generating one additional event for each external event. As a result, the number of atomic models, transition functions and events generated is computed as follows:

\begin{eqnarray}
\#{\rm Atomic} & = & (w-1) \cdot (d-1) + 1 \\
\#\delta_{\rm int} & = & \left((w-1)+\sum_{i=1}^{w-2}i \right) \cdot (d-1) + 1 \nonumber \\ 
 & = & \left(\frac{w^2 -w}{2}\right) \cdot (d-1) + 1 \\
\#\delta_{\rm ext} & = & \left((w-1)+\sum_{i=1}^{w-2}i \right) \cdot (d-1) + 1 \nonumber \\ 
 & = & \left(\frac{w^2 -w}{2}\right) \cdot (d-1) + 1 \\
\#{\rm Events} & = & \left((w-1)+\sum_{i=1}^{w-2}i \right) \cdot (d-1) + 1 \nonumber \\ 
 & = & \left(\frac{w^2 -w}{2}\right) \cdot (d-1) + 1 
\end{eqnarray}

\subsection{HO (Hi model with numerous Outputs) models}

\begin{figure}[ht]
\centering
\subfigure[DEVStone HO regular coupled component]{
\includegraphics[width=0.95\columnwidth]{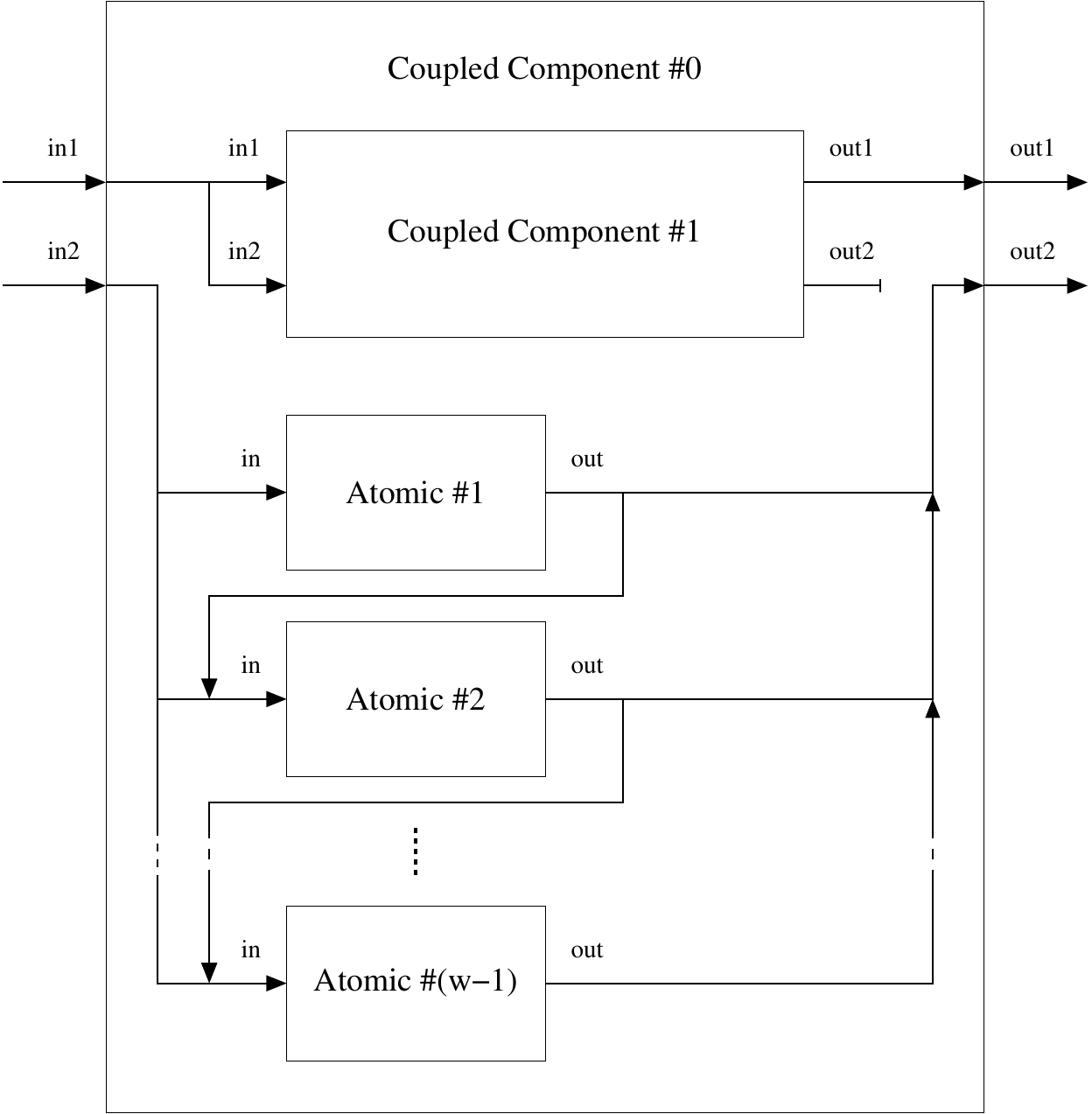}
\label{fig:HoRegularModel}
}
\subfigure[DEVStone HO deepest coupled component]{
\includegraphics[width=0.65\columnwidth]{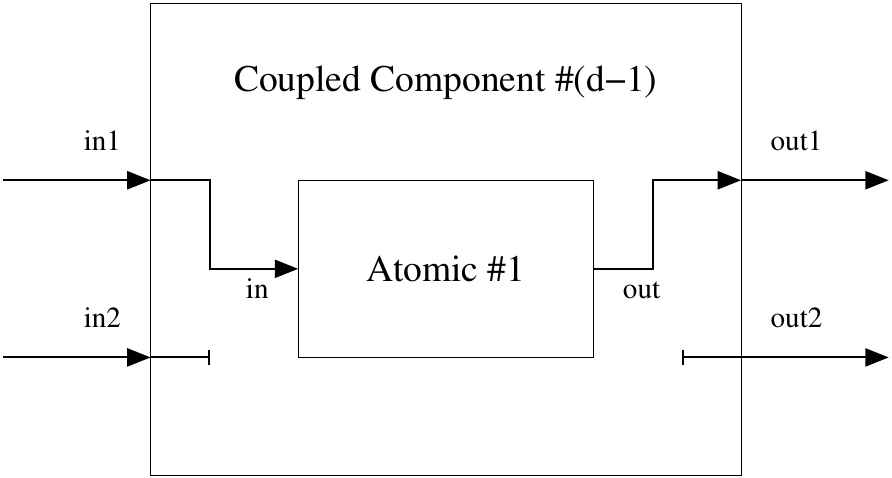}
\label{fig:HoDeepestModel}
}
\caption{DEVStone HO components}
\label{fig:HoModels}
\end{figure}

Figure~\ref{fig:HoModels} shows the general structure of a HO
model. HO has a more complex interconnection map with the same number
of atomic and coupled components. For example, HO coupled components
have two input and two output ports in each level. The main
differences with HI are that the second input port of each coupled model is connected to the input of each atomic model. Additionally, the output of each atomic model is connected to the second output of its parent coupled model.

It is worthwhile to mention that the number of atomic models, transition functions and events generated in HO models are exactly the same as in the HI model. However, the main difference is in both the execution time and memory footprint, which are higher due to the additional external input connections. Thus,

\begin{eqnarray}
\#{\rm Atomic} & = & (w-1) \cdot (d-1) + 1 \\
\#\delta_{\rm int} & = & \left((w-1)+\sum_{i=1}^{w-2}i \right) \cdot (d-1) + 1 \\
\#\delta_{\rm ext} & = & \left((w-1)+\sum_{i=1}^{w-2}i \right) \cdot (d-1) + 1  \\
\#{\rm Events} & = & \left((w-1)+\sum_{i=1}^{w-2}i \right) \cdot (d-1) + 1
\end{eqnarray}


\subsection{HOmod models}

\begin{figure}[ht]
\centering
\subfigure[DEVStone HOmod regular coupled component]{
\includegraphics[width=0.95\columnwidth]{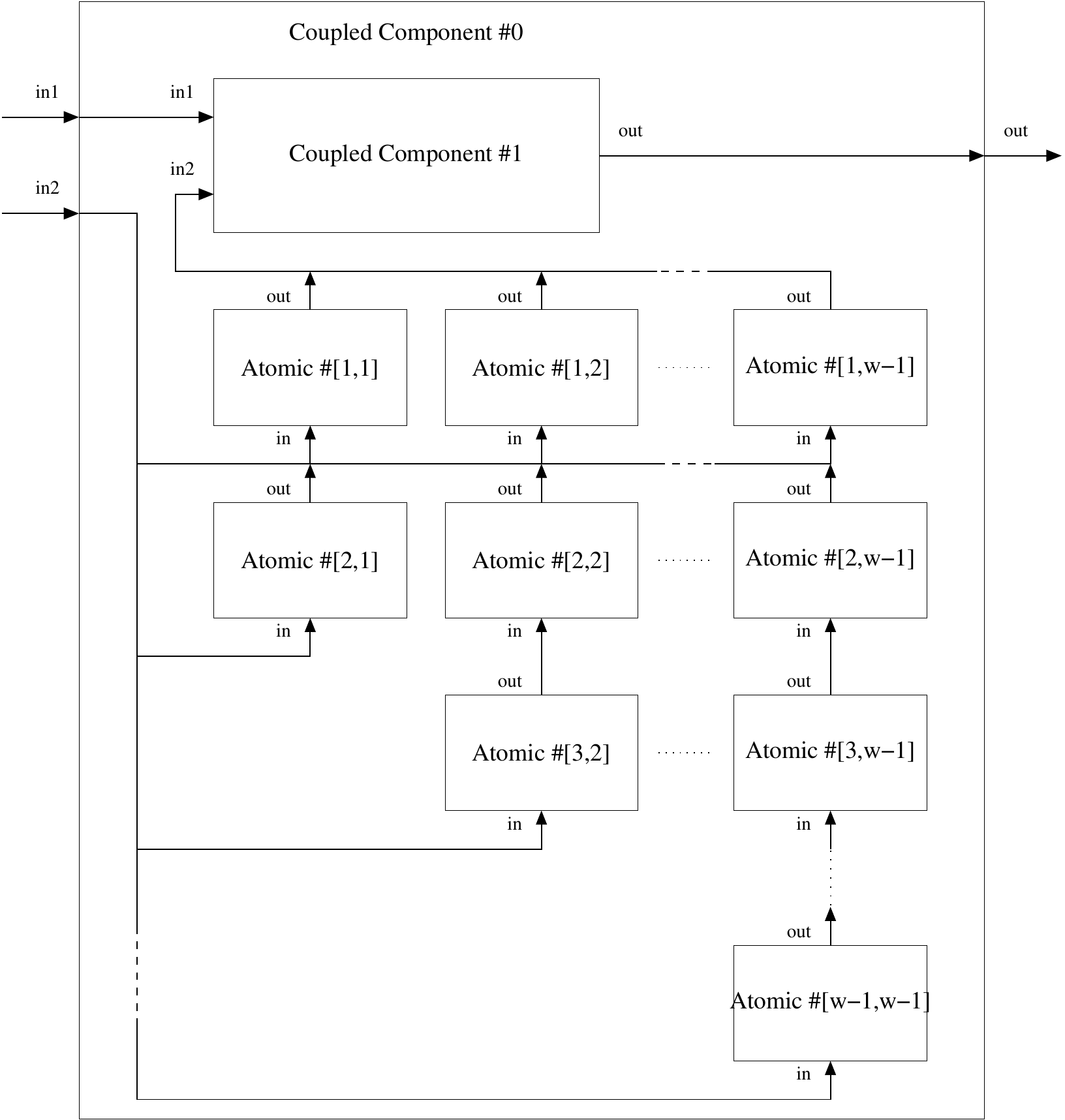}
\label{fig:HoModRegularModel}
}
\subfigure[DEVStone HOmod deepest coupled component]{
\includegraphics[width=0.65\columnwidth]{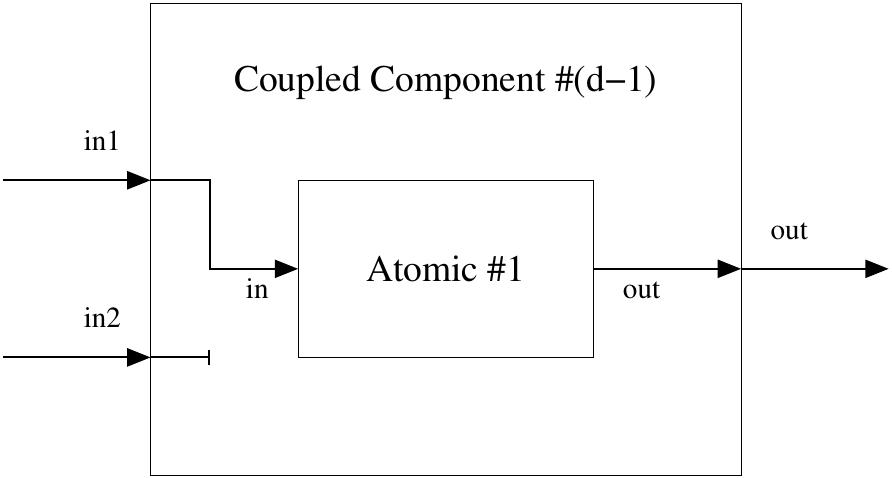}
\label{fig:HoModDeepestModel}
}
\caption{DEVStone HOmod components}
\label{fig:HoModModels}
\end{figure}

Figure~\ref{fig:HoModModels} depicts the structure of a HOmod DEVStone model. As usual, the deepest coupled model is formed by one single atomic model. The remaining coupled models are constituted by 1 coupled model, a chain of $w-1$ atomic models, and a set of $k = 1 \ldots w-1$ chains formed by $\sum_{i=1}^{k}i$ atomic models. The second external input port is connected to the whole first row and only to the first atomic component in the remaining rows. Additionally, all the atomic models in the second row are connected to the first row, which in turn send the whole output directly to the coupled component. Finally, each remaining atomic component is connected to its upper component.

The computation of the number of atomic modes is quite straightforward. 

\begin{eqnarray}
 \#{\rm Atomic} & = & \left((w-1) + \sum_{i=1}^{w-1}i \right)  \cdot (d-1) +1 \label{eq:homod1}
\end{eqnarray}

However, the calculation of the number of transition functions is hard. After an exhaustive mathematical analysis we have determined that:

\begin{eqnarray}
 \#{\rm Atomic} & = & \left((w-1) + \sum_{i=1}^{w-1}i \right)  \cdot (d-1) +1 \\
 \#\delta{\rm int} & = & (d-1) \cdot (w-1)^2 + \nonumber \\
  &  & + \left( (d-1) + (w-1) \cdot \sum_{i=1}^{d-2}i \right) \nonumber \\
  &  & \times \left( (w-1) + \sum_{i=1}^{w-1}i \right) + 1 \\
 \#\delta{\rm ext} & = & (d-1) \cdot (w-1)^2 \nonumber \\
  & & + \left( (d-1) + (w-1) \cdot \sum_{i=1}^{d-2}i \right) \nonumber \\
  & & \times \left( (w-1) + \sum_{i=1}^{w-1}i \right) + 1
\end{eqnarray}

Similarly, the computation of the number of events follows a recursive equation, defined below:

\begin{eqnarray}
  \#{\rm Events} & = & \sum_{l=1}^{d-1} ( \sum_{c=1}^{K_l+w-1} ( W_1 \times \sum_{i=1}^{w}{P_l}^{c-i+1} \nonumber \\
  & & + \sum_{i=1}^{w} ( W_i \cdot {P_l}^{c-i+1} ) ) ) + 1
\end{eqnarray}

\noindent where:

\begin{eqnarray}
W_i  & = & \left\{ \begin{array}{rl}
 w-i &\mbox{ if $w-i\ge 0$} \\
  0 &\mbox{ otherwise}
       \end{array} \right.
\end{eqnarray}
\begin{eqnarray}
K_l  & = & \left\{ \begin{array}{rl}
 1 &\mbox{ if $l=1$} \\
  K_{l-1}+W_1 &\mbox{ if $l>1$}
       \end{array} \right.
\end{eqnarray}

\noindent and

\begin{eqnarray}
 P_1^1 & = & 1 \\
 P_l^j & = & 0 \ {\rm if } \ 1 > j > K_l \\
  P_{l}^j & = & (w-1) \times \sum_{i=1}^{w} P_{l-1}^{j-i+1} \label{eq:homod2}
\end{eqnarray}

As can be seen, the complexity of the equations describing the metrics of HOmod is high. The inclusion of these equations in a simulator is hard, and the theoretical analysis becomes prohibitive. For these reasons, we have defined a new DEVStone benchmark named HOmem that, providing the same computational effort than HOmod into the different simulation frameworks, shows a straightforward mathematical formulation.


\subsection{HOmem models}

As stated above, we propose the inclusion of a new model in the DEVStone benchmark called HOmem. HOmem is basically proposed as a mechanism to increment the traffic of events with respect to HO, equivalently to HOmod, but with a simpler structure and mathematical description.

\begin{figure}[ht]
\centering
\subfigure[DEVStone HOmem regular coupled component]{
\includegraphics[width=0.95\columnwidth]{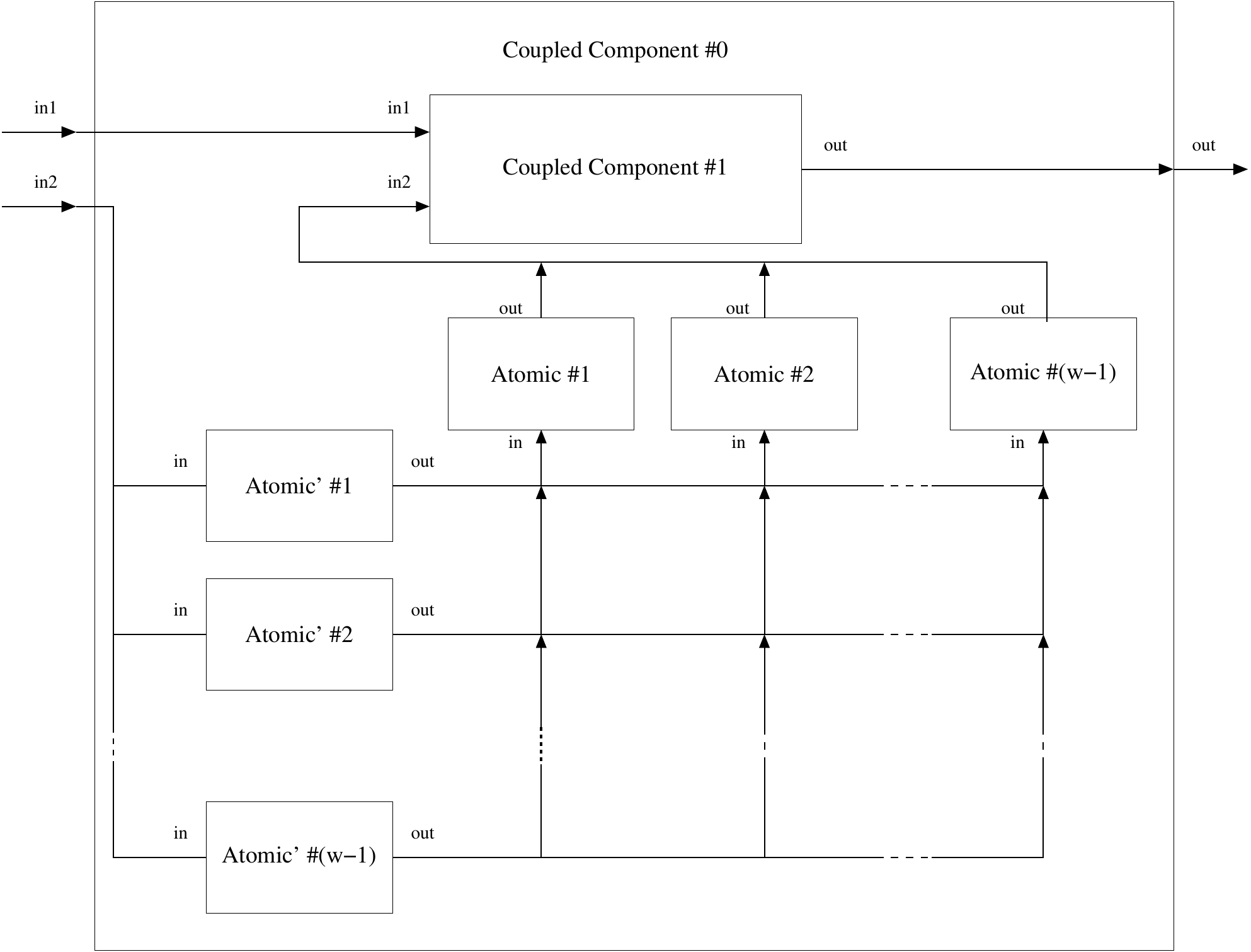}
\label{fig:HoMemRegularModel}
}
\subfigure[DEVStone HOmem deepest coupled component]{
\includegraphics[width=0.65\columnwidth]{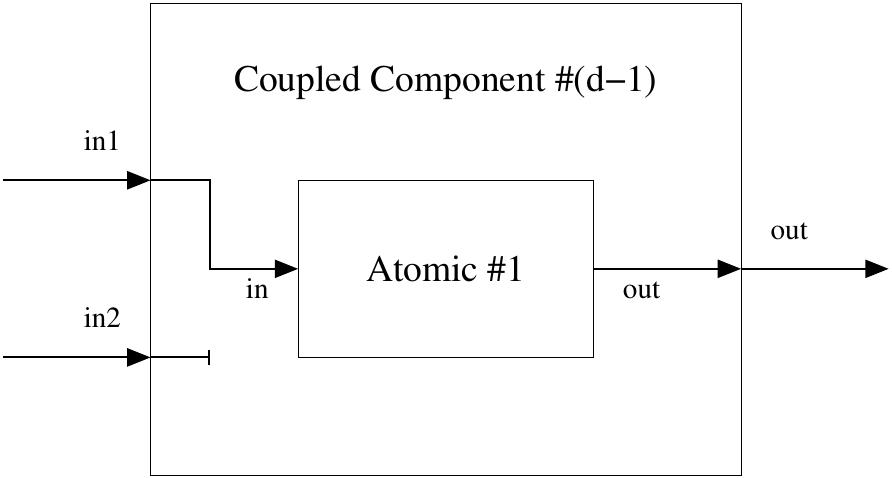}
\label{fig:HoMemDeepestModel}
}
\caption{DEVStone HOmem components}
\label{fig:HoMemModels}
\end{figure}

Figure~\ref{fig:HoMemModels} shows the structure of the HOmem DEVStone benchmark. As can be seen, the deepest coupled model is identical to HOmod. As for the remaining coupled models, each one is formed by $1$ coupled model and $2\cdot(w-1)$ atomic models. The second $w-1$ chain receives the input through external input connections, and propagates these inputs to the first chain of $w-1$ atomic models. These, in turn, send all the inputs received to the coupled model. 

The number of transition functions are easy to compute, since it is equal to the number of atomic models. However, to calculate the number of events it must be taken into account that each single event is sent $w-1$ times to the whole second chain of atomic models. This grows exponentially with the depth of the model, in the following form:

\begin{eqnarray}\label{eq:homem1}
\#{\rm Atomic} & = & 2\cdot (w-1) \cdot (d-1) + 1 \\ 
\#\delta_{\rm int} & = & 2\cdot (w-1) \cdot (d-1) + 1 \\
\#\delta_{\rm ext} & = & 2\cdot (w-1) \cdot (d-1) + 1 \\
\#{\rm Events} & = & \sum_{l=1}^{d-1} \left((w-1)^{2\cdot l} + (w-1)^{2\cdot l -1} \right) \nonumber \\
 & & + 1 \label{eq:homem2}
\end{eqnarray}

Experimental results show that this straightforward specification leads to similar execution time and memory footprint, when compared to HOmod.

\section{Experimental Methodology}\label{sec:ExpMethodology}
Once the DEVStone equations have been analytically derived, we compare CPU execution time and memory footprint over a total of five well known simulation engines using all the benchmarks presented above. Our aim is to show an exhaustive comparison and a standard procedure to evaluate the performance of any new discrete event simulator. 

We first provide a detailed description of the experimental set-up used in this research. 

All the benchmarks presented above (LI, HI, HO, HOmod and HOmem) were executed using five simulation engines: aDEVS 2.8.1, CD++ 2.45 (a CD++ branch with support for PDEVS), DEVSJAVA 3.1, xDEVS 1.20151013 and PyPDEVS 2.2.4. Table \ref{tab:Structures} shows the programming language and the main data structures used in each simulation engine. As stated above, CD++ and aDEVS are C++ implementations, DEVSJAVA and xDEVS have been implemented using Java, whereas PyPDEVS is a Python simulation engine. As Table \ref{tab:Structures} shows, aDEVS and xDEVS use generic classes, whereas PyPDEVS uses duck typing. To store events, aDEVS and CD++ use standard C++ arrays. On the contrary, DEVSJAVA, xDEVS and PyPDEVS use dynamic data structures, like linked lists or dictionaries. Finally, to store components and implement the simulation scheduler, all the frameworks use dynamic data structures such as sets or linked lists.

\begin{table*}[ht]
\centering
\begin{tabular}{l | ccccc}
\hline
 & \textbf{aDEVS} & \textbf{CD++} &  \textbf{DEVSJAVA} & \textbf{xDEVS} & \textbf{PyPDEVS} \\
\hline
\textbf{Programming Language}  & C++        & C++        & Java      & Java            & Python      \\
\textbf{Generics}              & Yes        & No         & No        & Yes             & Duck typing \\
\textbf{Events container}      & array/port & array/port & Hashtable & LinkedList/port & dictionary  \\
\textbf{Components container}  & std::set   & std::list  & HashSet   & LinkedList      & list        \\
\hline
\end{tabular}
\caption{Main data structures used in the simulation engines\label{tab:Structures}}
\end{table*}

We tested all these simulation engines in two different machines: a 48 GB AMD Opteron 6272 @ 2.1 GHz (abbreviated as AMD) and a 64 GB Intel Xeon 2670 @ 2.6GHz ``Sandy Bridge'' (abbreviated as Intel), in both cases under a GNU/Linux Debian 8 Operating System. aDEVS and CD++ were compiled using the \texttt{gcc -O3} optimization level.

In all the test cases, only one external event was injected, generating the total number of transition functions and the total number of events given in the previous equations. As demonstrated in~\cite{Wainer2011}~\cite{VanTendeloo2014}~\cite{Vicino2015}, the previous metric just scaled linearly with the number of external events.

Each benchmark type was generated for different values of width and depth. These values were defined for running different trials with all the five simulators. We  looked for a good trade-off between wall-clock simulation time and memory footprint, since these are the metrics measured in all the simulations. Table~\ref{tab:Range} shows these intervals, where each row represent a DEVStone benchmark type, in relation with the width and depth, each  described by the minimum value, the step size, and the maximum value used to generate a full range for these parameters. For example, the smallest LI model is a $2\times 1$ model, where $\mathrm{width}=2$ and $\mathrm{depth}=1$. The biggest model, on the other hand, is a $1502\times 1501$ model. 

Finally, each simulation is repeated 10 times for each simulator, benchmark, size, and hardware platforms. Simulation wall-clock time and memory footprint are averaged over these 10 trials. Although no significant deviations were appreciated, we kept this number of trials to avoid spurious deviations. Table~\ref{tab:Range} shows in the last column the total number of simulations performed.

\begin{table*}[ht]
\centering
\begin{tabular}{l | ccc | ccc | c}
\hline
 & \multicolumn{3}{c|}{\textbf{Width}} & \multicolumn{3}{c|}{\textbf{Depth}} &  \\
\textbf{Benchmark} & Min. & Step & Max. & Min. & Step & Max. & \textbf{\# Simulations}\\
\hline
LI & 2 & 100 & 1502 & 1 & 100 & 1501 & 25600 \\
HI & 2 & 100 & 1102 & 1 & 100 & 1101 & 14400 \\
HO & 2 & 100 & 1102 & 1 & 100 & 1101 & 14400 \\
HOmod & 2 & 1 & 10 & 1 & 1 & 10 & 9000 \\
HOmem & 2 & 1 & 10 & 1 & 1 & 10 & 9000 \\
\hline
\end{tabular}
\caption{Parameters configuration\label{tab:Range}}
\end{table*}

\section{Results}\label{sec:Results} 

\subsection{CPU comparison}

\begin{table*}[ht]
\centering
\begin{tabular}{l | cc | cc | cc }
\hline
 & \multicolumn{2}{c|}{\textbf{LI}} & \multicolumn{2}{c|}{\textbf{HI}} &  \multicolumn{2}{c}{\textbf{HO}} \\
\textbf{Simulator} & AMD & Intel & AMD & Intel & AMD & Intel \\
\hline
\multirow{2}{*}{aDEVS} & $2.5 \times 10^0$ & $\mathbf{2.1 \times 10^0}$ & $1.0 \times 10^3$ & $1.0 \times 10^3$ & $1.2 \times 10^3$ & $1.2 \times 10^3$ \\
 & 1.19 & 1.19 & 1.11 & 1.11 & 1.11 & 1.11 \\
\hline
\multirow{2}{*}{CD++} & $\infty$ & $\infty$ & $6.3 \times 10^3$ & $\mathbf{5.1 \times 10^3}$ & $7.0 \times 10^3$ & $\mathbf{4.5 \times 10^3}$ \\
 & $\infty$ & $\infty$ & 3.46 & 3.46 & 3.69 & 3.69 \\
\hline
\multirow{2}{*}{DEVSJAVA} & $\infty$ & $\infty$ & $6.6 \times 10^4$ & $\mathbf{4.0 \times 10^4}$ & $\infty$ & $\infty$  \\
 & $\infty$ & $\infty$ & 4.23 & 4.21 & $\infty$ & $\infty$ \\
\hline
\multirow{2}{*}{xDEVS} & $3.8 \times 10^0$ & $\mathbf{2.6 \times 10^0}$ & $9.3 \times 10^2$ & $\mathbf{4.6 \times 10^2}$ & $1.0 \times 10^3$ & $\mathbf{5.0 \times 10^2}$  \\
 & 1.95 & 2.07 & 1.88 & 1.84 & 1.94 & 1.84 \\
\hline
\multirow{2}{*}{PyPDEVS} & $\infty$ & $\infty$ & $\infty$ & $\infty$ & $\infty$ & $\infty$  \\
 & $\infty$ & $\infty$ & $\infty$ & $\infty$ & $\infty$ & $\infty$ \\
\hline
\end{tabular}
\caption{Execution time (seconds) and memory footprint (GiB) of the larger models executed by the five simulation engines and in both AMD and Intel servers\label{tab:AmdVsIntel}}
\end{table*}

Table~\ref{tab:AmdVsIntel} shows a comparison in execution time (in seconds, measured inside the simulator to avoid the loading time of the model) and memory footprint (in GiB, measured using the general GNU \texttt{time} command) for the five simulation engines and the largest models of the DEVStone models tested in this work, i.e., LI 1502$\times$1501, HI 1102$\times$1101, and HO 1102$\times$1101. HOmod 10$\times$10 and HOmem 10$\times$10 are not included because no simulator was able to finish them, at least during the 48 hours we run these tests. The same happended in all those cases in Table~\ref{tab:AmdVsIntel} marked with $\infty$. As can be seen, only aDEVS and xDEVS were able to finish all the models in Table~\ref{tab:AmdVsIntel}, followed by CD++, which was not able to load the largest LI model. Regarding memory footprint, there is not much difference between both servers. However, in terms of execution time, the best server in almost all cases was the 64 GB Intel Xeon 2670 @ 2.6GHz ``Sandy Bridge'', since between both servers, this one has the fastest processor and memory. Thus, simulation results are coherent with the server used, i.e., the faster the processor and the greater the memory size, the faster the simulation. Memory footprint is independent of the server, since it only depends on the internal structure of the DEVStone model.

As can be seen in Table~\ref{tab:AmdVsIntel}, some simulators were not able to execute the model because the system was unable to handle the memory requirements. To tackle these issues in the remaining analysis, the wall clock execution time is limited to 1200 seconds and the memory footprint to 4 GiB, enough to perform our more than 70000 simulations in a reasonable amount of time, also obtaining significant values to compare. Thus, in the following, every experiment with time or memory greater than the aforementioned values is truncated to 1200 seconds or 4 GiB, respectively.

\subsection{Execution time}

\begin{figure*}[ht]
\centering
\subfigure[aDEVS - LI]{
\includegraphics[width=0.45\textwidth, height=3cm]{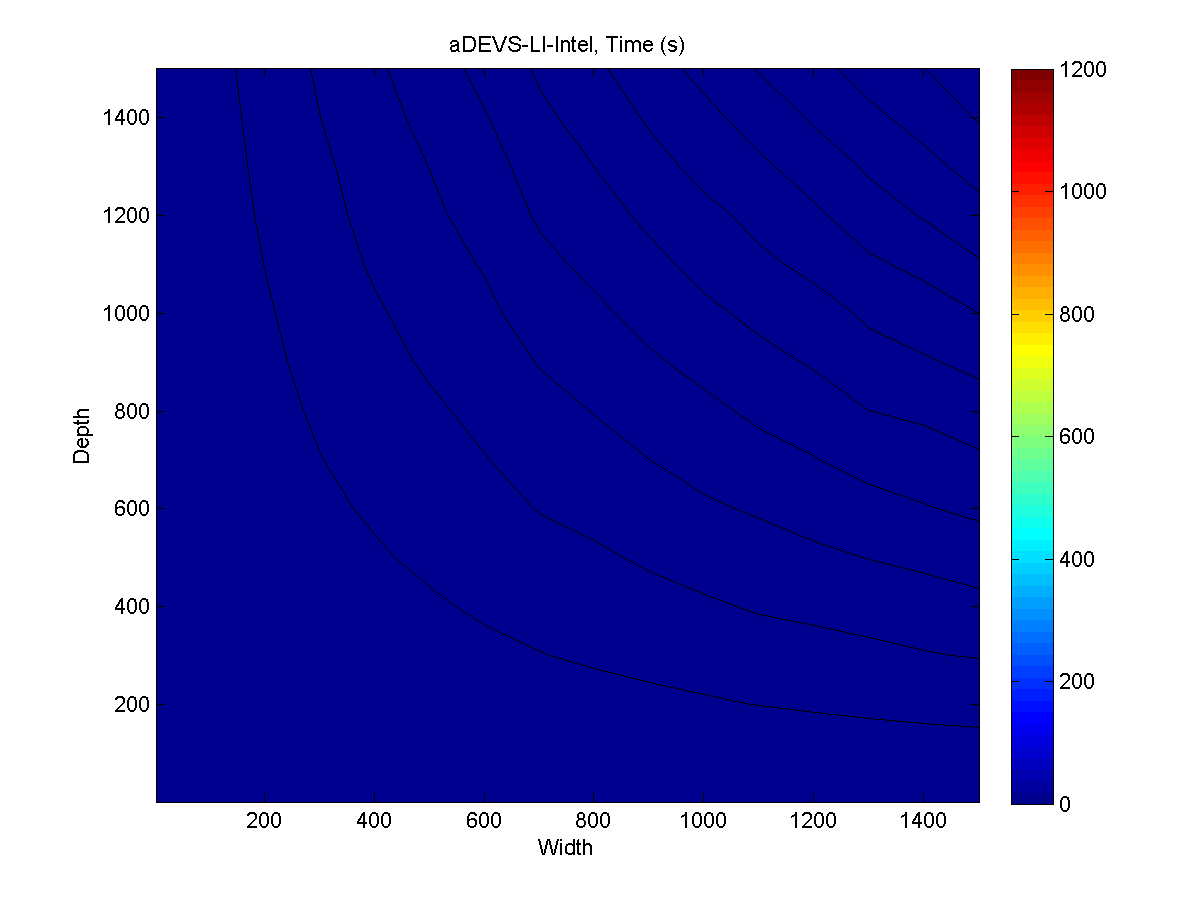}
\label{fig:aDevsLiIntelTime}
}
\subfigure[aDEVS - HI]{
\includegraphics[width=0.45\textwidth, height=3cm]{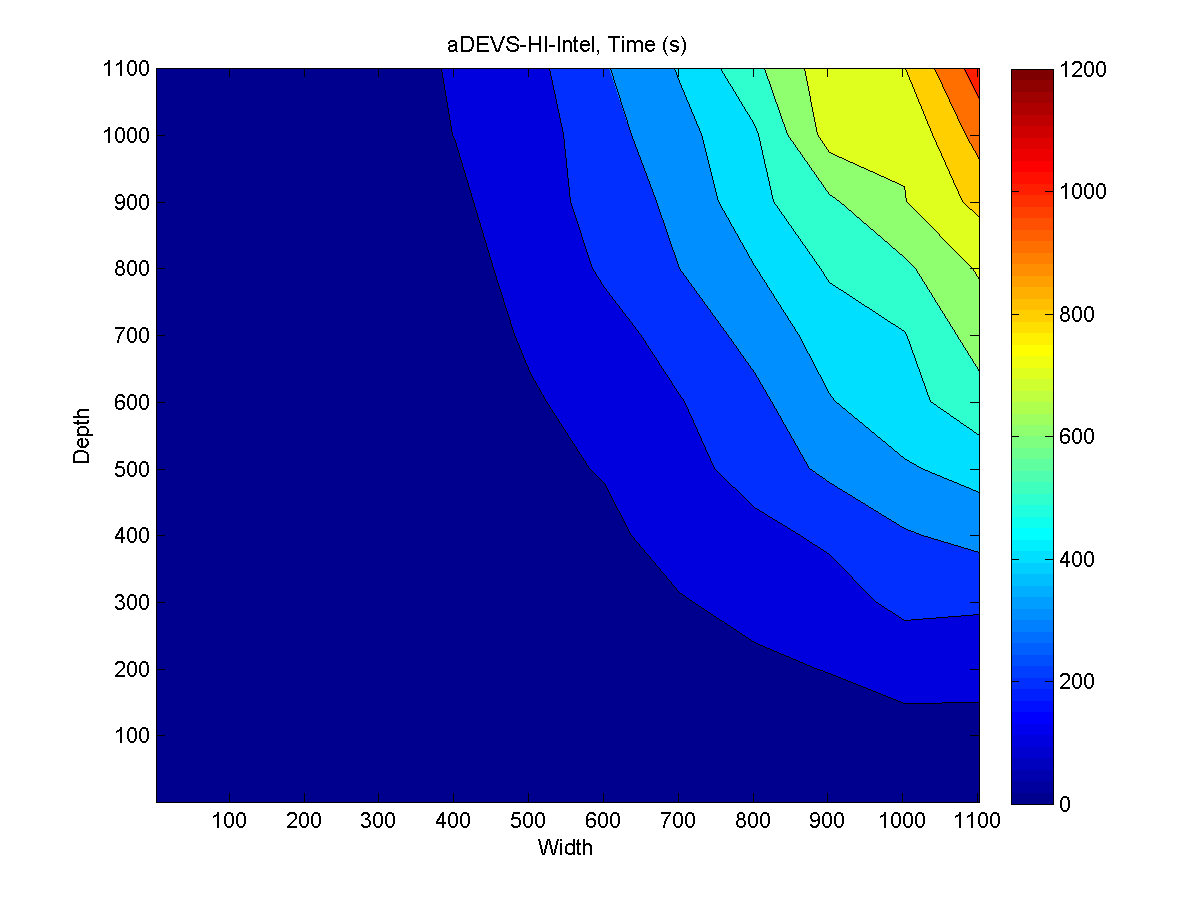}
\label{fig:aDevsHiIntelTime}
}
\subfigure[CD++ - LI]{
\includegraphics[width=0.45\textwidth, height=3cm]{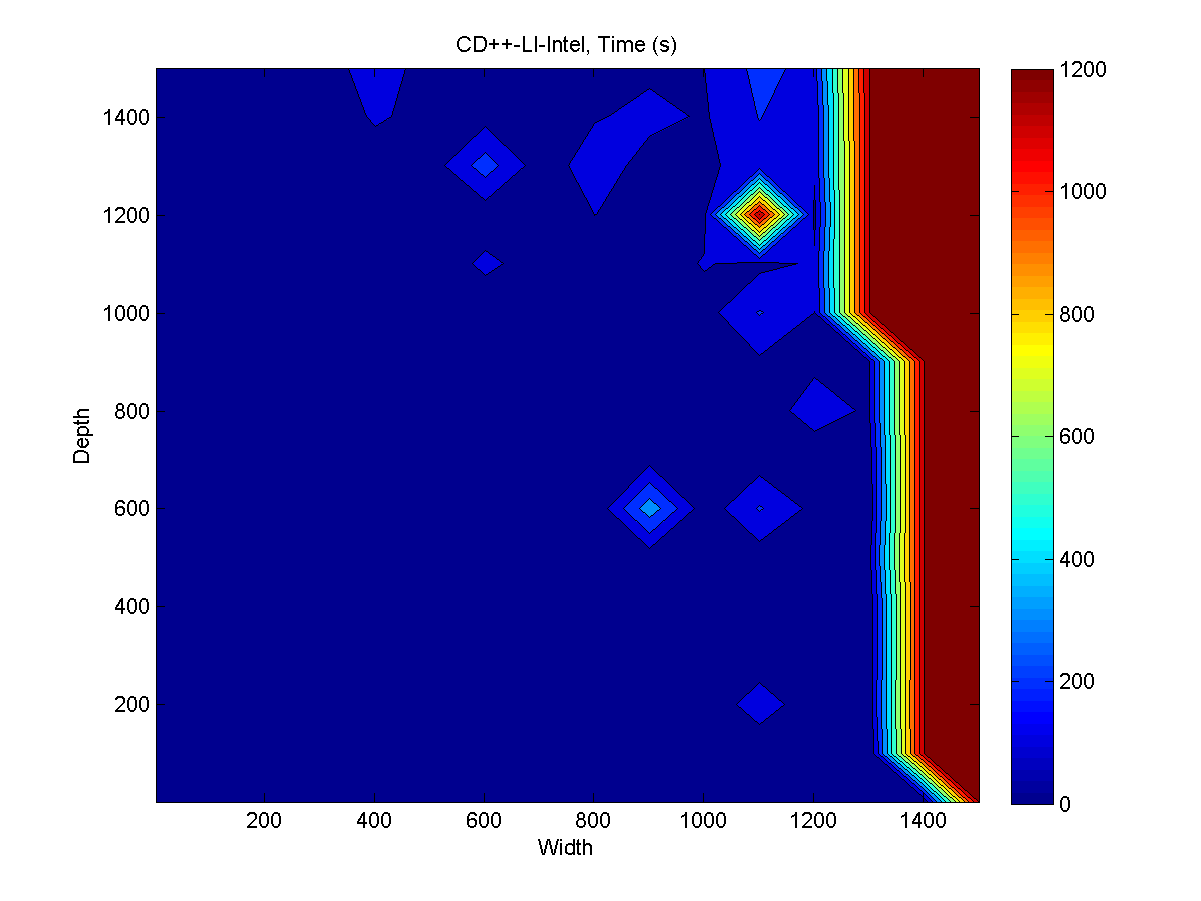}
\label{fig:CdPpLiIntelTime}
}
\subfigure[CD++ - HI]{
\includegraphics[width=0.45\textwidth, height=3cm]{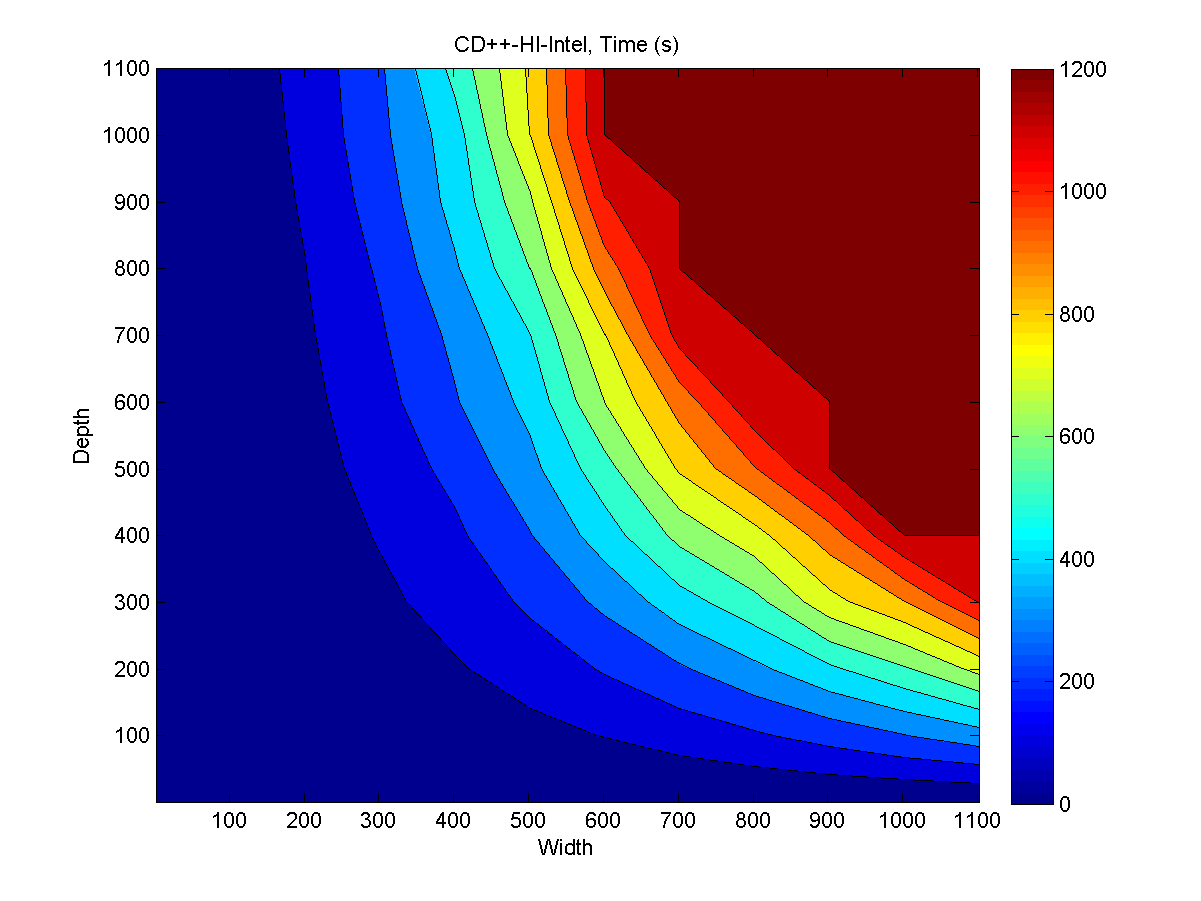}
\label{fig:CdPpHiIntelTime}
}
\subfigure[DEVSJAVA - LI]{
\includegraphics[width=0.45\textwidth, height=3cm]{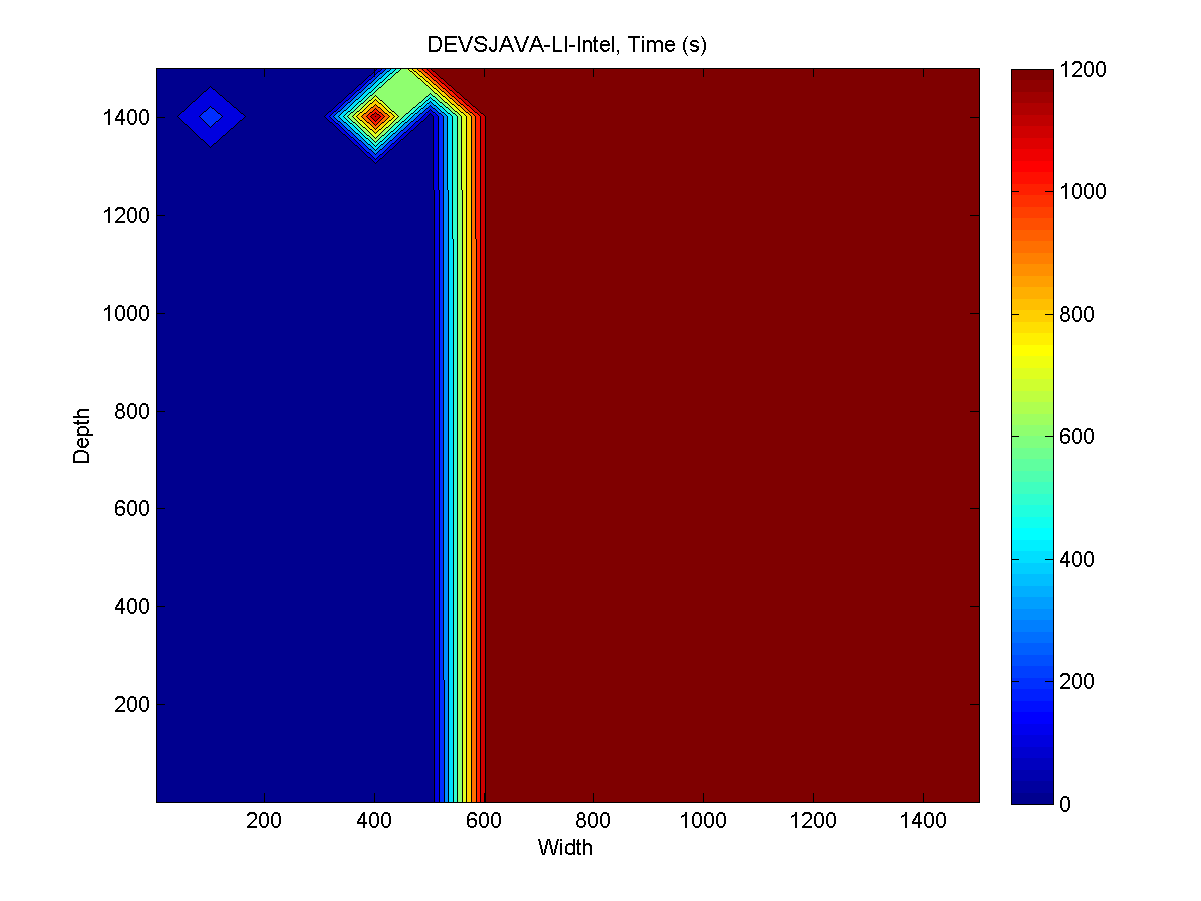}
\label{fig:DevsJavaLiIntelTime}
}
\subfigure[DEVSJAVA - HI]{
\includegraphics[width=0.45\textwidth, height=3cm]{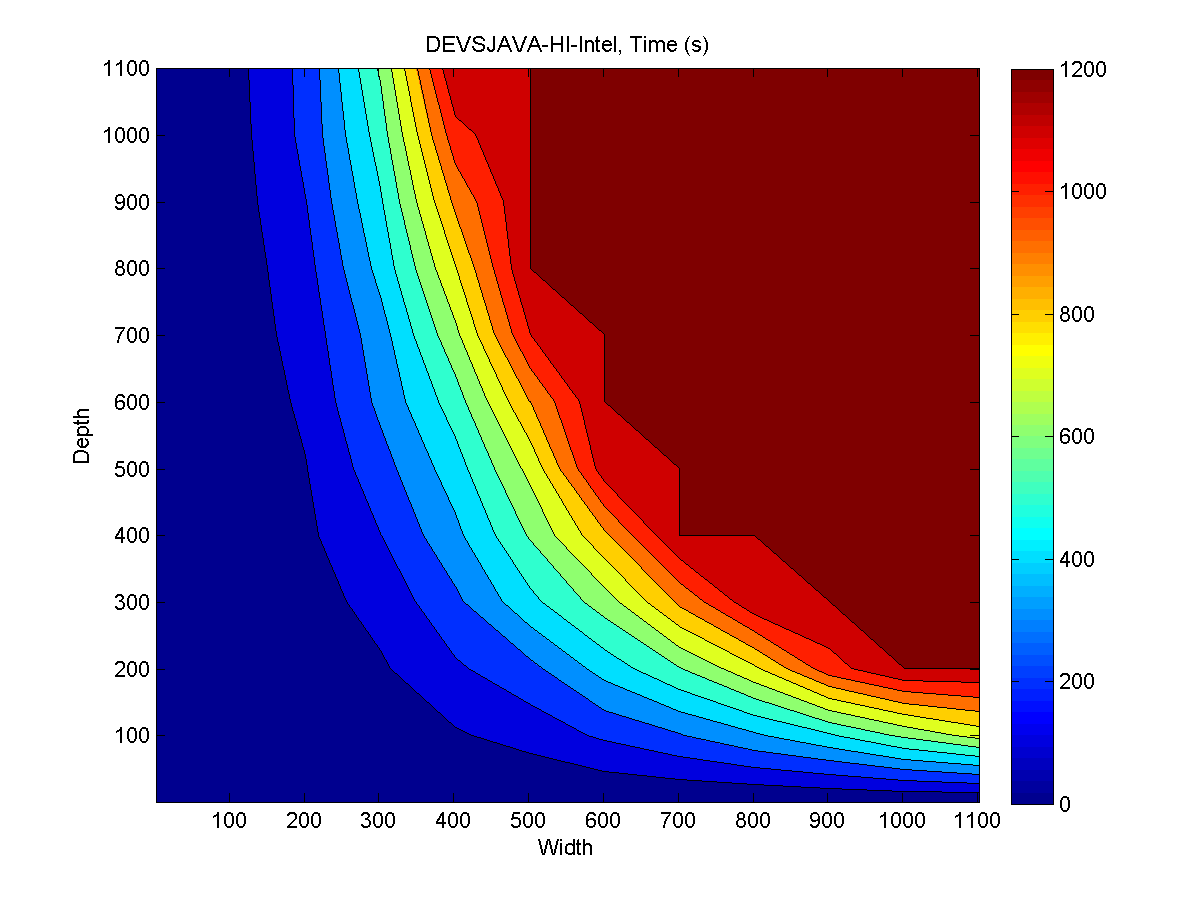}
\label{fig:DevsJavaHiIntelTime}
}
\subfigure[xDEVS - LI]{
\includegraphics[width=0.45\textwidth, height=3cm]{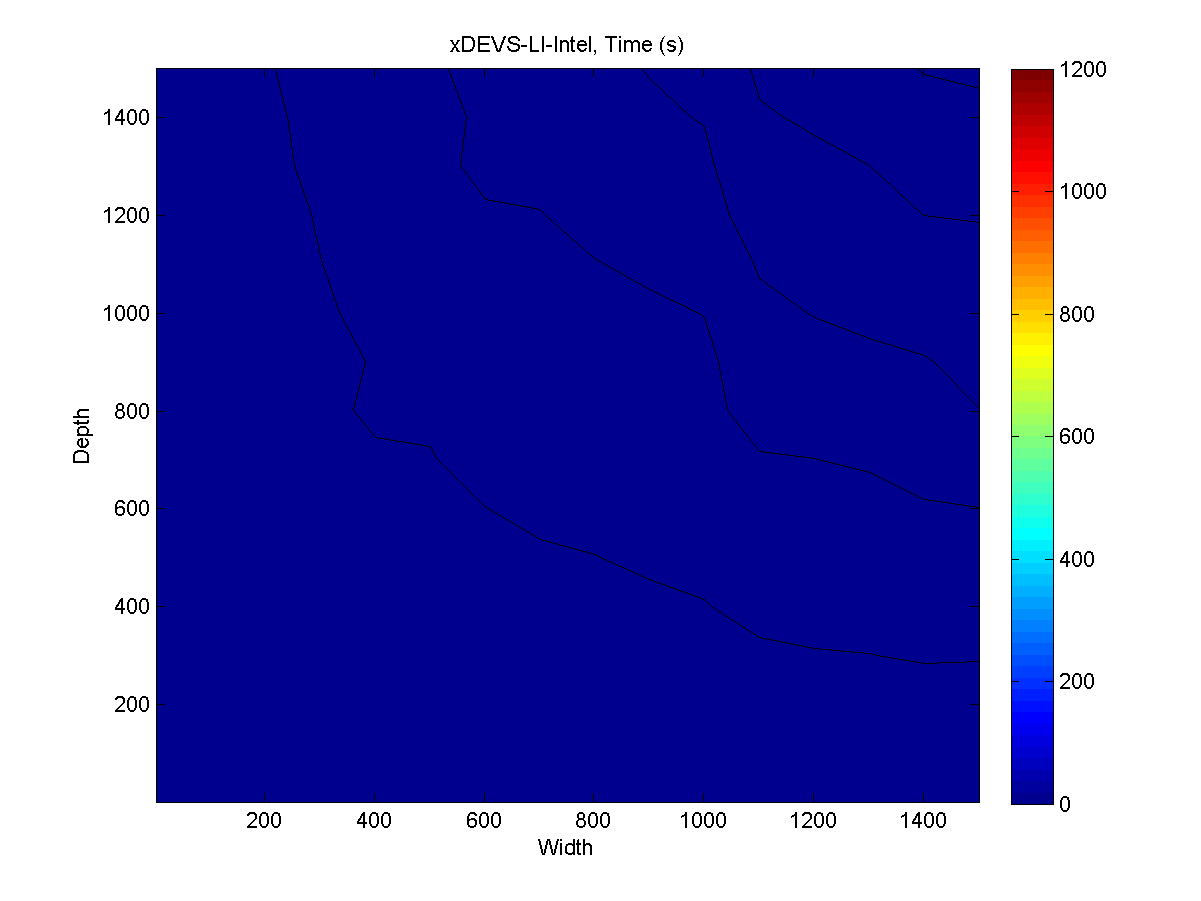}
\label{fig:xDEVSLiIntelTime}
}
\subfigure[xDEVS - HI]{
\includegraphics[width=0.45\textwidth, height=3cm]{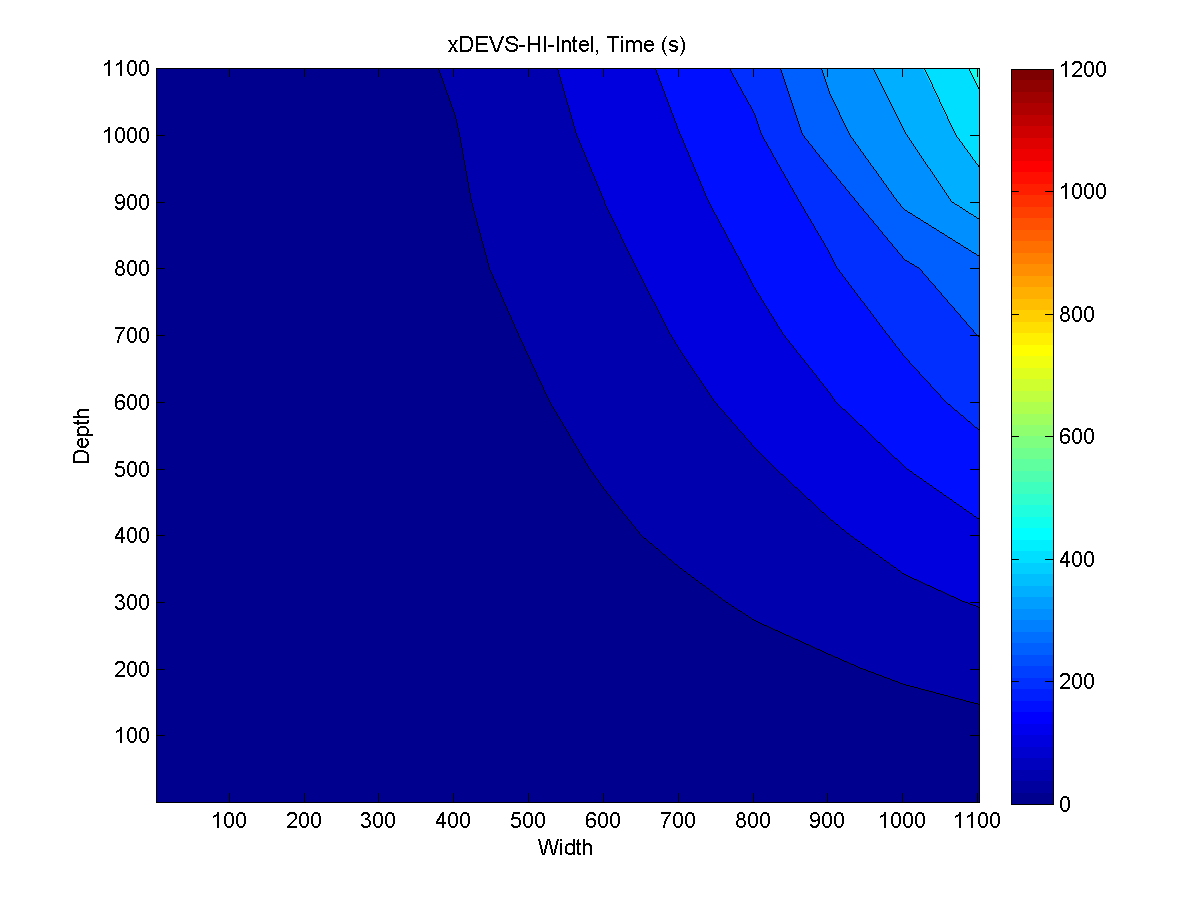}
\label{fig:xDEVSHiIntelTime}
}
\subfigure[PyPDEVS - LI]{
\includegraphics[width=0.45\textwidth, height=3cm]{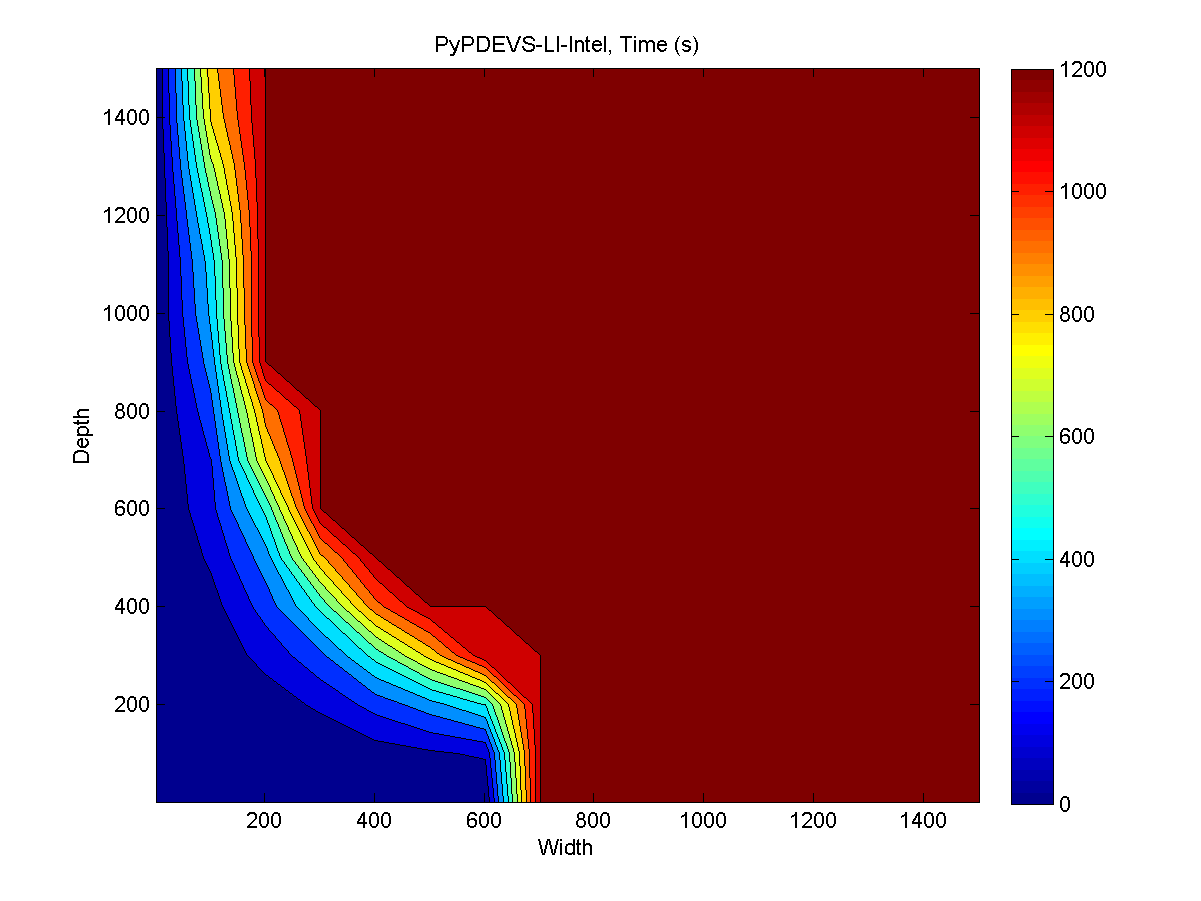}
\label{fig:PyDevsLiIntelTime}
}
\subfigure[PyPDEVS - HI]{
\includegraphics[width=0.45\textwidth, height=3cm]{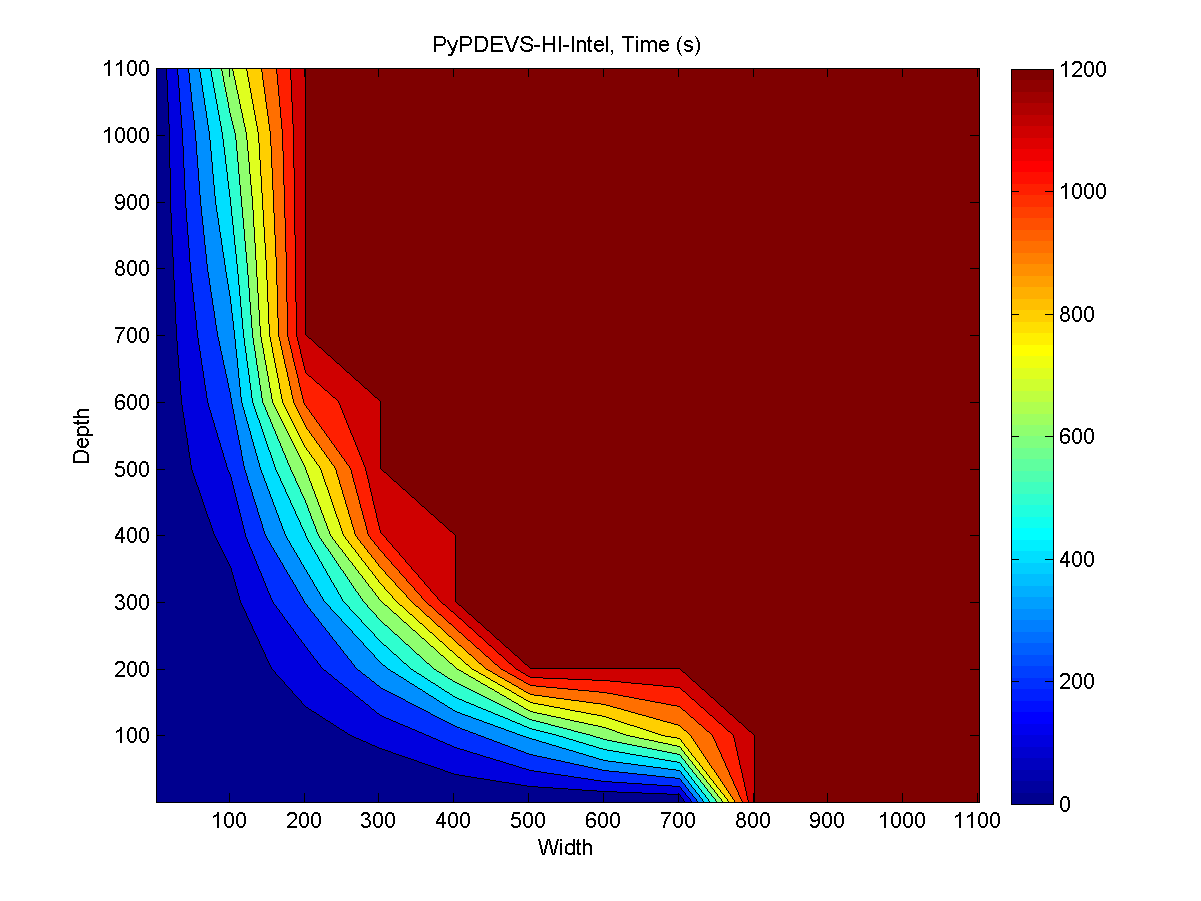}
\label{fig:PyDevsHiIntelTime}
}
\caption{Execution time of LI and HI models}
\label{fig:AllLiHiIntelTime}
\end{figure*}

Figure \ref{fig:AllLiHiIntelTime} shows the contour maps of the different execution times needed by all the five simulators in both LI and HI models. Blue regions mean low execution time, whereas red regions mean high execution time.

CD++, DEVSJAVA and PyPDEVS saturated the execution time of 1200 seconds multiple times in both models. aDEVS and xDEVS, on the contrary, reached the best results. 

Regarding the LI model, the ordered list of simulators, from best to worst contour maps is: aDEVS, xDEVS, CD++, DEVSJAVA and PyPDEVS. 

With respect to the HI model, the list is: xDEVS, aDEVS, CD++, DEVSJAVA and PyPDEVS. 

\begin{figure*}[ht]
\centering
\subfigure[aDEVS - HO]{
\includegraphics[width=0.45\textwidth, height=3cm]{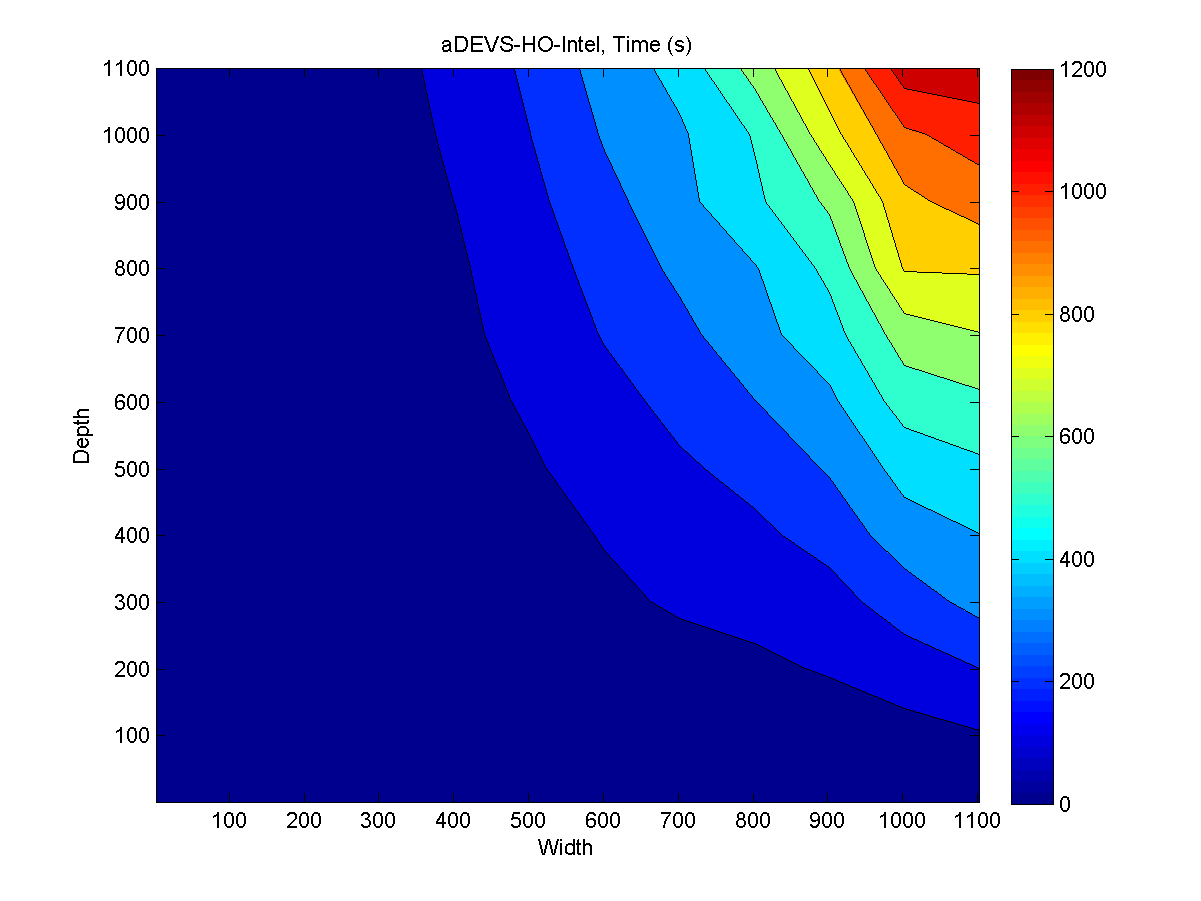}
\label{fig:aDevsHoIntelTime}
}
\subfigure[aDEVS - HOmem]{
\includegraphics[width=0.45\textwidth, height=3cm]{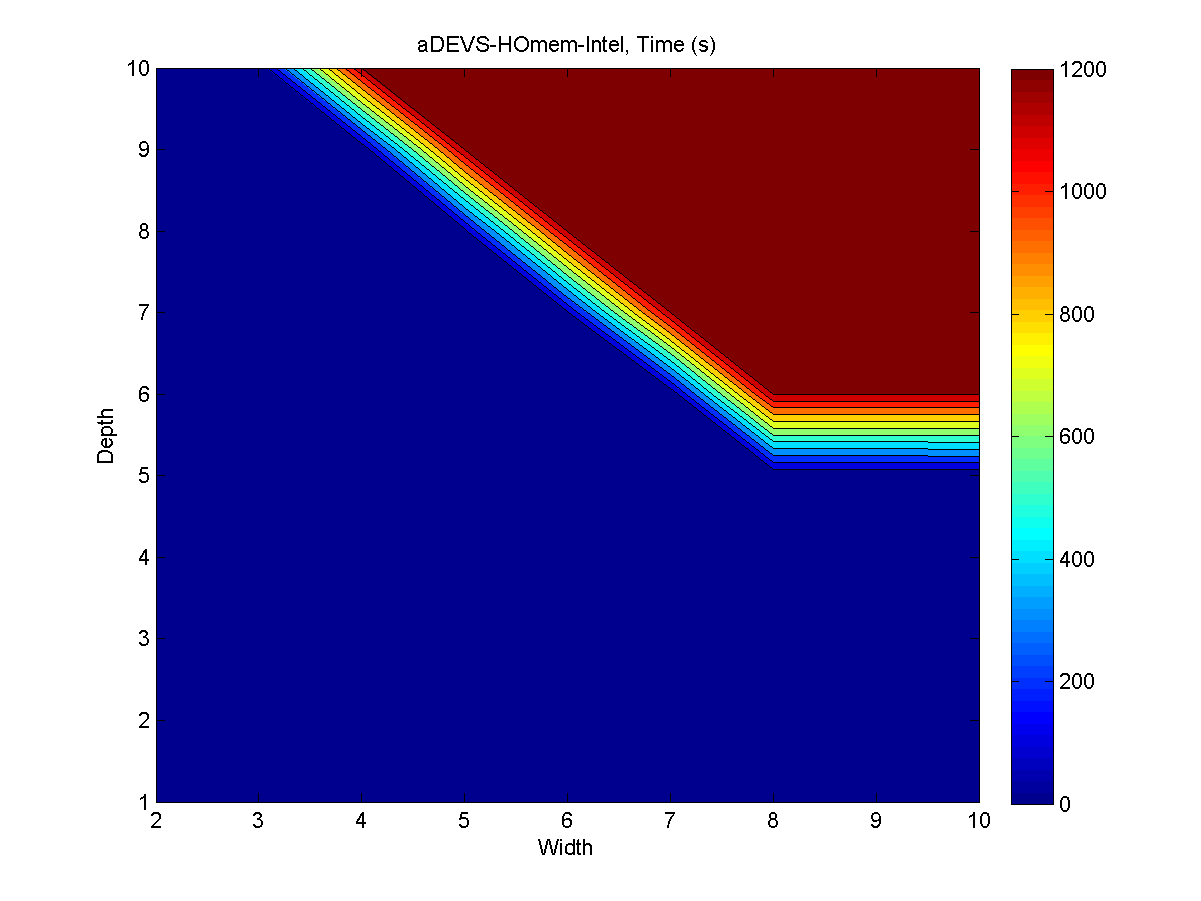}
\label{fig:aDevsHoMemIntelTime}
}
\subfigure[CD++ - HO]{
\includegraphics[width=0.45\textwidth, height=3cm]{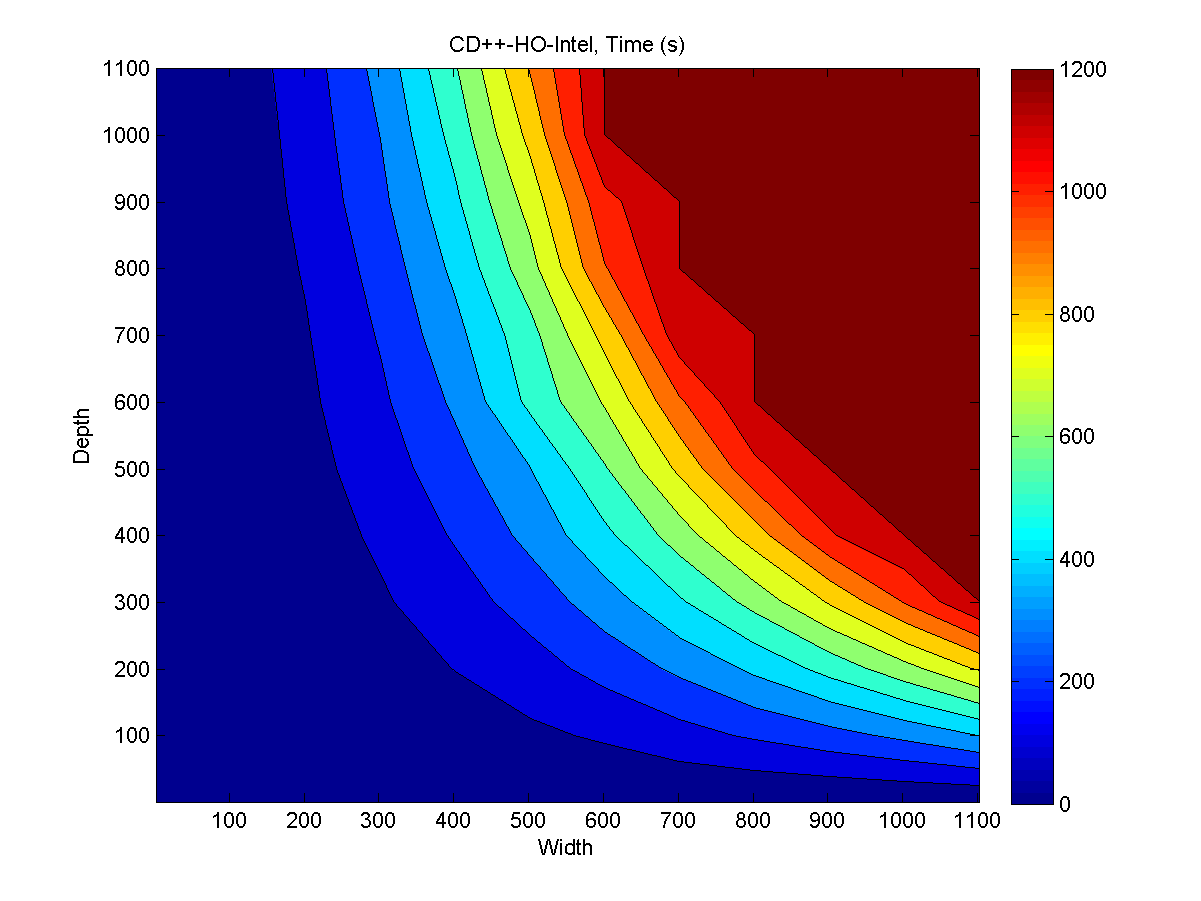}
\label{fig:CdPpHoIntelTime}
}
\subfigure[CD++ - HOmem]{
\includegraphics[width=0.45\textwidth, height=3cm]{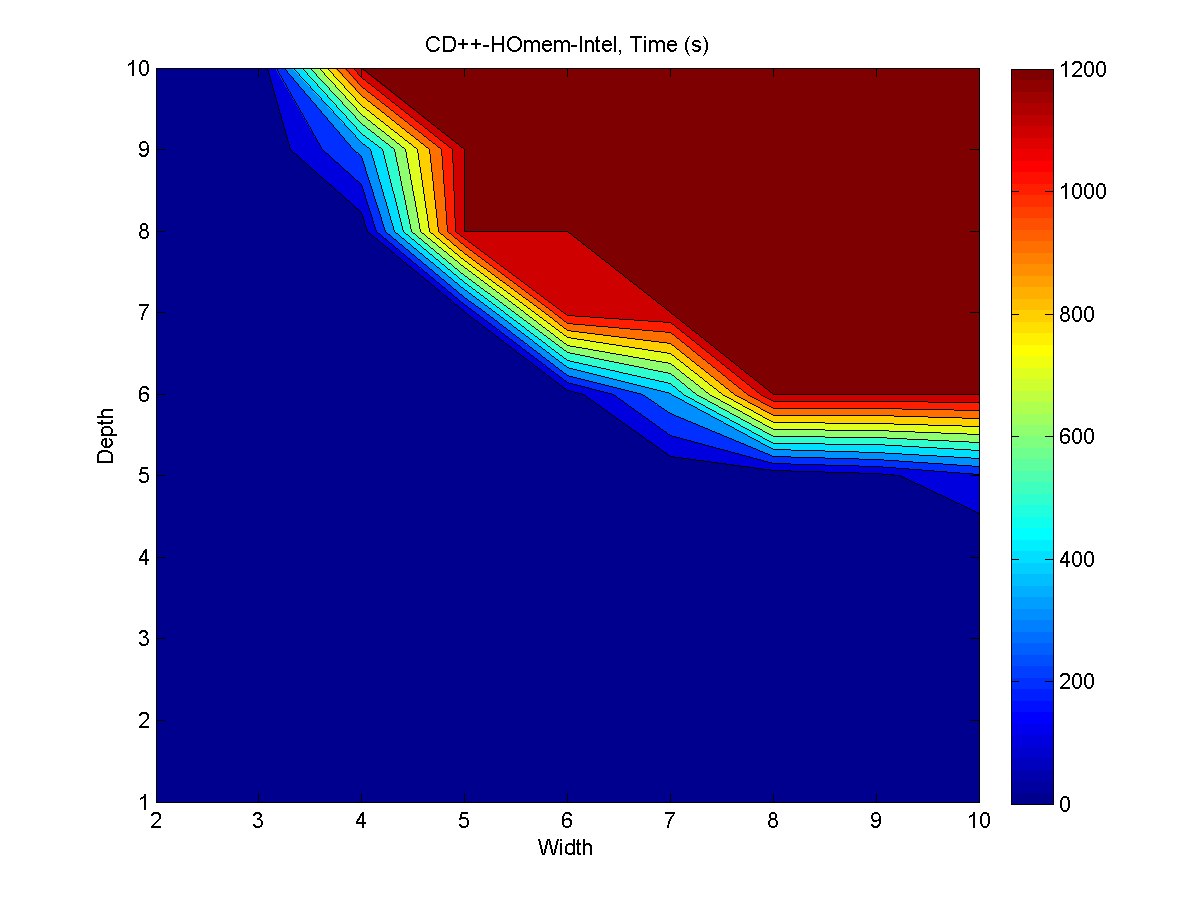}
\label{fig:CdPpHoMemIntelTime}
}
\subfigure[DEVSJAVA - HO]{
\includegraphics[width=0.45\textwidth, height=3cm]{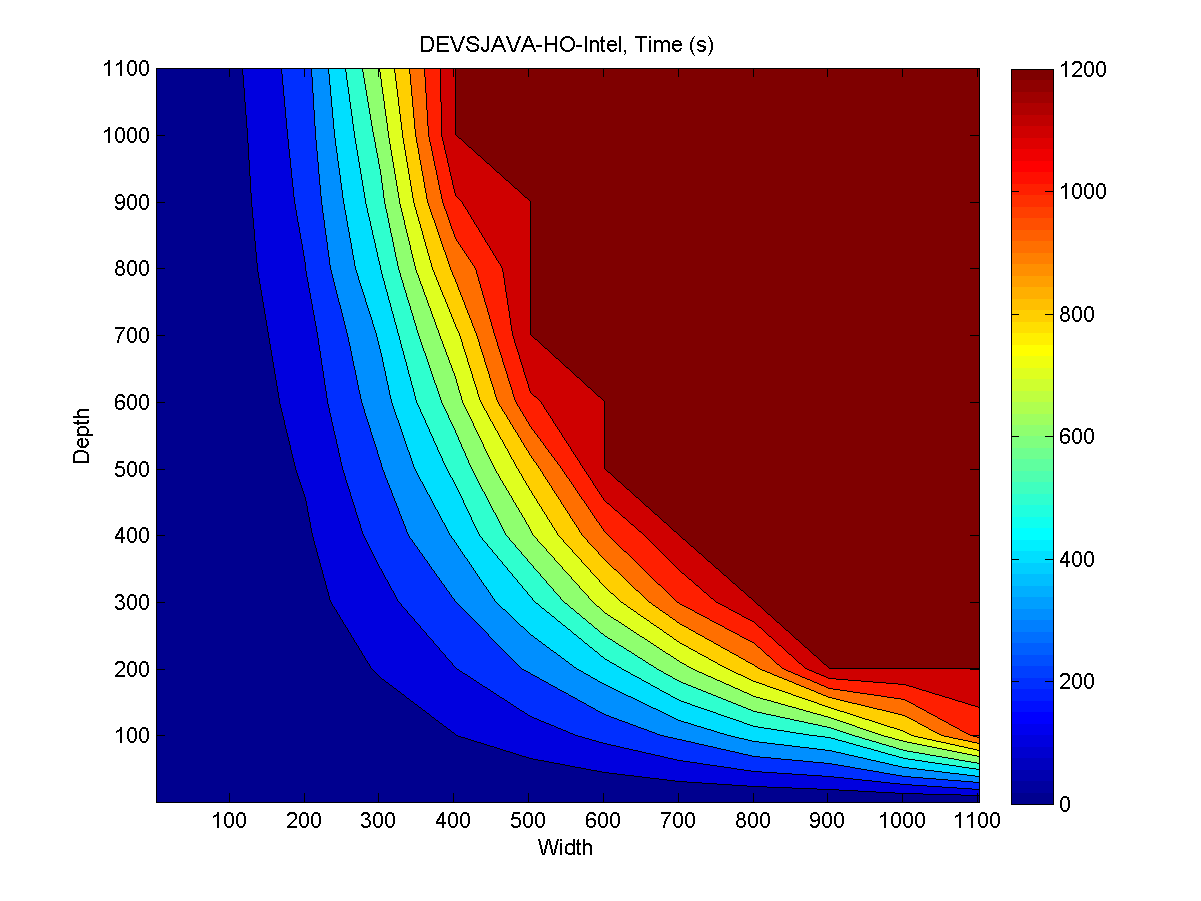}
\label{fig:DevsJavaHoIntelTime}
}
\subfigure[DEVSJAVA - HOmem]{
\includegraphics[width=0.45\textwidth, height=3cm]{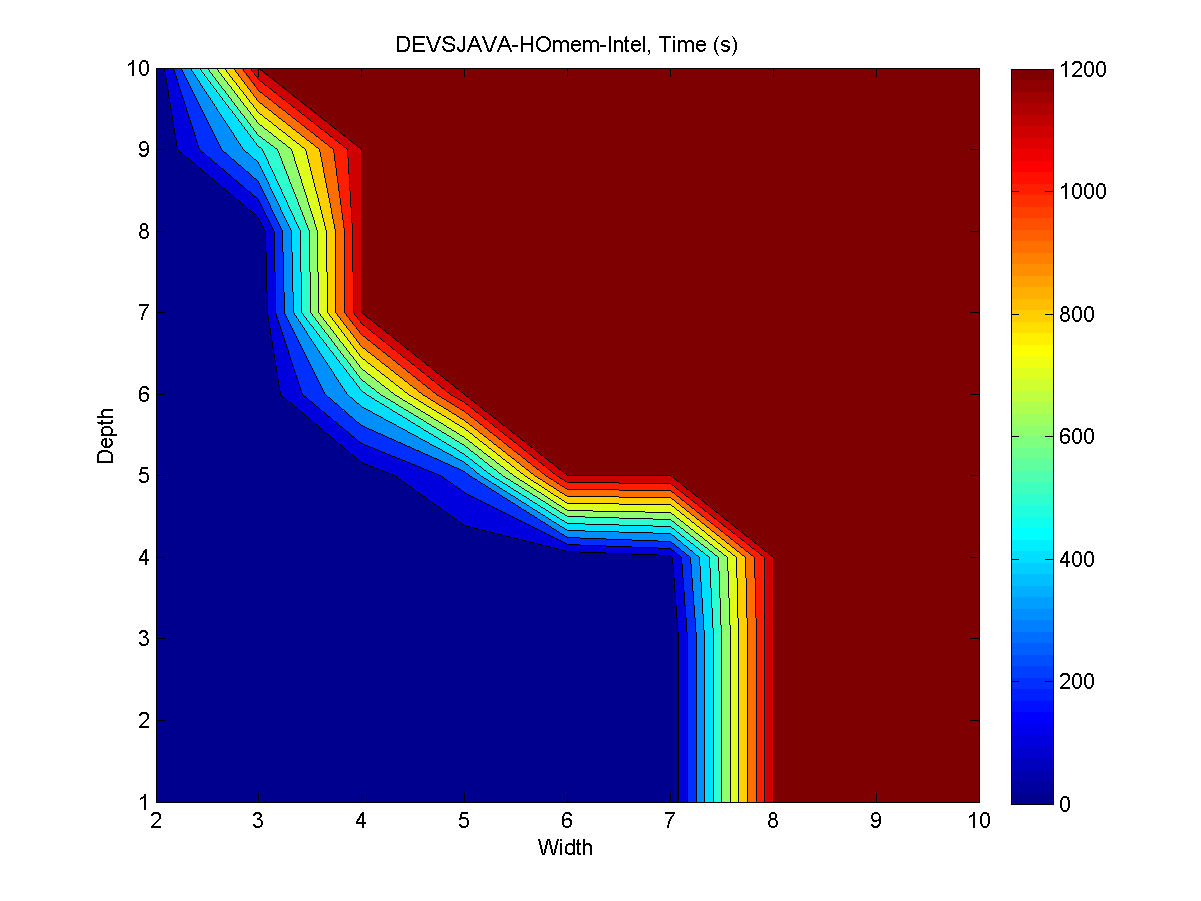}
\label{fig:DevsJavaHoMemIntelTime}
}
\subfigure[xDEVS - HO]{
\includegraphics[width=0.45\textwidth, height=3cm]{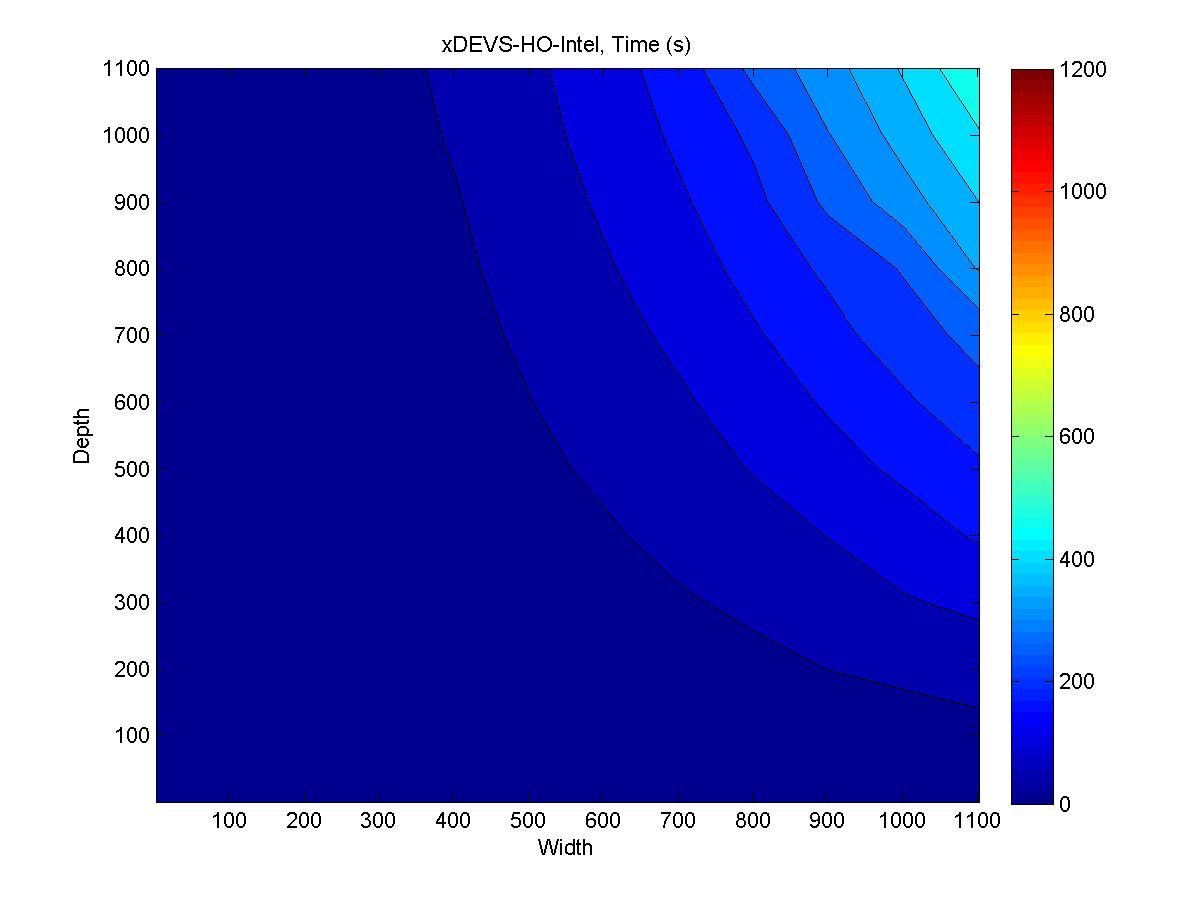}
\label{fig:xDEVSHoIntelTime}
}
\subfigure[xDEVS - HOmem]{
\includegraphics[width=0.45\textwidth, height=3cm]{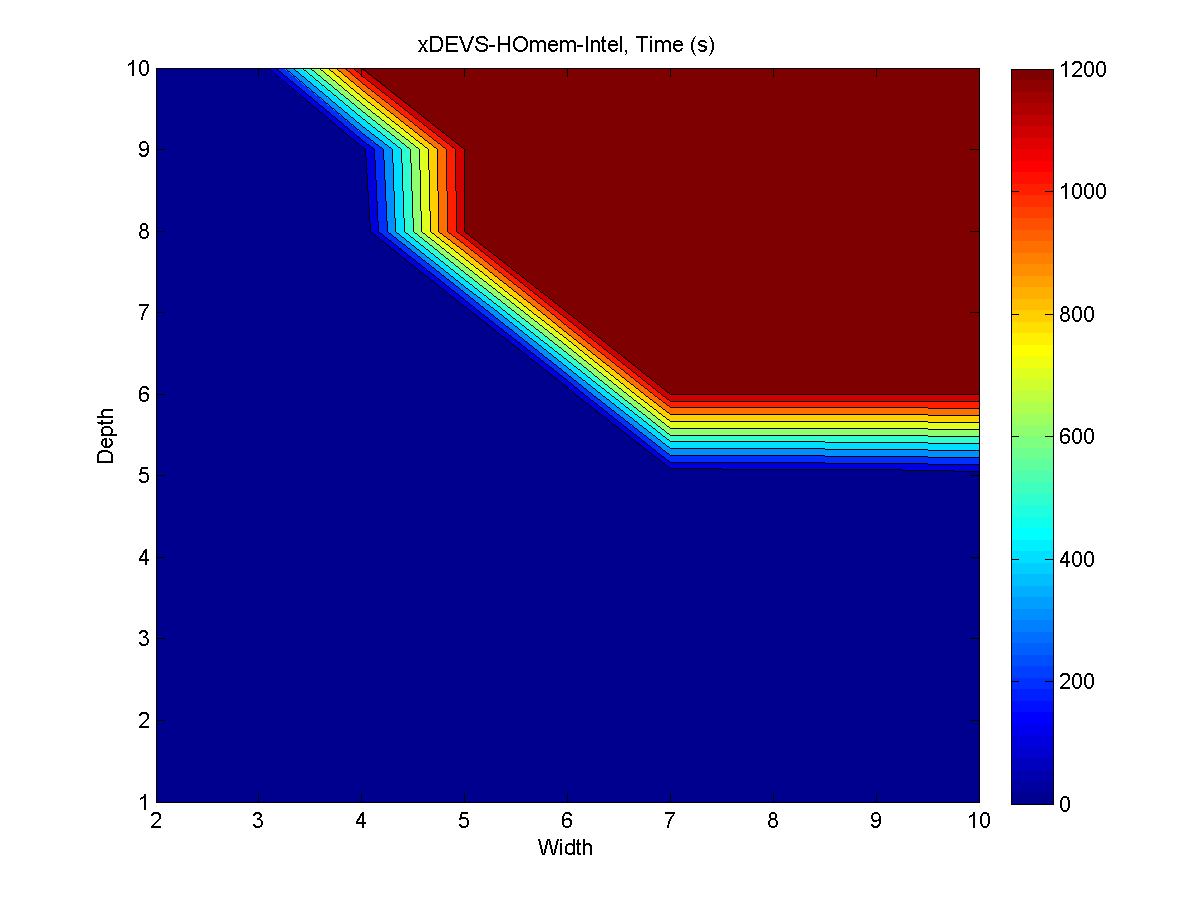}
\label{fig:xDEVSHoMemIntelTime}
}
\subfigure[PyPDEVS - HO]{
\includegraphics[width=0.45\textwidth, height=3cm]{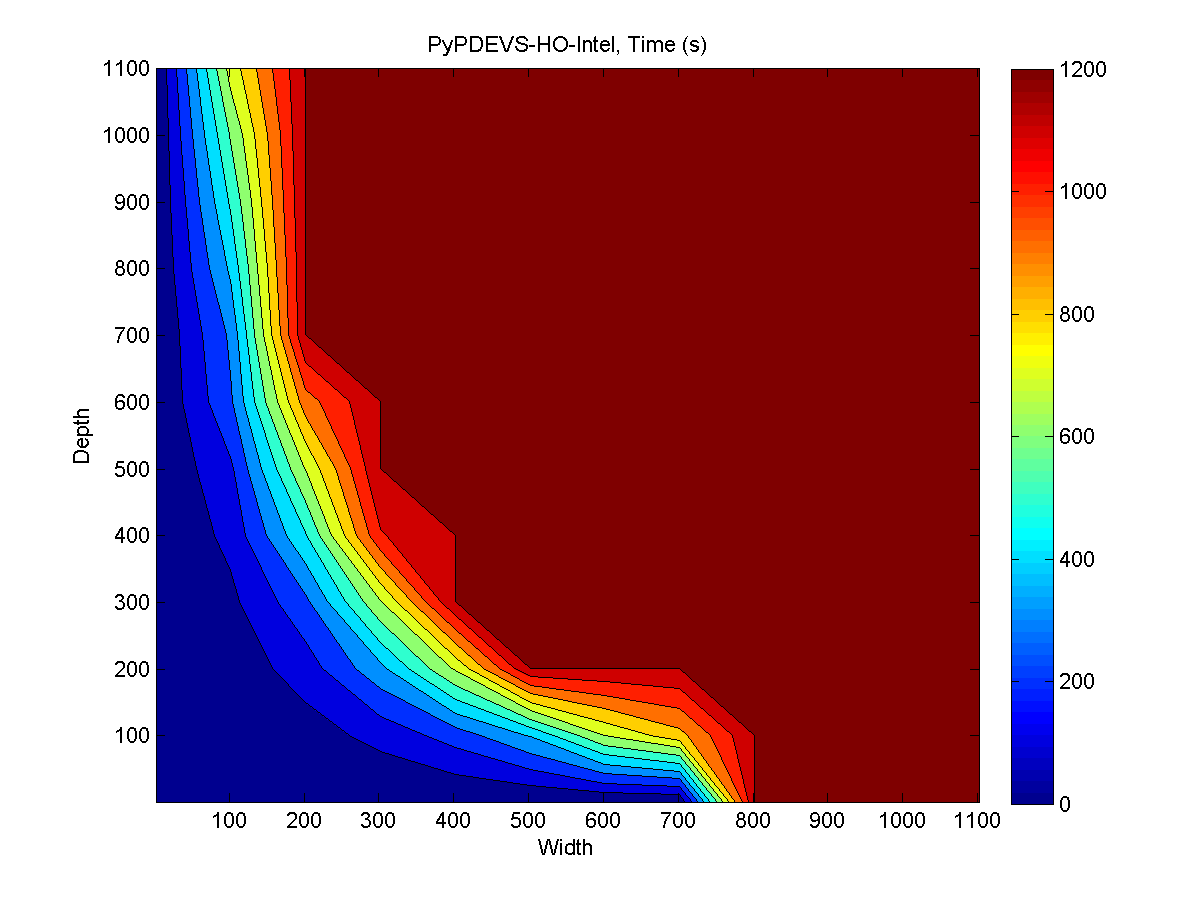}
\label{fig:PyDevsHoIntelTime}
}
\subfigure[PyPDEVS - HOmem]{
\includegraphics[width=0.45\textwidth, height=3cm]{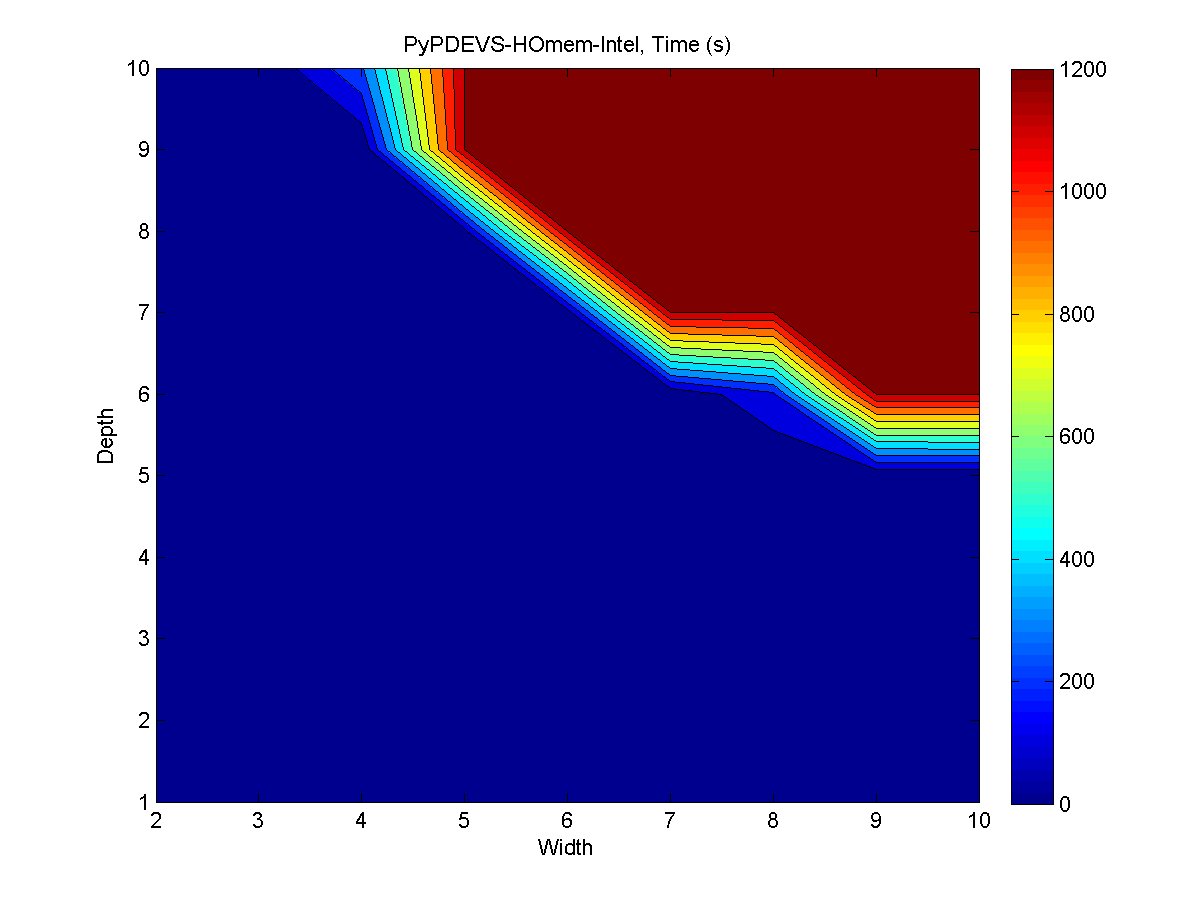}
\label{fig:PyDevsHoMemIntelTime}
}
\caption{Execution time of HO and HOmem models}
\label{fig:AllHoHoMemIntelTime}
\end{figure*}

Continuing with this analysis, Figure \ref{fig:AllHoHoMemIntelTime} shows the same contour maps, this time in HO and HOmem models. 

Regarding the HO model, xDEVS obtained best execution times, specially as width and depth were increased. For low values of width and depth, aDEVS was better than xDEVS. Once again, CD++, DEVAJAVA and PyPDEVS saturated the execution time limit of 1200 seconds.

With respect to HOmod and HOmem, all the simulators reached the maximum execution time quite soon, with relatively small models. Moreover, in the case of HOmod, only two simulators, aDEVS and xDEVS, were able to load all the models in memory, before the execution of the simulation. In fact, this is due to the intrinsic complexity of the HOmod benchmark, which includes many more atomic models than HOmem. HOmem is simpler in structure than HOmod, and all the simulation engines are able to load it. Once the simulation starts, HOmod and HOmem offer similar execution time and memory footprint, as shown in Section \ref{sec:finalComparison}.

We do not show a comparison between all the simulators in HOmod because only aDEVS and xDEVS were able to run a significant number of HOmod instances. These experiments are shown in the comparison between aDEVS and xDEVS.

\subsection{Memory footprint}

As mentioned before, memory footprint is the memory high-water mark of a process. The comparison of all the five simulators were performed constraining the execution time to 1200 seconds and the memory footprint to 4 GiB. The set of five simulators compared in this paper have been developed using different programming languages: aDEVS and CD++ in C++, DEVSJAVA and xDEVS in JAVA, and PyPDEVS in Python. Since JAVA and Python use their own virtual machines, it is expected that these simulators have a higher memory footprint. However, our experimental results showed some exceptions in this regard. 

\begin{figure*}[ht]
\centering
\subfigure[aDEVS - LI]{
\includegraphics[width=0.45\textwidth, height=3cm]{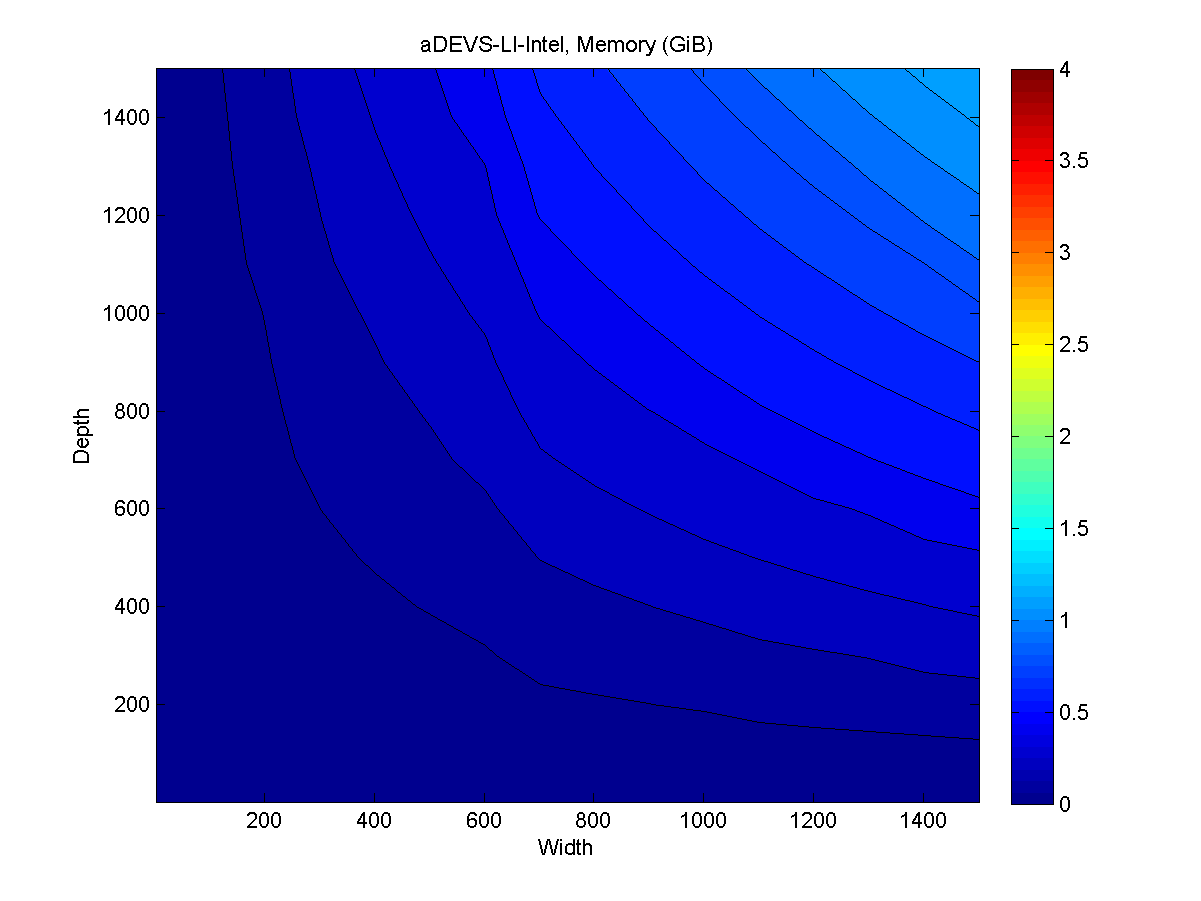}
\label{fig:aDevsLiIntelMemory}
}
\subfigure[aDEVS - HI]{
\includegraphics[width=0.45\textwidth, height=3cm]{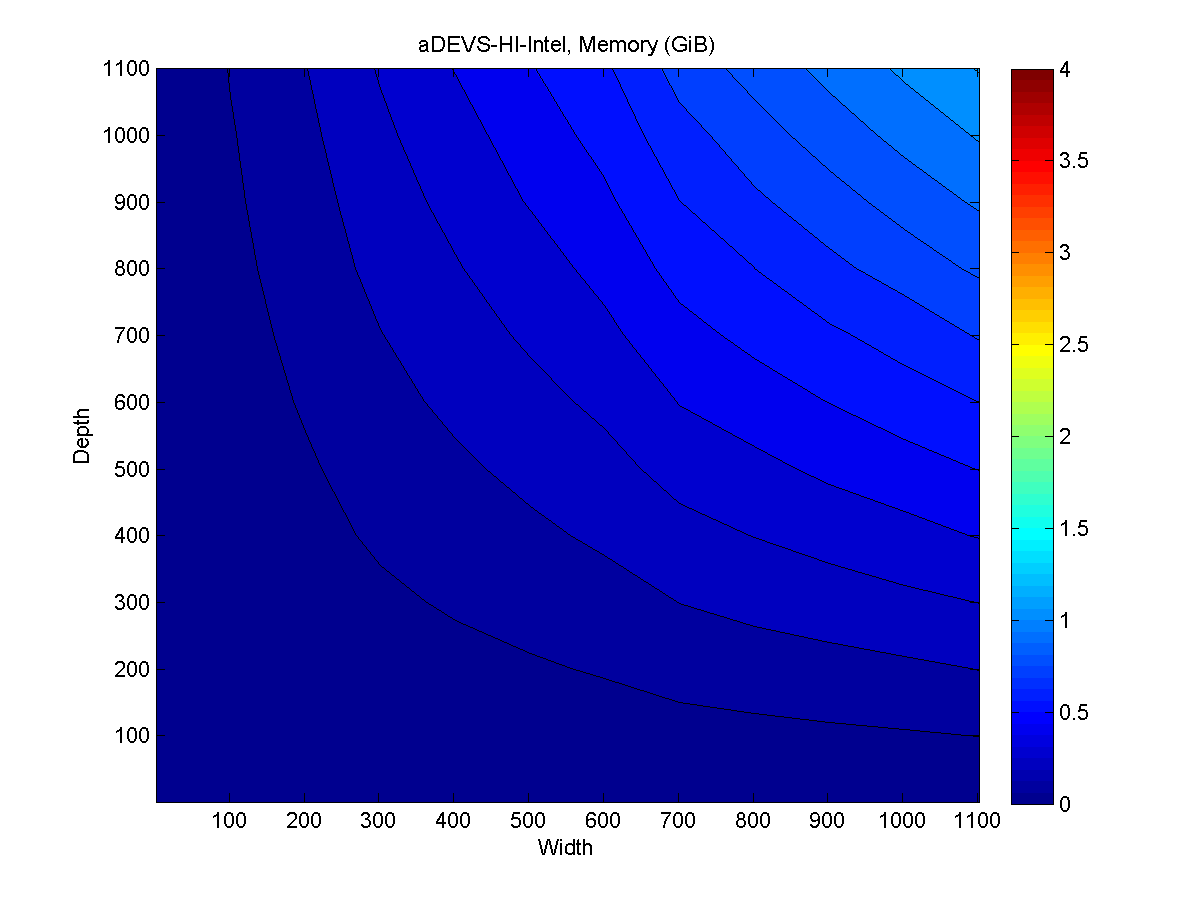}
\label{fig:aDevsHiIntelMemory}
}
\subfigure[CD++ - LI]{
\includegraphics[width=0.45\textwidth, height=3cm]{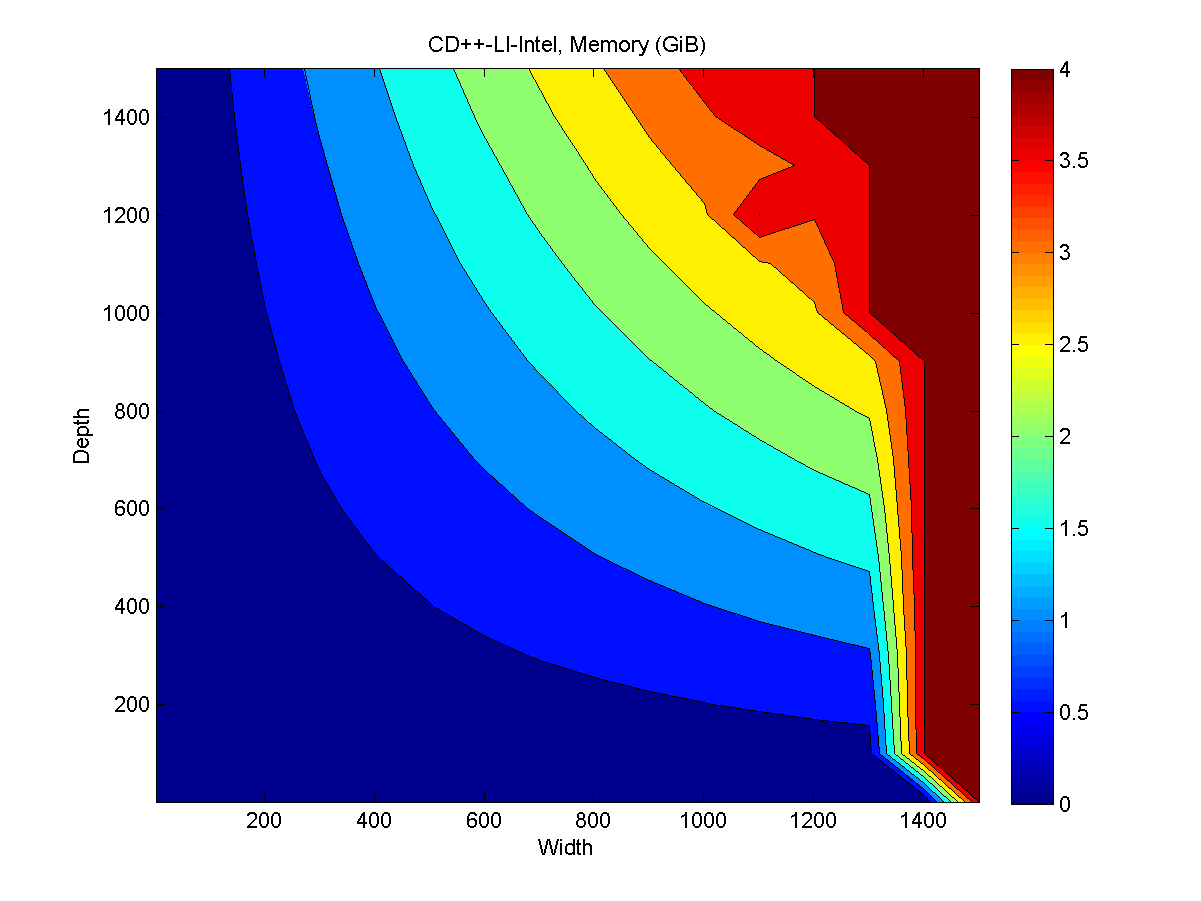}
\label{fig:CdPpLiIntelMemory}
}
\subfigure[CD++ - HI]{
\includegraphics[width=0.45\textwidth, height=3cm]{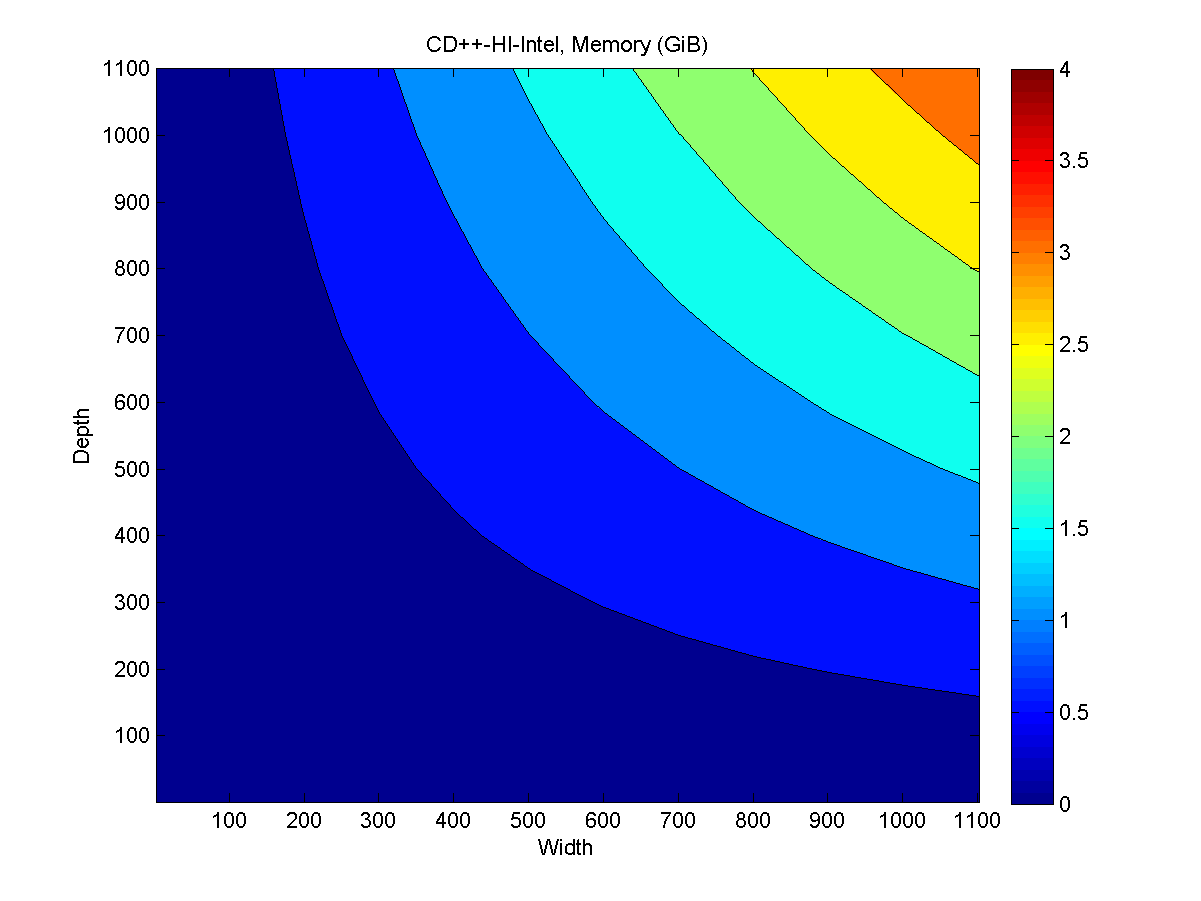}
\label{fig:CdPpHiIntelMemory}
}
\subfigure[DEVSJAVA - LI]{
\includegraphics[width=0.45\textwidth, height=3cm]{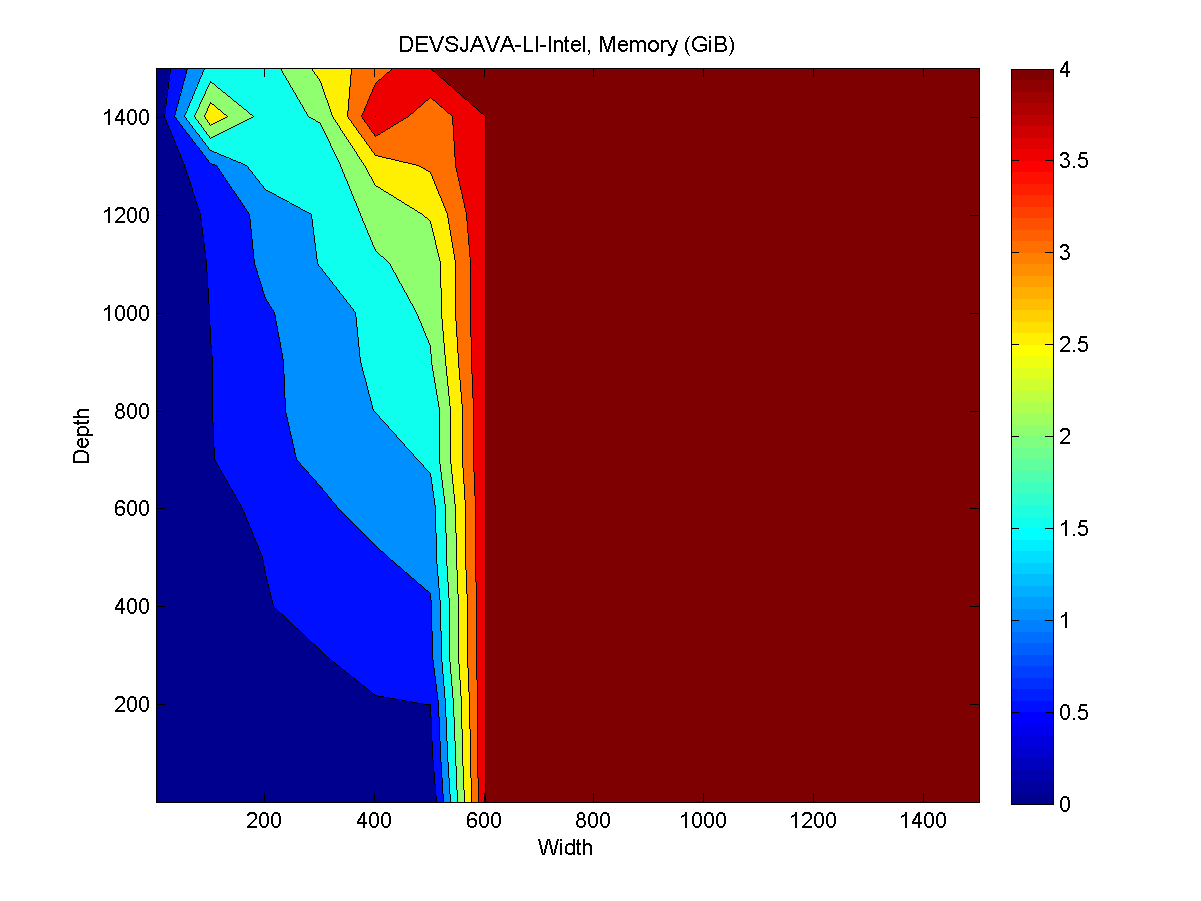}
\label{fig:DevsJavaLiIntelMemory}
}
\subfigure[DEVSJAVA - HI]{
\includegraphics[width=0.45\textwidth, height=3cm]{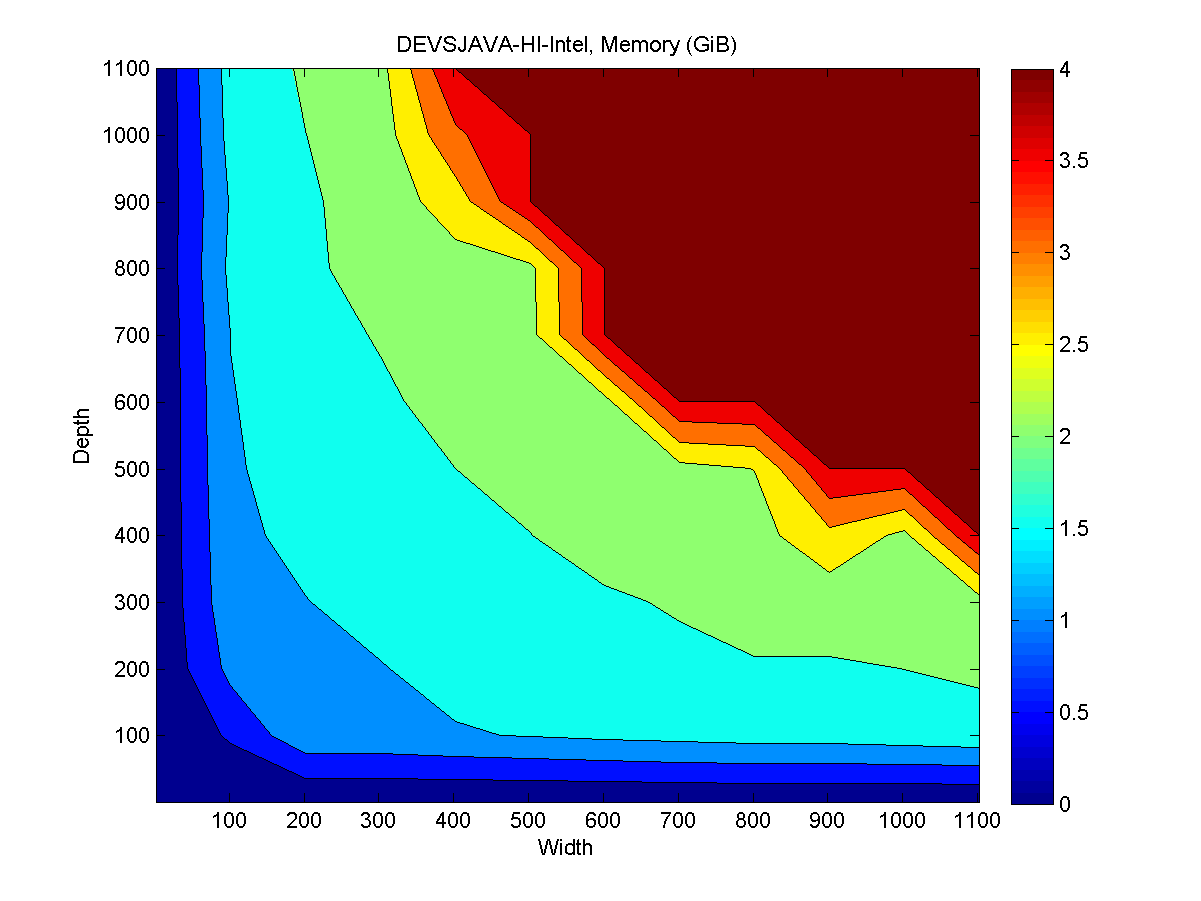}
\label{fig:DevsJavaHiIntelMemory}
}
\subfigure[xDEVS - LI]{
\includegraphics[width=0.45\textwidth, height=3cm]{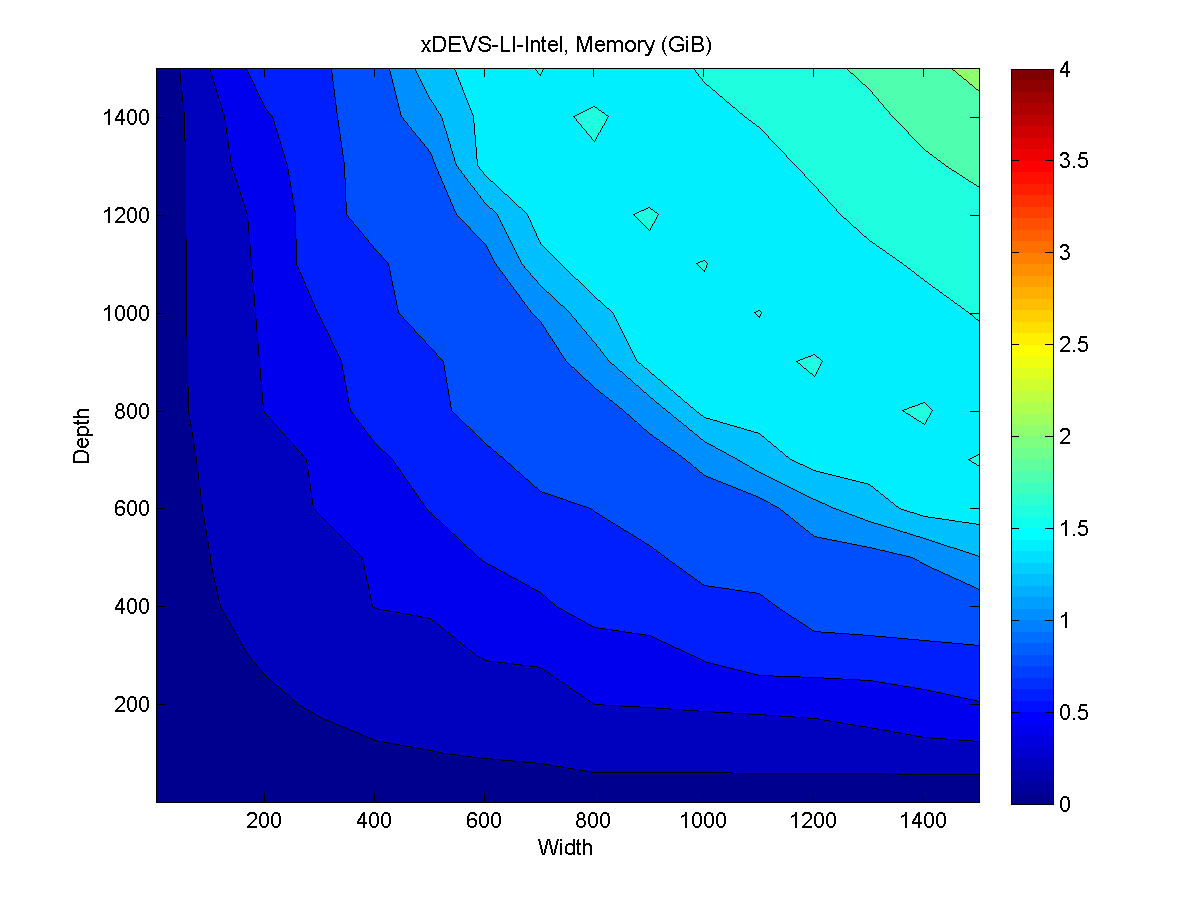}
\label{fig:xDEVSLiIntelMemory}
}
\subfigure[xDEVS - HI]{
\includegraphics[width=0.45\textwidth, height=3cm]{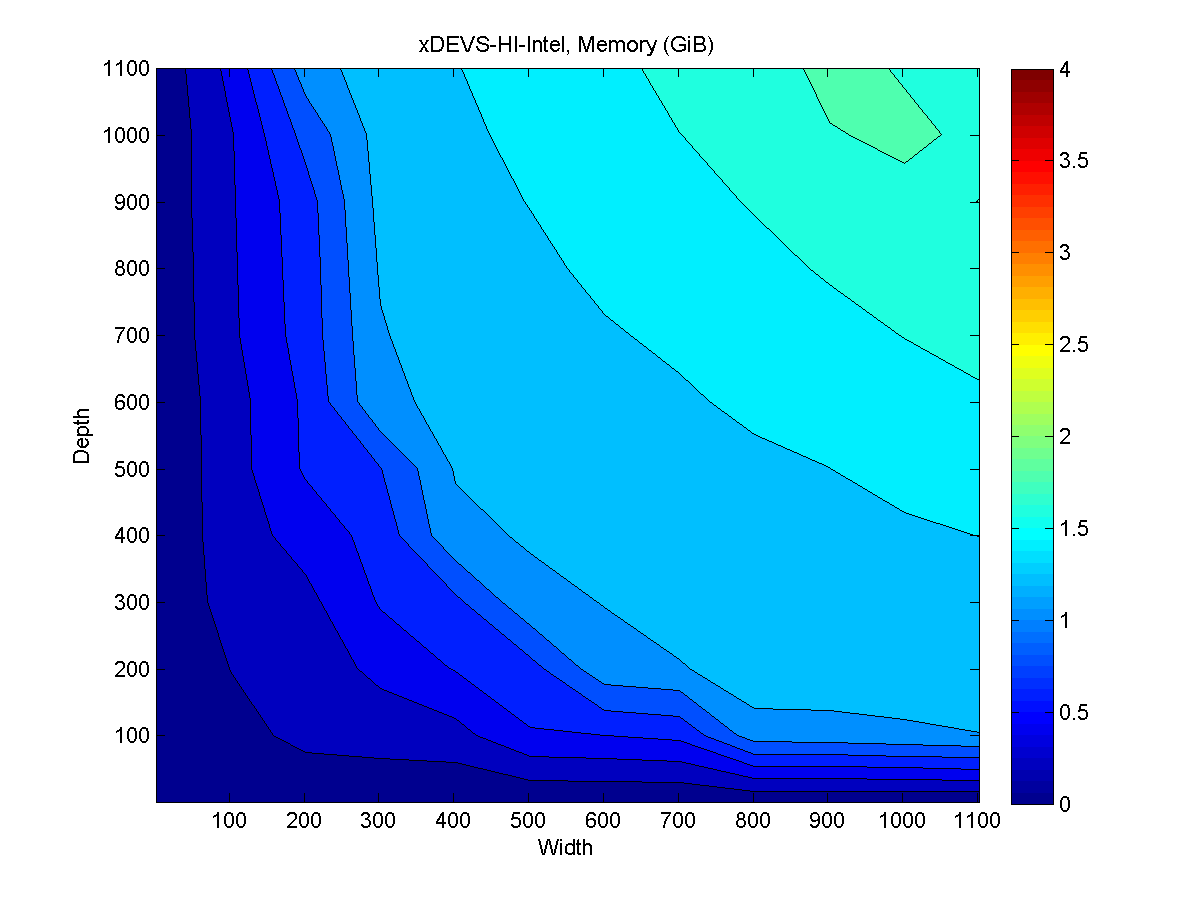}
\label{fig:xDEVSHiIntelMemory}
}
\subfigure[PyPDEVS - LI]{
\includegraphics[width=0.45\textwidth, height=3cm]{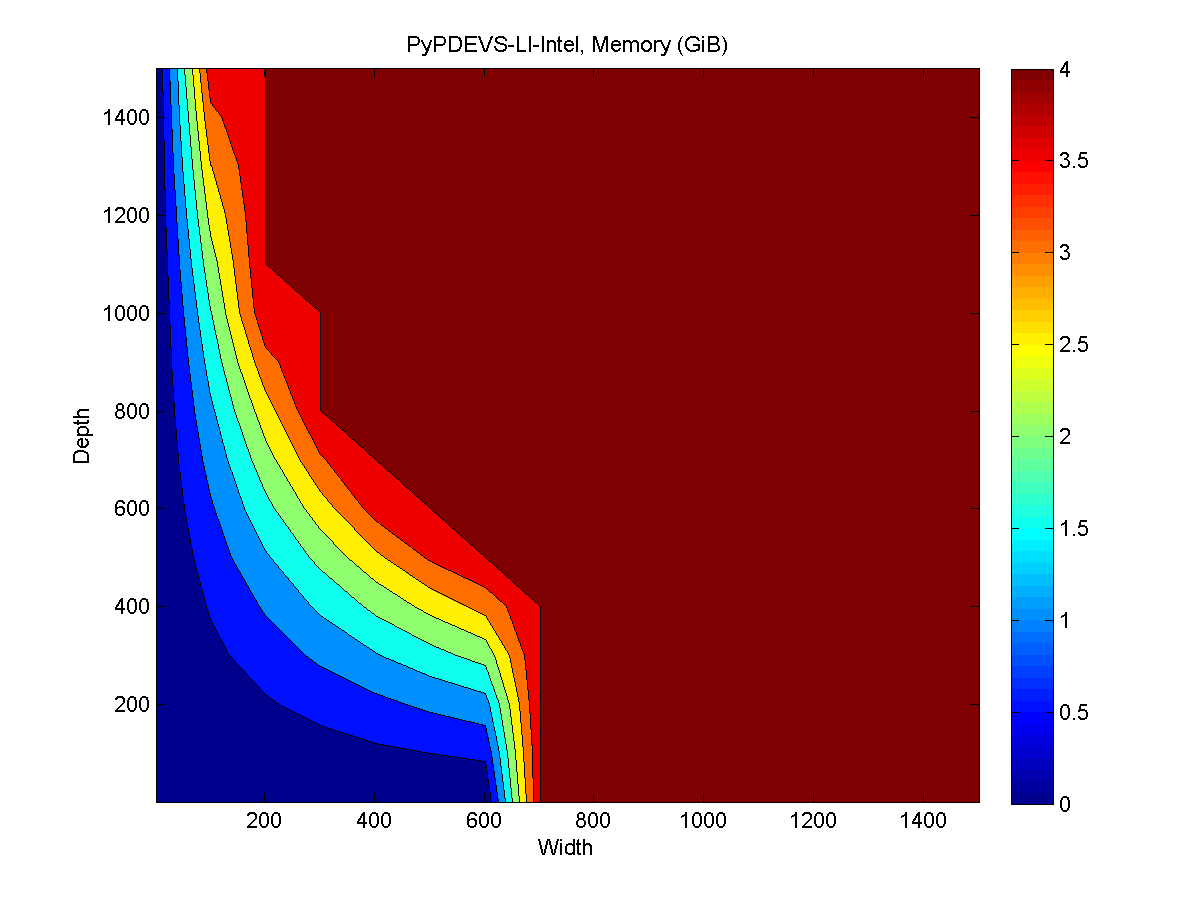}
\label{fig:PyDevsLiIntelMemory}
}
\subfigure[PyPDEVS - HI]{
\includegraphics[width=0.45\textwidth, height=3cm]{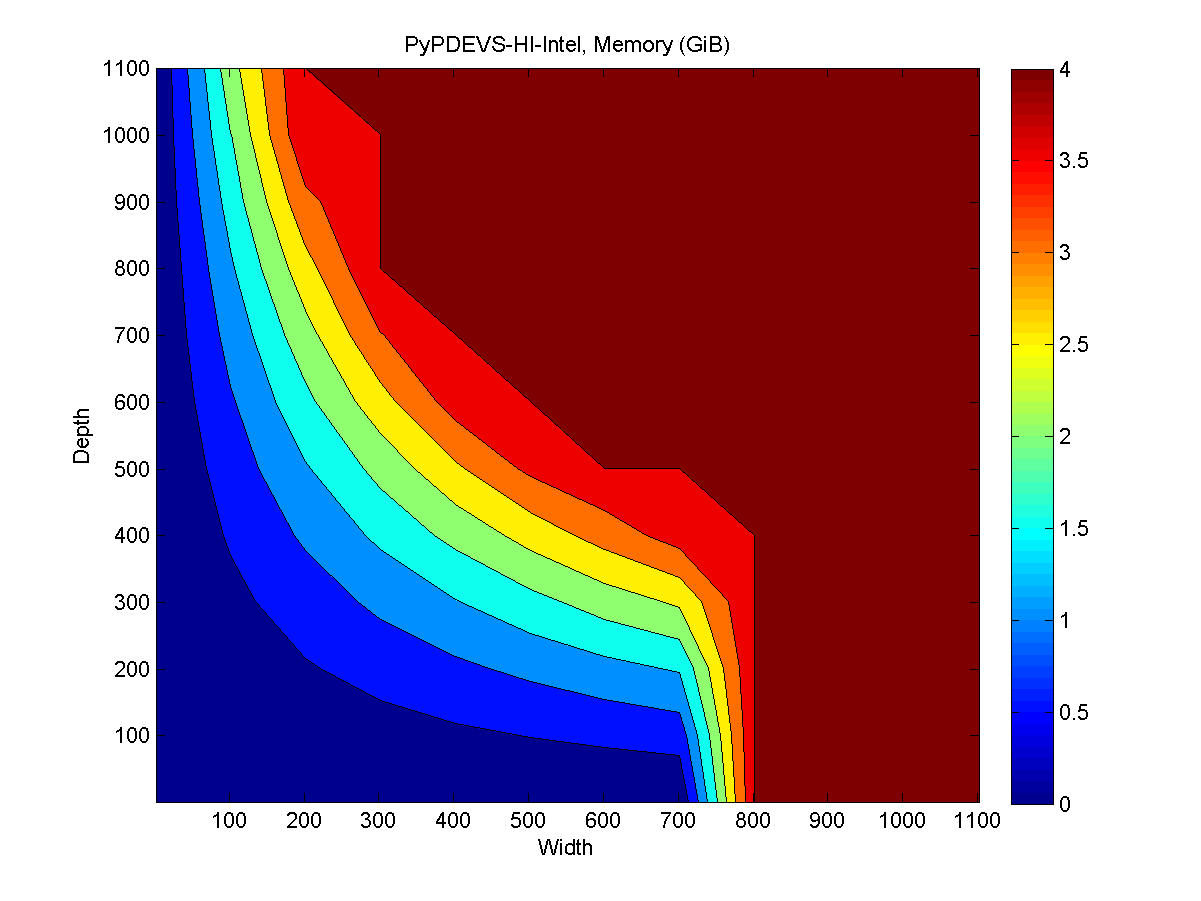}
\label{fig:PyDevsHiIntelMemory}
}
\caption{Memory footprint of LI and HI models}
\label{fig:AllLiHiIntelMemory}
\end{figure*}

Figure \ref{fig:AllLiHiIntelMemory} shows the memory footprint reached by the five simulators in LI and HI models. As can be seen, DEVSJAVA and specially PyPDEVS reached the memory limit quite soon. aDEVS had by far the lowest memory usage. However, between CD++ and xDEVS, the latter obtained less memory footprint even when the Java Virtual Machine must be loaded into memory. This is because CD++ uses a complex structure to store the model, as is evident when CD++ is completely saturated once width and depth is greater than 1200.

\begin{figure*}[ht]
\centering
\subfigure[aDEVS - HO]{
\includegraphics[width=0.45\textwidth, height=3cm]{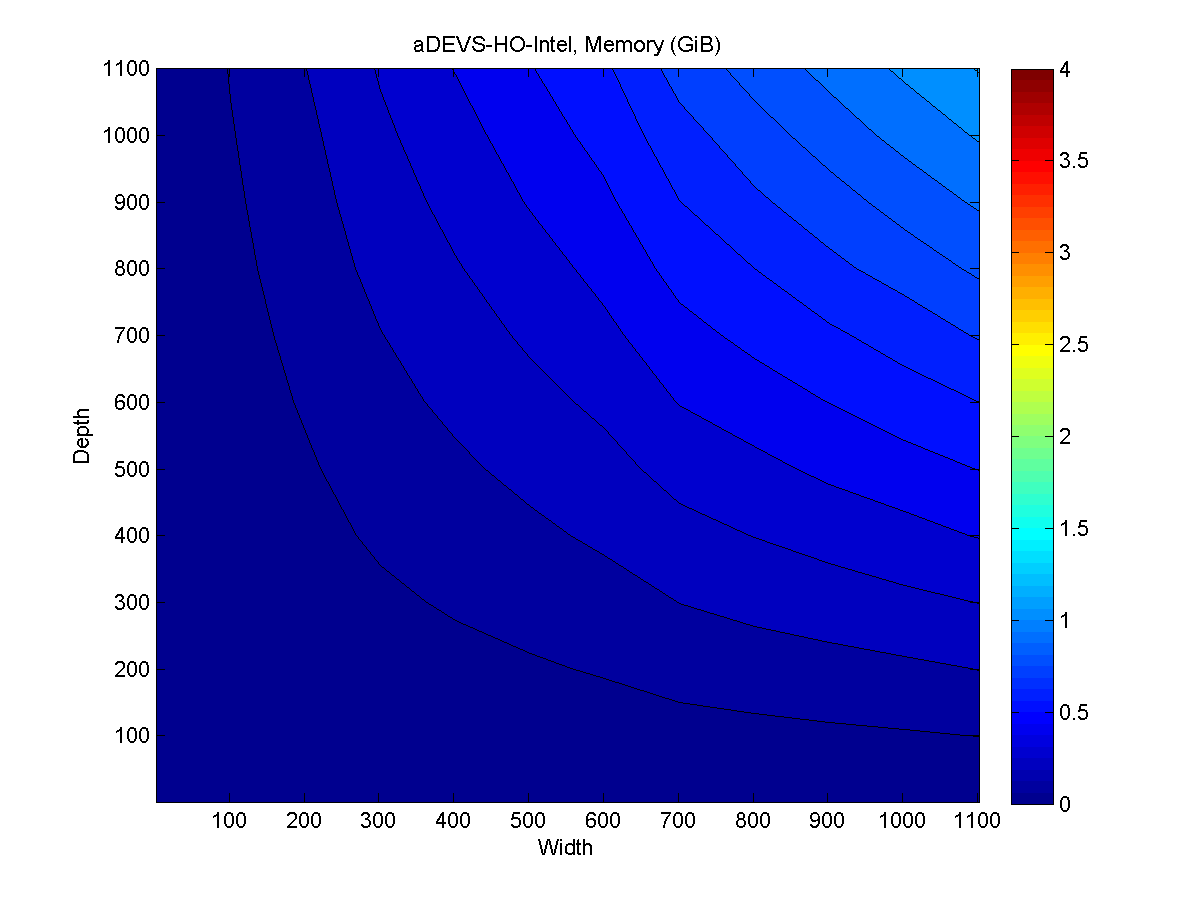}
\label{fig:aDevsHoIntelMemory}
}
\subfigure[aDEVS - HOmem]{
\includegraphics[width=0.45\textwidth, height=3cm]{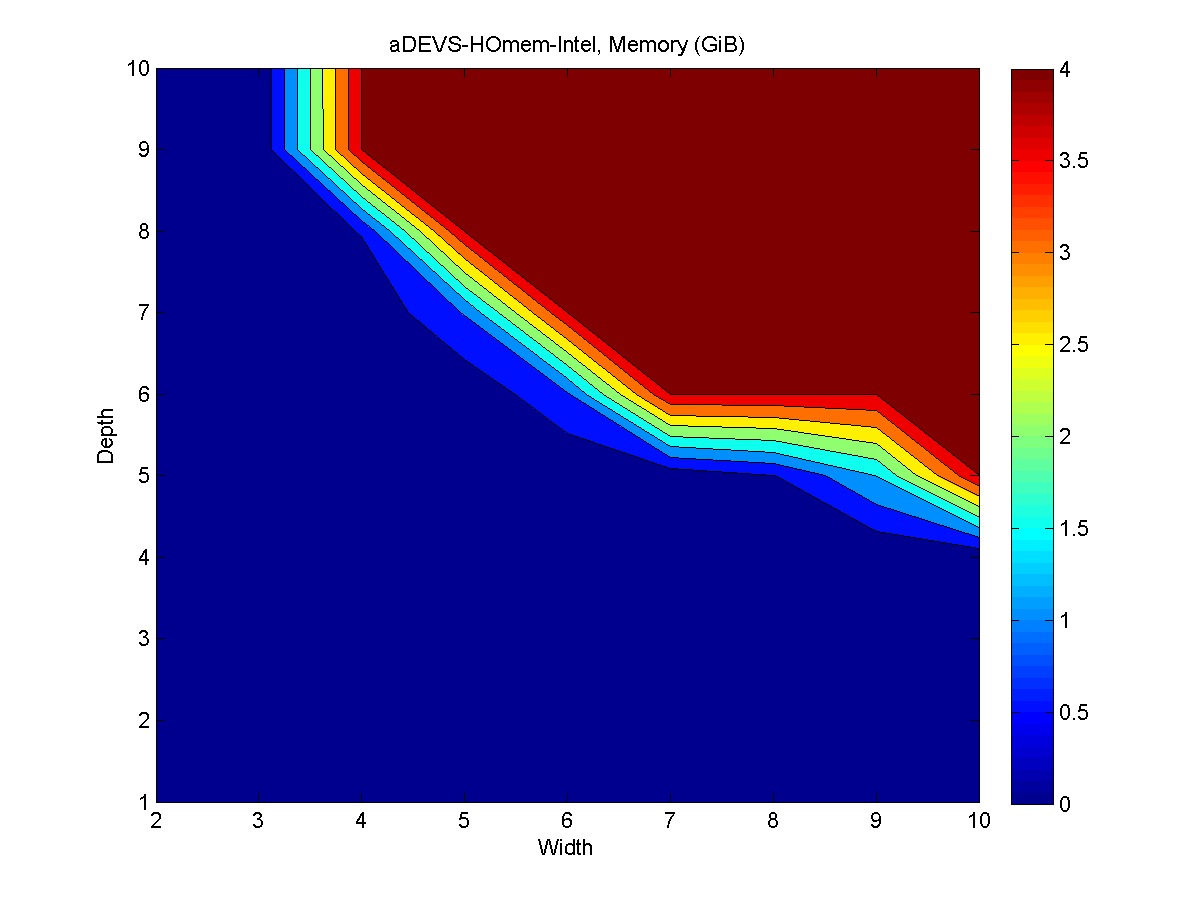}
\label{fig:aDevsHoMemIntelMemory}
}
\subfigure[CD++ - HO]{
\includegraphics[width=0.45\textwidth, height=3cm]{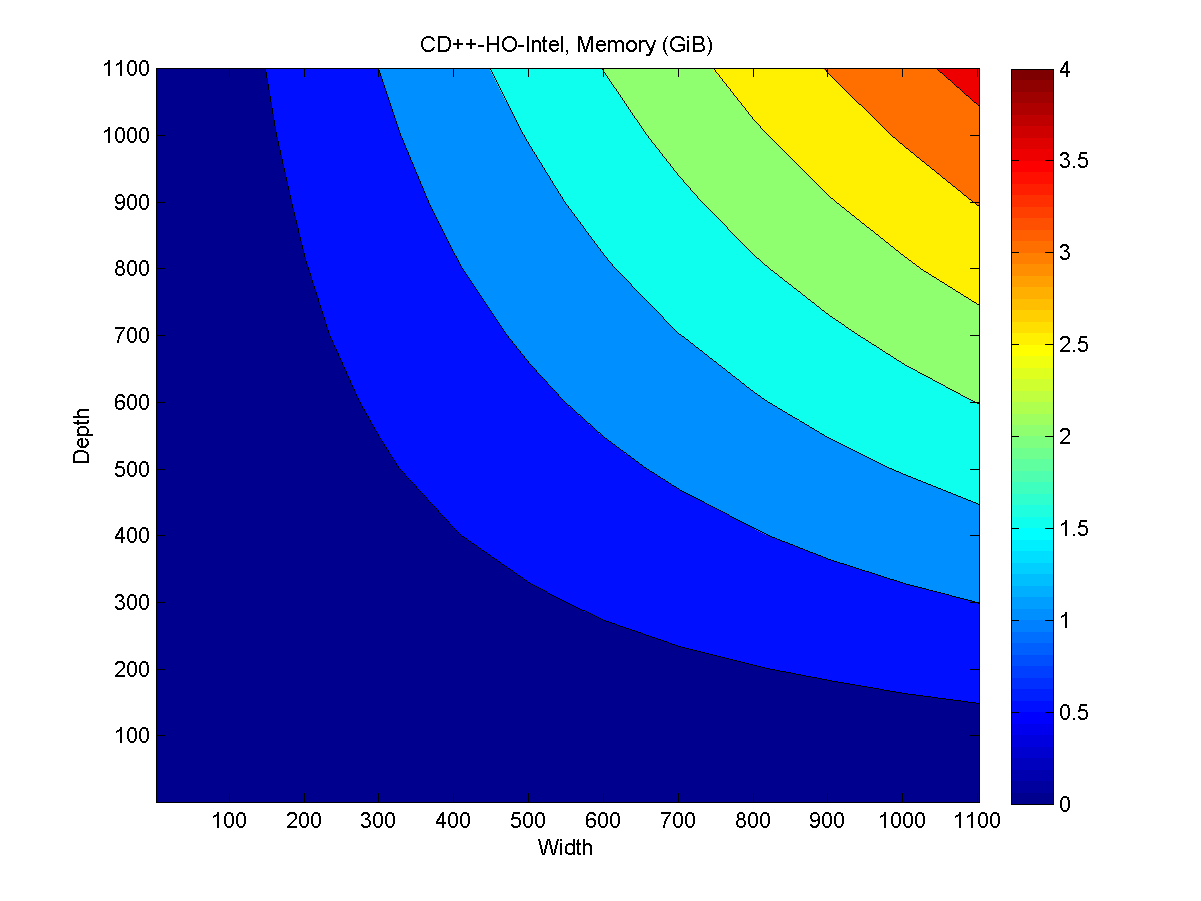}
\label{fig:CdPpHoIntelMemory}
}
\subfigure[CD++ - HOmem]{
\includegraphics[width=0.45\textwidth, height=3cm]{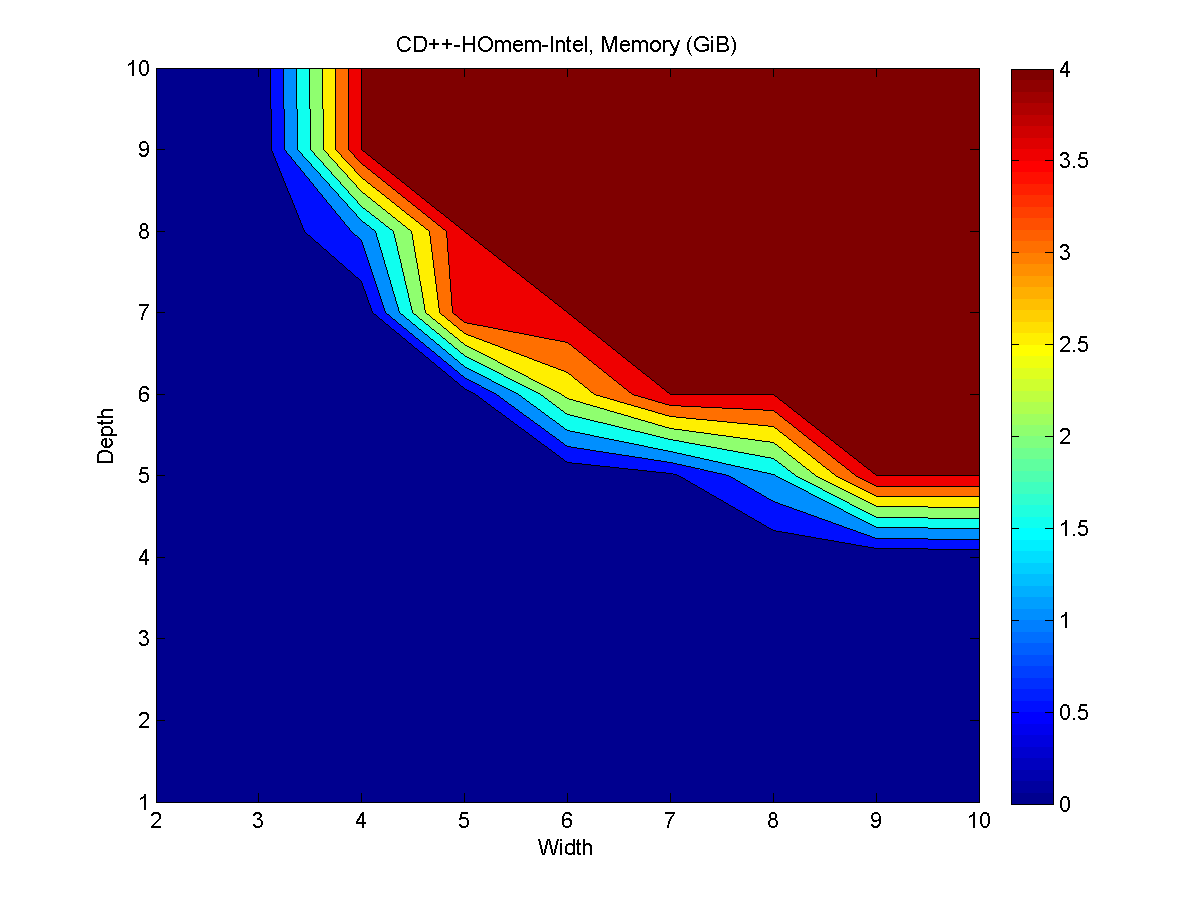}
\label{fig:CdPpHoMemIntelMemory}
}
\subfigure[DEVSJAVA - HO]{
\includegraphics[width=0.45\textwidth, height=3cm]{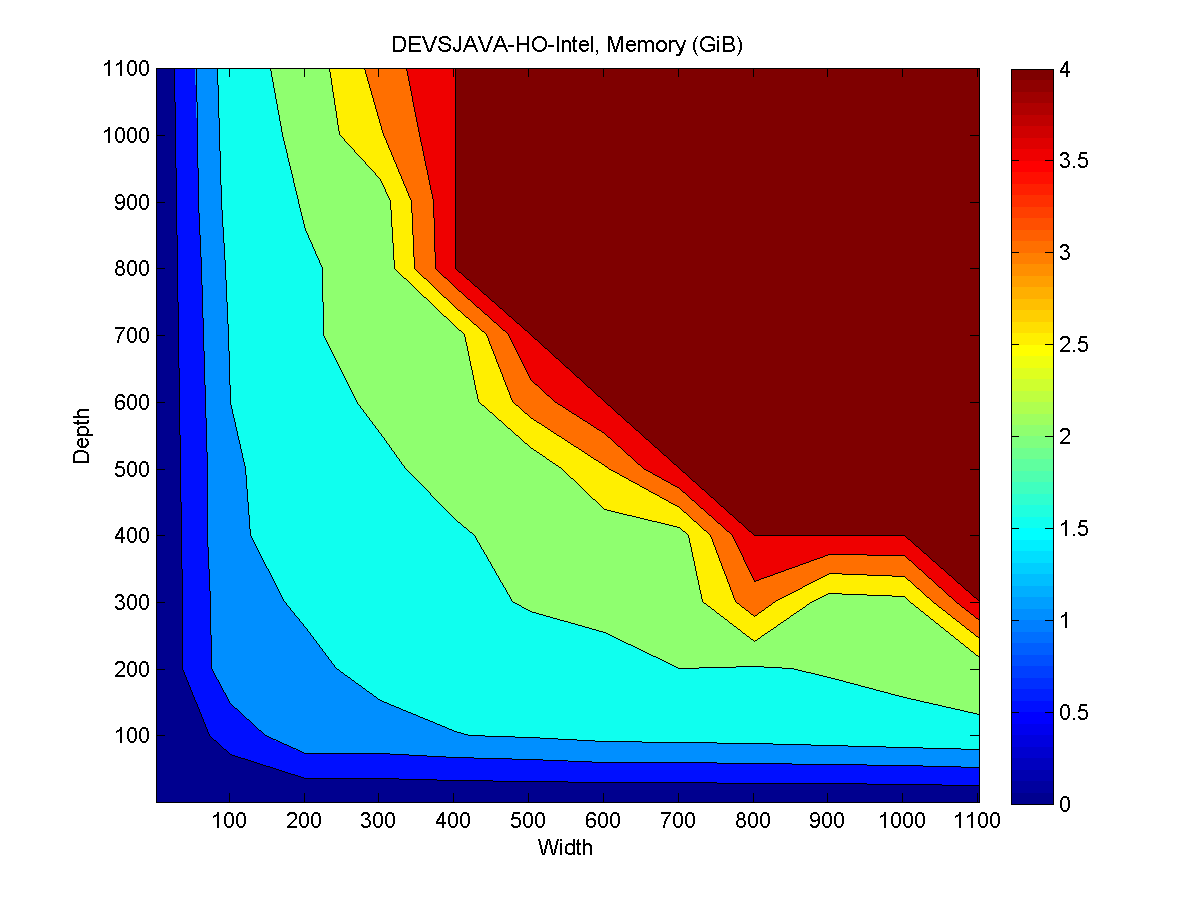}
\label{fig:DevsJavaHoIntelMemory}
}
\subfigure[DEVSJAVA - HOmem]{
\includegraphics[width=0.45\textwidth, height=3cm]{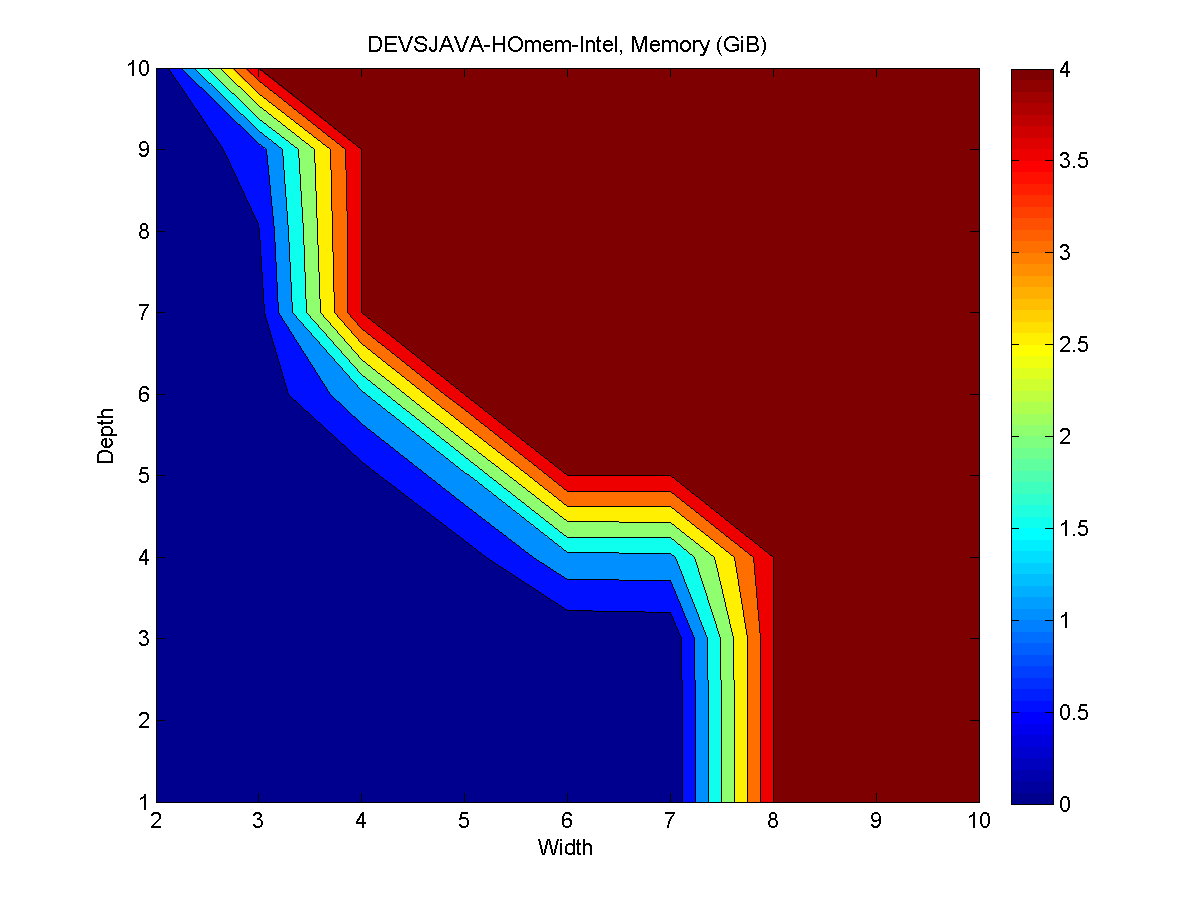}
\label{fig:DevsJavaHoMemIntelMemory}
}
\subfigure[xDEVS - HO]{
\includegraphics[width=0.45\textwidth, height=3cm]{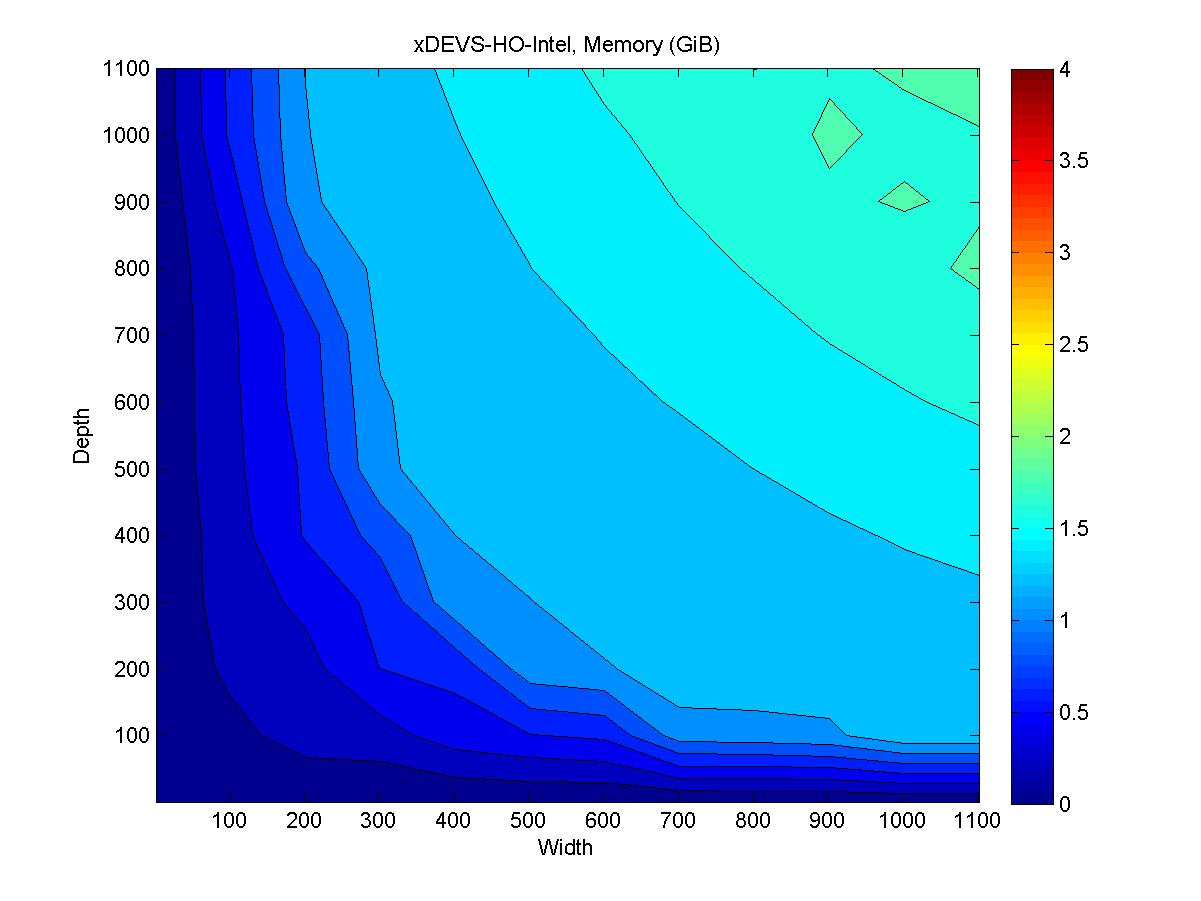}
\label{fig:xDEVSHoIntelMemory}
}
\subfigure[xDevs - HOmem]{
\includegraphics[width=0.45\textwidth, height=3cm]{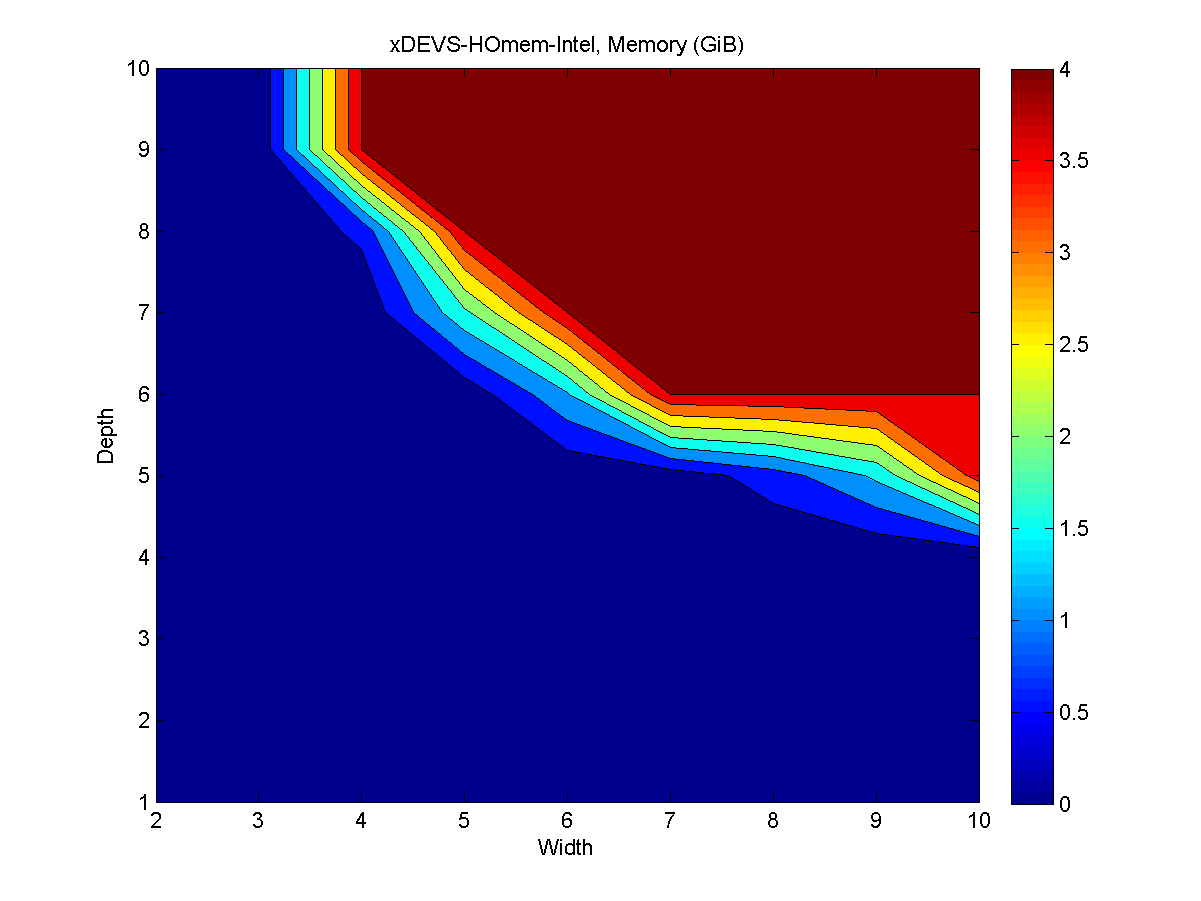}
\label{fig:xDEVSHoMemIntelMemory}
}
\subfigure[PyPDEVS - HO]{
\includegraphics[width=0.45\textwidth, height=3cm]{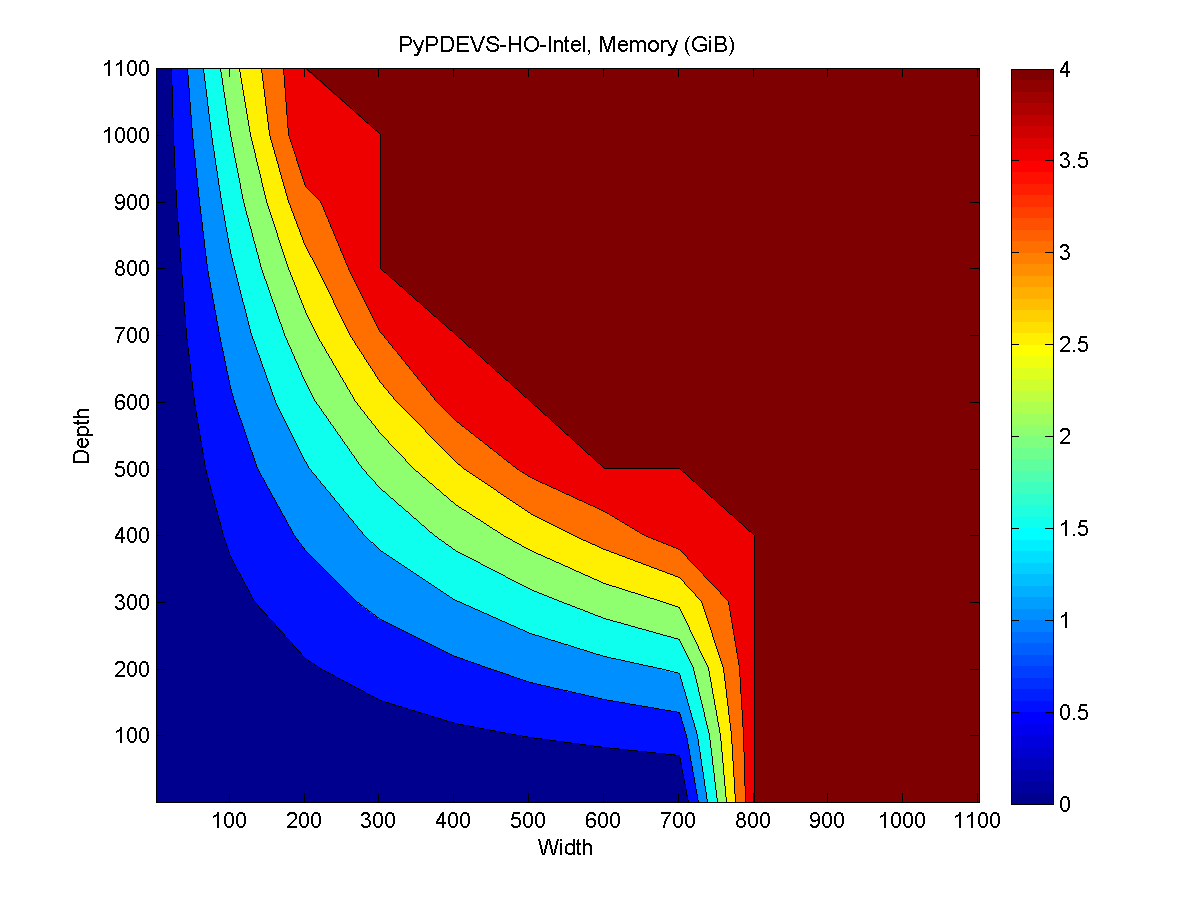}
\label{fig:PyDevsHoIntelMemory}
}
\subfigure[PyPDEVS - HOmem]{
\includegraphics[width=0.45\textwidth, height=3cm]{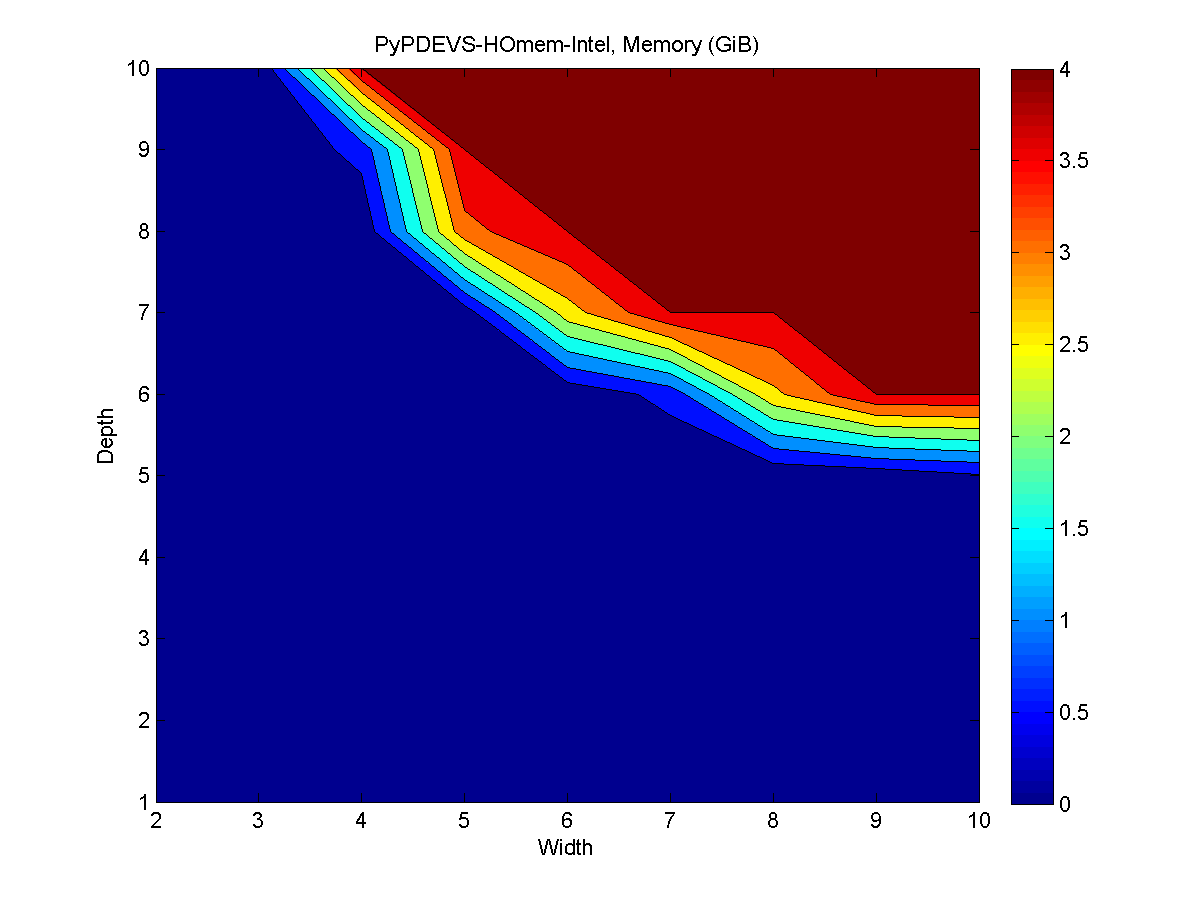}
\label{fig:PyDevsHoMemIntelMemory}
}
\caption{Memory footprint of HO and HOmem models}
\label{fig:AllHoHoMemIntelMemory}
\end{figure*}

Now, Figure \ref{fig:AllHoHoMemIntelMemory} shows the memory footprint reached by the five simulators in HO and HOmem models.

Regarding HO, the situation is almost identical to the HI model. aDEVS and xDEVS are still the two best simulators.

With respect to HOmem, all the five simulators reached the memory limit quite soon. As in the execution time analysis, DEVSJAVA was the first to leave the model, for \emph{width} greater than six. Surprisingly, PyPDEVS offered comparable results to aDEVS, CD++ and xDEVS in HOmem and HOmod, the more complex models.

As in the previous section, we do not show a comparison between all the simulators in HOmod because only aDEVS and xDEVS were able to run a significant number of HOmod instances.

As a conclusion, we may say that, regarding memory footprint, aDEVS is by far the best DEVS simulator between those analyzed in this paper. In the case of execution time, xDEVS is better as the complexity of the model increases, until the cases of HOmod and HOmem, where the complexity of both models cannot determine a classification with clarity. In the following, we investigate the performance of aDEVS and xDEVS simulators in finer details, as well as the similarities between HOmod and HOmem.

\subsection{Comparison between aDEVS and xDEVS}\label{sec:finalComparison}

Firstly, we show the difference in execution time and memory footprint obtained by both simulators in LI, HI, HO, HOmod and HOmem models. 

\begin{figure*}[ht]
\centering
\subfigure[LI: xDEVS - aDEVS]{
\includegraphics[width=0.45\textwidth, height=3cm]{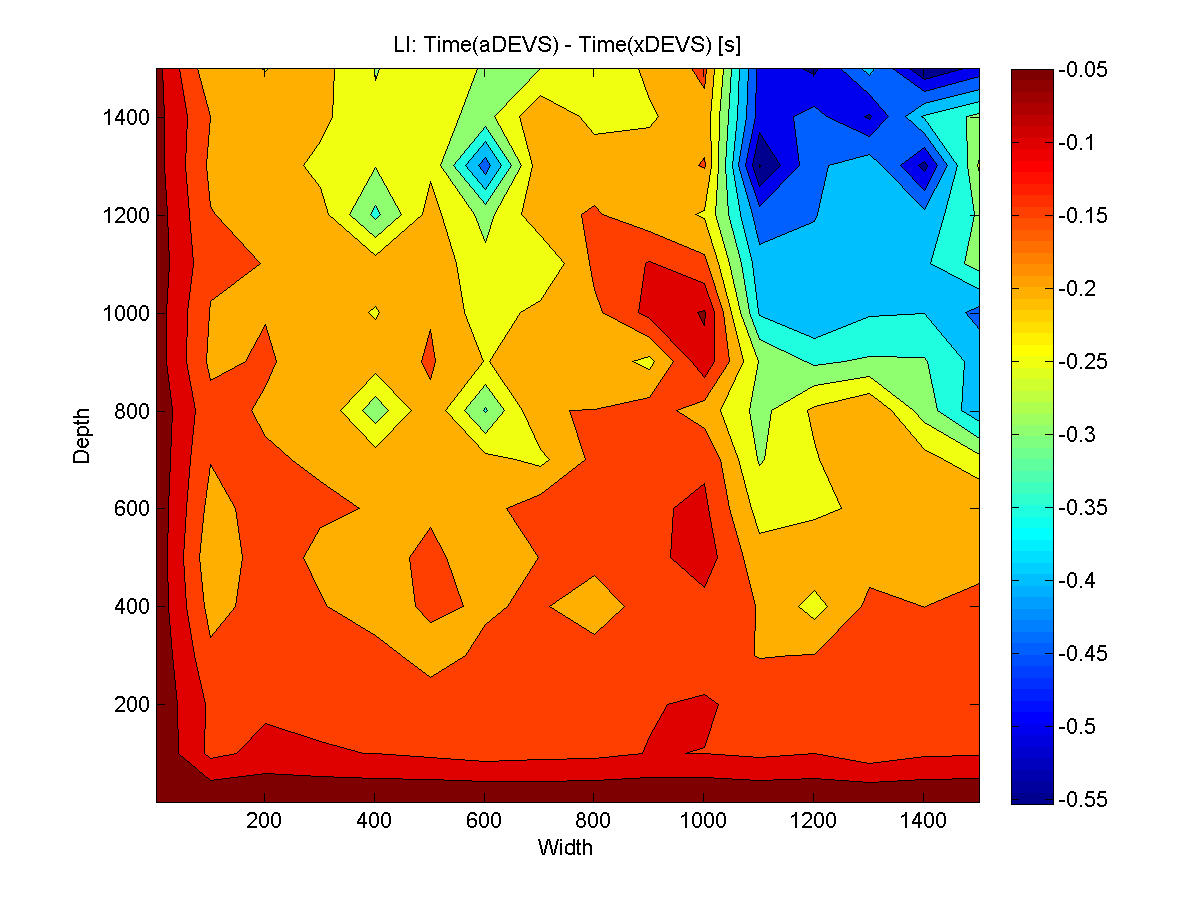}
\label{fig:ComparisonLiIntelTime}
}
\subfigure[HI: xDEVS - aDEVS]{
\includegraphics[width=0.45\textwidth, height=3cm]{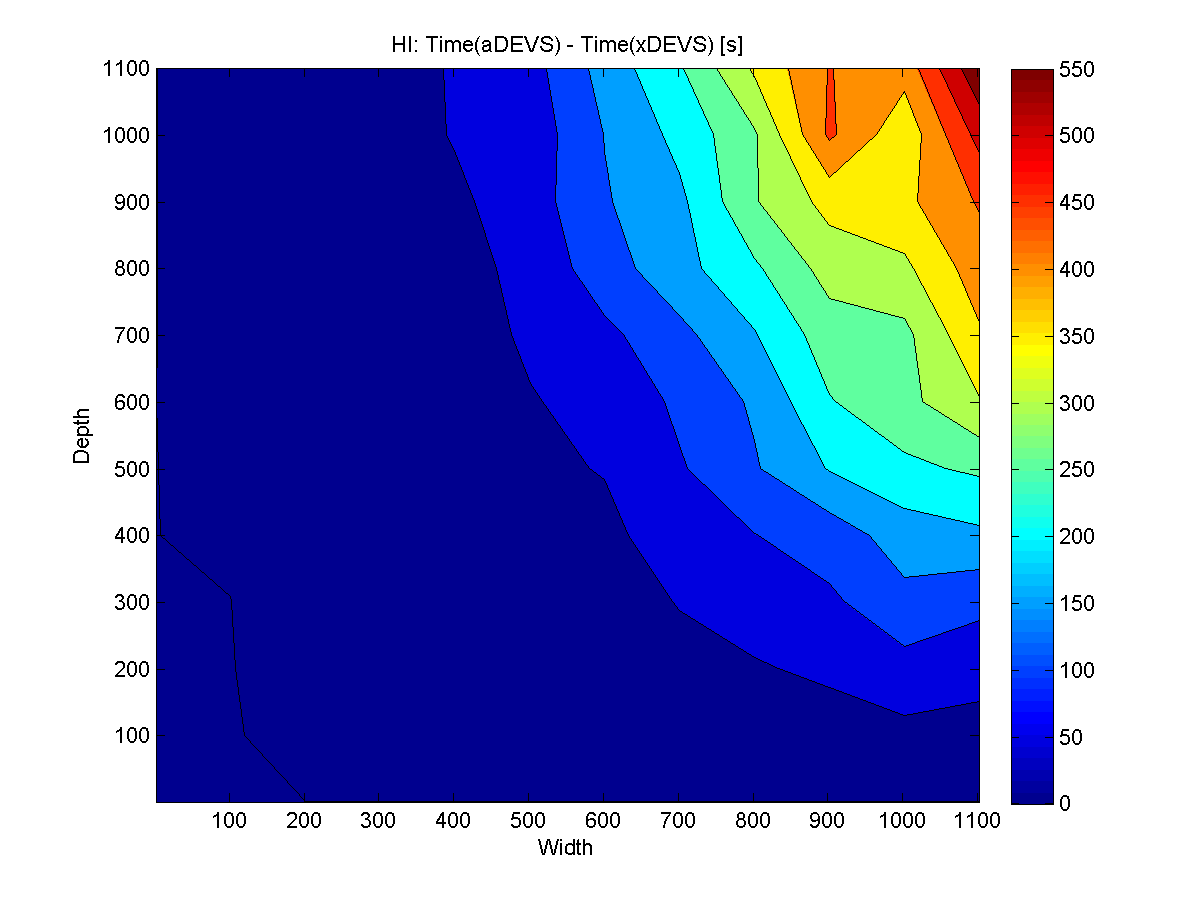}
\label{fig:ComparisonHiIntelTime}
}
\subfigure[HO: xDEVS - aDEVS]{
\includegraphics[width=0.45\textwidth, height=3cm]{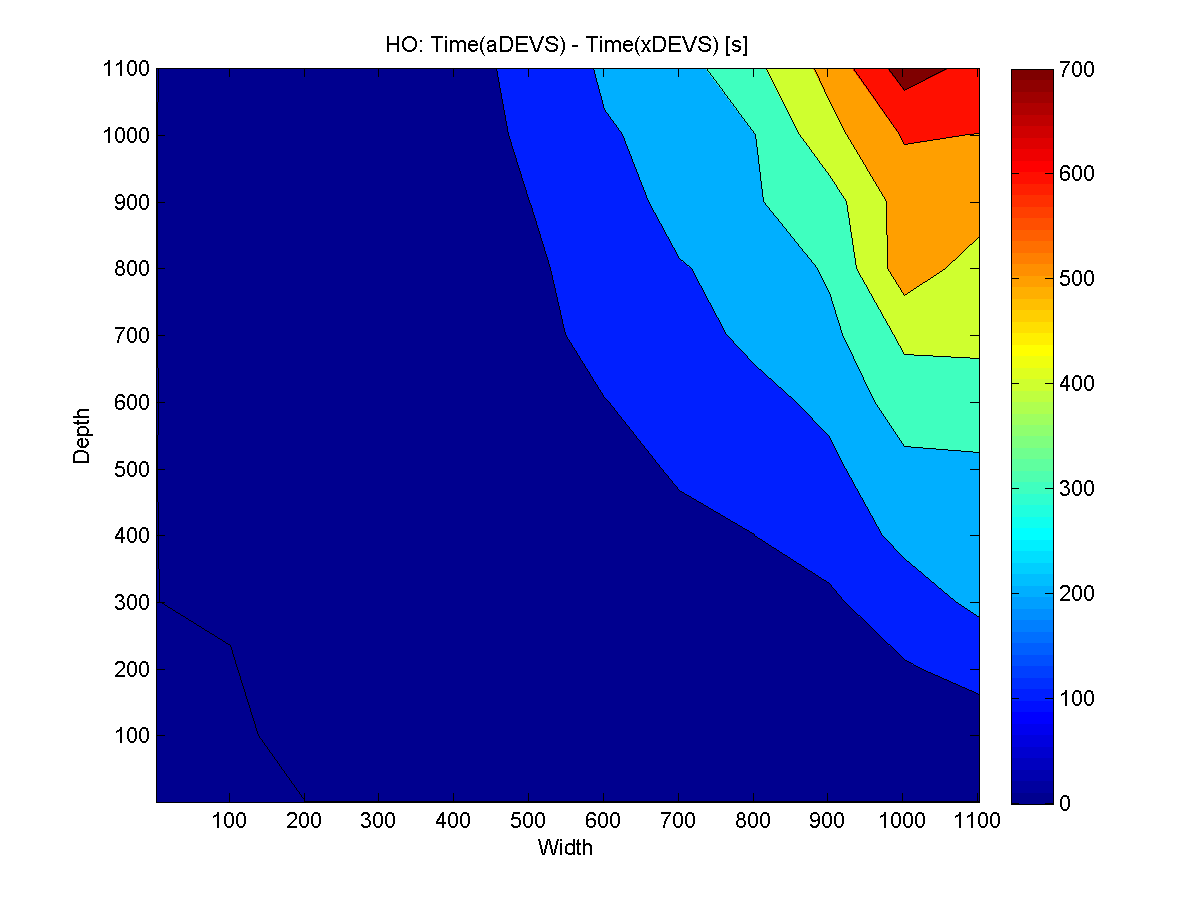}
\label{fig:ComparisonHoIntelTime}
}
\subfigure[HOmod: xDEVS - aDEVS]{
\includegraphics[width=0.45\textwidth, height=3cm]{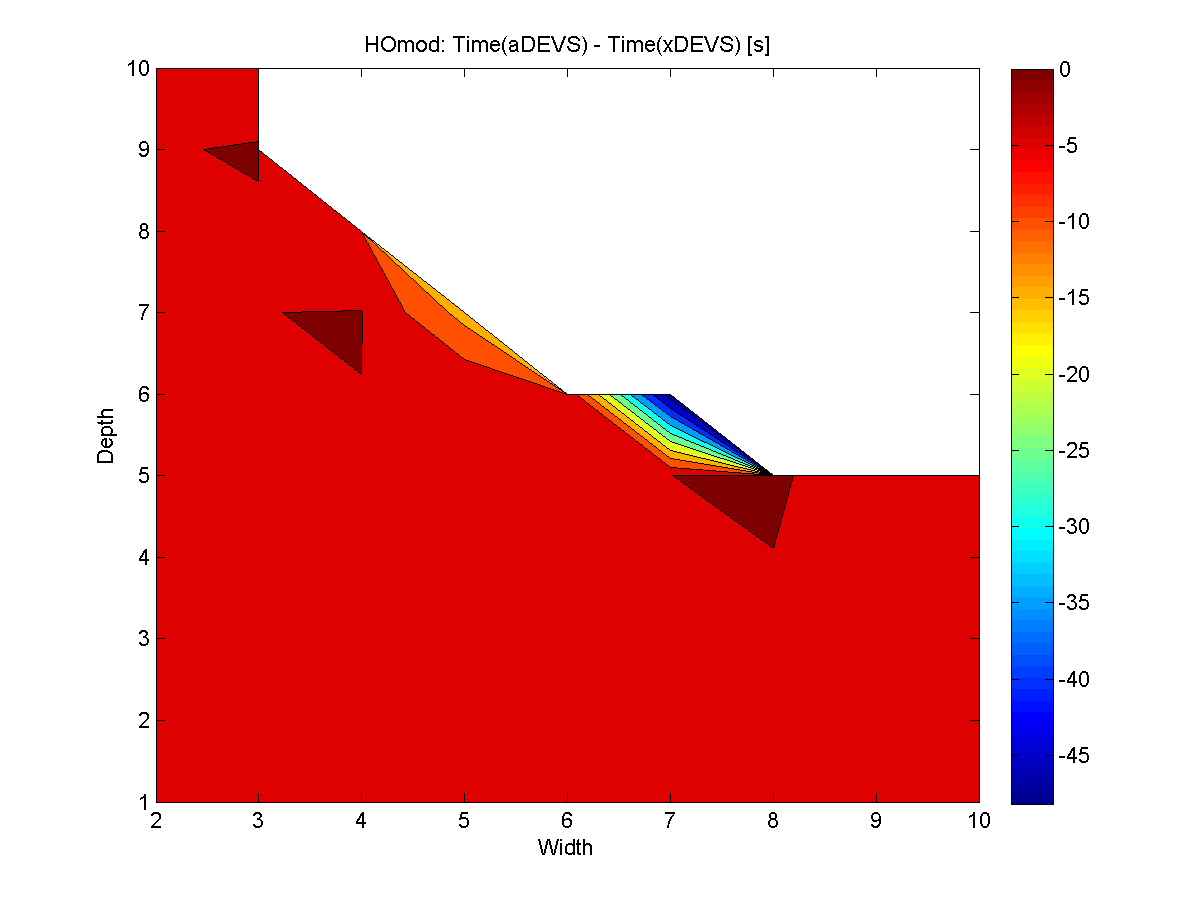}
\label{fig:ComparisonHoModIntelTime}
}
\subfigure[HOmem: xDEVS - aDEVS]{
\includegraphics[width=0.45\textwidth, height=3cm]{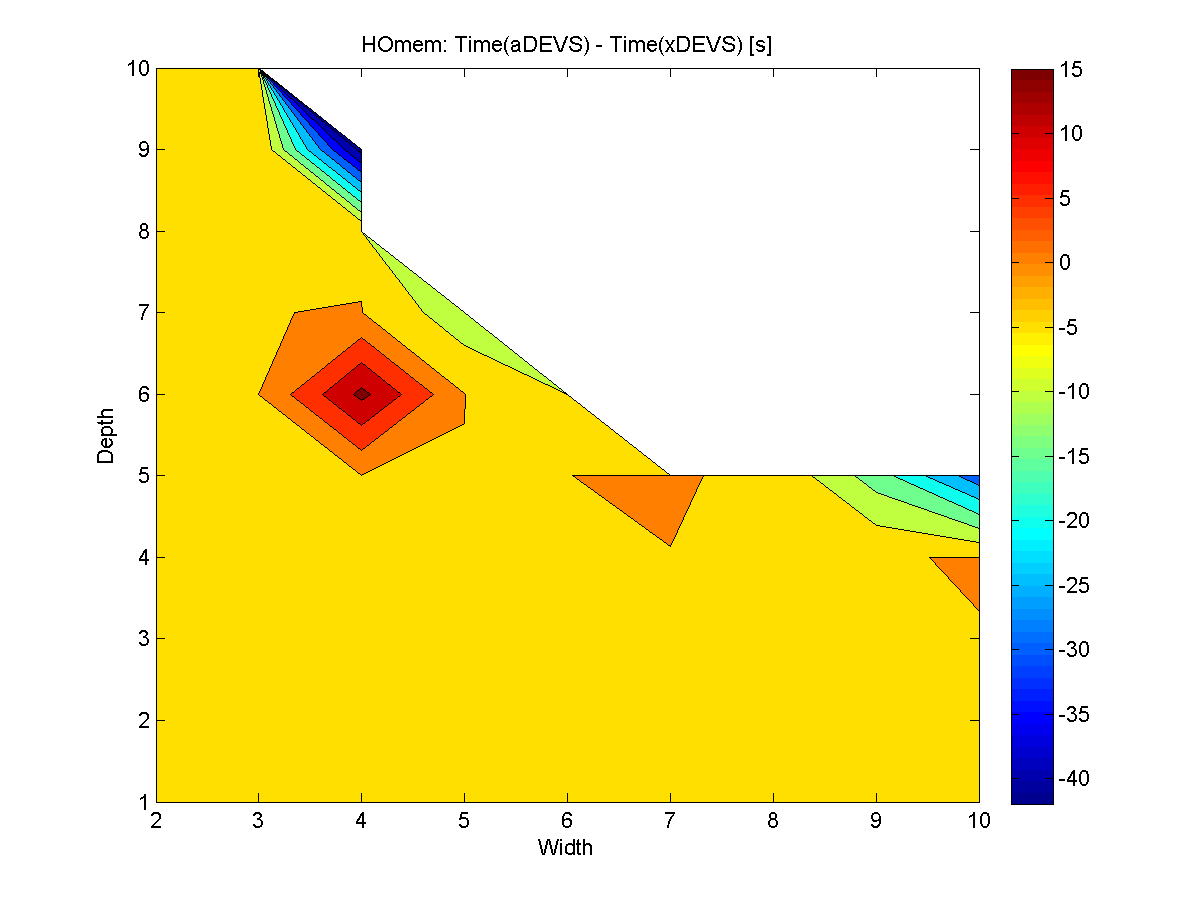}
\label{fig:ComparisonHoMemIntelTime}
}
\caption{Execution time comparison of LI, HI, HO, HOmod and HOmem models computed as Time(xDEVS) - Time(aDEVS)}
\label{fig:Comparison}
\end{figure*}

Figure \ref{fig:Comparison} depicts five contour maps. Each one represent the difference, in execution time, of xDEVS minus aDEVS. 

In the case of models with lower complexity, like the LI model in Figure \ref{fig:ComparisonLiIntelTime}, the difference is small (2.5 seconds vs 2.1 seconds according to Table \ref{tab:AmdVsIntel}) and in favor of aDEVS. With respect to HOmod and HOmem, the difference fundamentally varies from -10 seconds to 10 seconds, with more cases in favor of aDEVS. However, these two models remain indecisive since they show sparse maps.

The analysis of Figures \ref{fig:ComparisonHiIntelTime} and \ref{fig:ComparisonHoIntelTime} is much clearer. As the model complexity is increased, the difference is higher, in favor of xDEVS (up to 700 seconds faster in the case of HO).

We now compare both simulator in the HOmod and HOmem DEVStone model. aDEVS and xDEVS were the only two simulators that were able to simulate a significant number of HOmod models.

\begin{figure*}[ht]
\centering
\subfigure[aDEVS - HOmod (Time)]{
\includegraphics[width=0.45\textwidth, height=3cm]{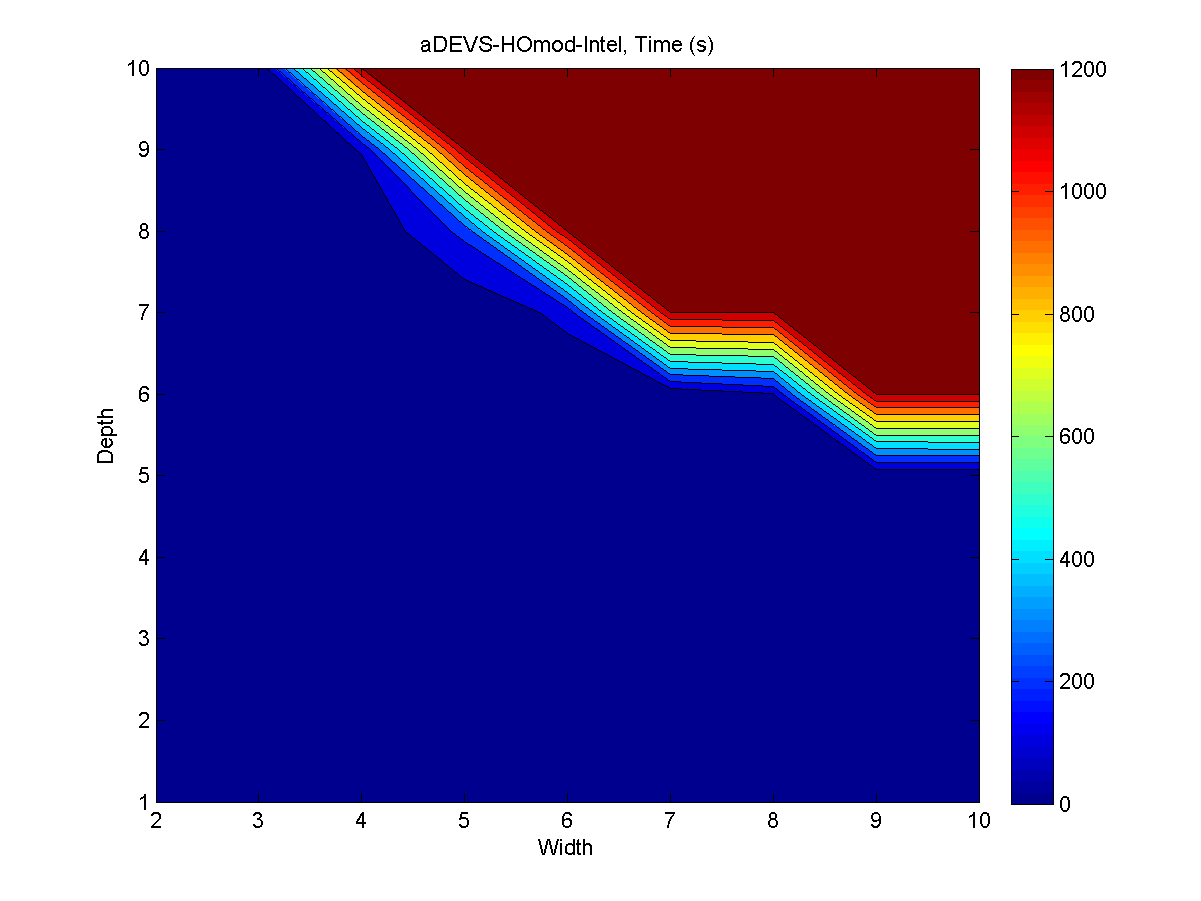}
\label{fig:aDevsHoModIntelTime}
}
\subfigure[aDEVS - HOmod (Memory)]{
\includegraphics[width=0.45\textwidth, height=3cm]{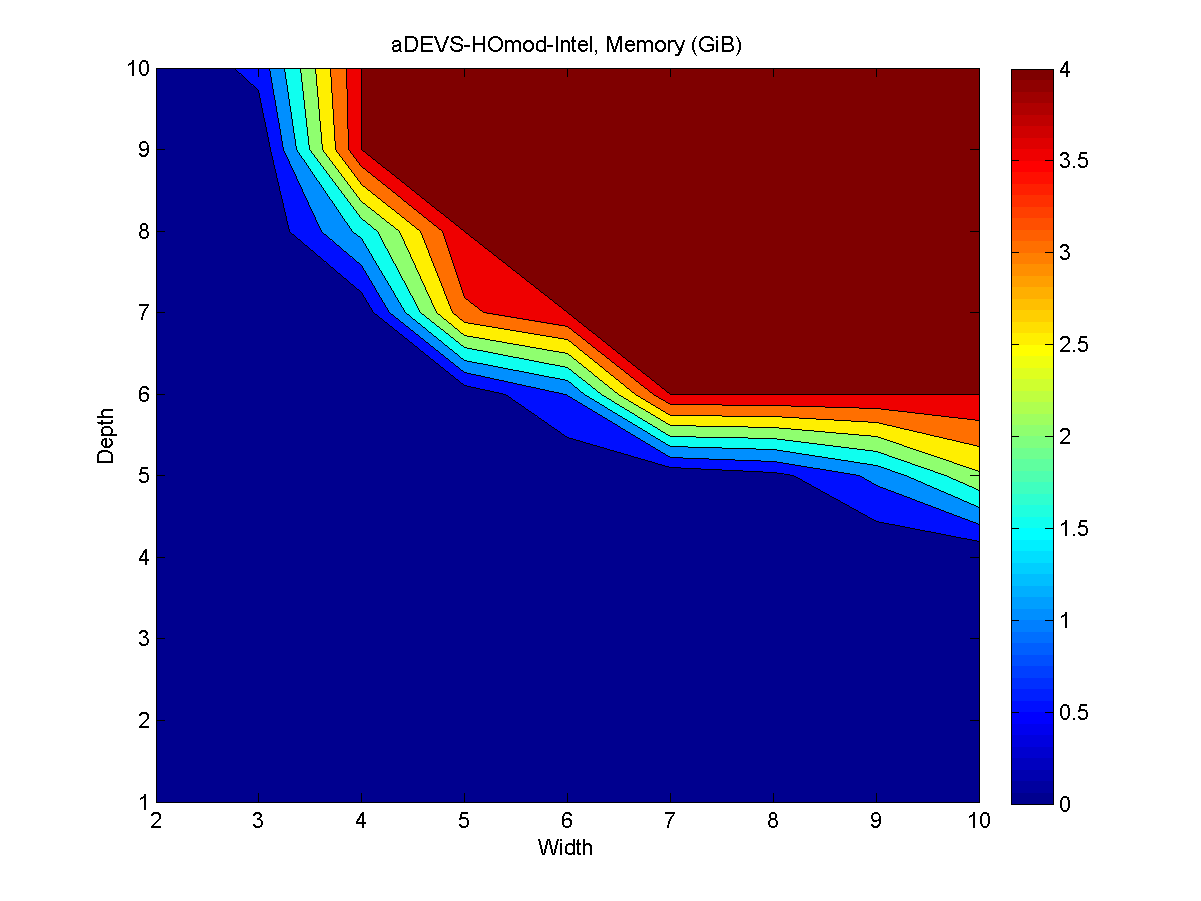}
\label{fig:aDevsHoModIntelMemory}
}
\subfigure[xDEVS - HOmod (Time)]{
\includegraphics[width=0.45\textwidth, height=3cm]{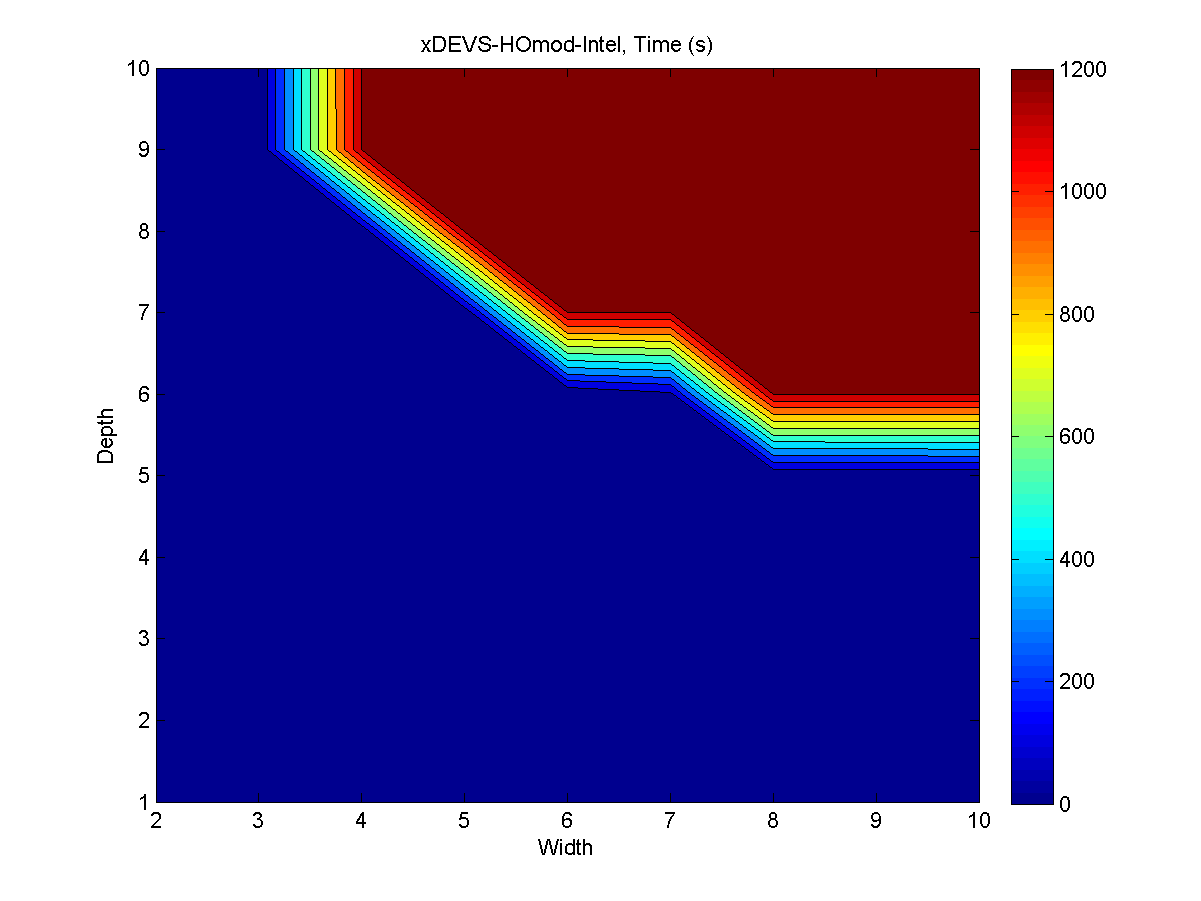}
\label{fig:xDEVSHoModIntelTime}
}
\subfigure[xDEVS - HOmod (Memory)]{
\includegraphics[width=0.45\textwidth, height=3cm]{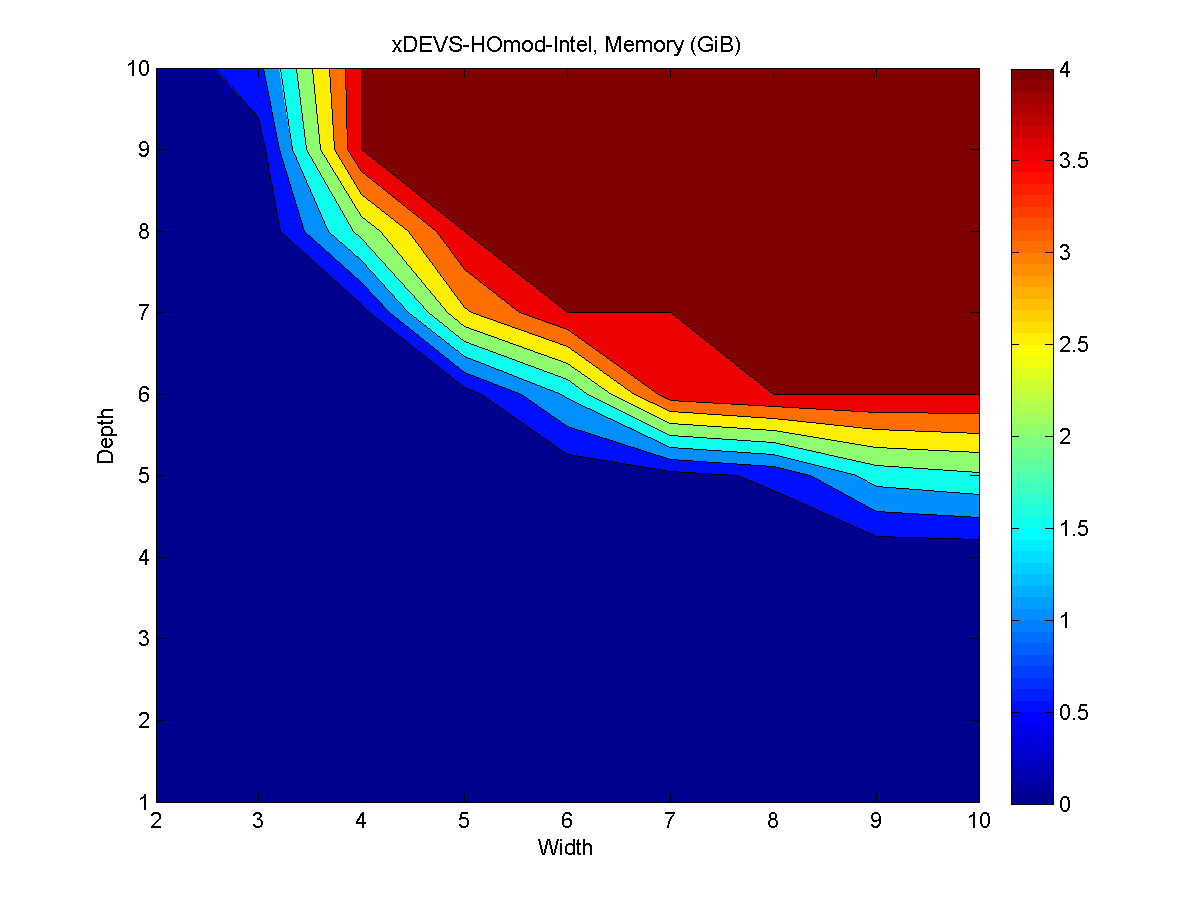}
\label{fig:xDEVSHoModIntelMemory}
}
\subfigure[aDEVS - HOmem (Time)]{
\includegraphics[width=0.45\textwidth, height=3cm]{aDEVS_HOmem_intel_Time.png}
\label{fig:aDevsHoMemIntelTime2}
}
\subfigure[aDEVS - HOmem (Memory)]{
\includegraphics[width=0.45\textwidth, height=3cm]{aDEVS_HOmem_intel_Memory.png}
\label{fig:aDevsHoMemIntelMemory2}
}
\subfigure[xDEVS - HOmem (Time)]{
\includegraphics[width=0.45\textwidth, height=3cm]{xDEVS_HOmem_intel_Time.png}
\label{fig:xDEVSHoMemIntelTime2}
}
\subfigure[xDEVS - HOmem (Memory)]{
\includegraphics[width=0.45\textwidth, height=3cm]{xDEVS_HOmem_intel_Memory.png}
\label{fig:xDEVSHoMemIntelMemory2}
}
\caption{Execution time and memory footprint of HOmod and HOmem models given by aDEVS and xDEVS}
\label{fig:AllHoModIntel}
\end{figure*}

Figure \ref{fig:AllHoModIntel} depicts both the execution time and memory footprint reached by aDEVS and xDEVS in HOmod and HOmem models. In both cases, contour maps are practically Yes/No maps, where, after a given \emph{width} and \emph{depth} both simulators immediately reach the limit in execution time and memory footprint. These ``saturation'' values in HOmod are reached ``sooner'' (in terms of $w$ and $d$) than the corresponding values in the HOmem model. We prove here that HOmem offers the same results than HOmod with a more straightforward mathematical formulation, after a comparison of equations \eqref{eq:homem1}-\eqref{eq:homem2} against equations \eqref{eq:homod1}-\eqref{eq:homod2}.

\section{Conclusions}\label{sec:Conclusions}
The Discrete Event System Specification formalism (DEVS) has been widely used to conceive, design, model and develop a great variety of systems. DEVS has been implemented in various languages and platforms over the years. The DEVStone benchmark defines a set of models with varied structure and behavior, and was designed to evaluate the performance of DEVS-based simulators.

The key contributions of this work are the following. We have added a new model to the benchmark, called HOmem, which shows identical qualitative behavior than HOmod but with a more manageable mathematical formulation. As HOmod, HOmem is also intensive on both execution time and memory usage. We have added the study of memory footprint in DEVStone, deriving the equations needed to compute the number of events triggered inside the model and per each single injected external event. We have also recalculated the number of transition functions triggered in all the DEVStone benchmarks. Finally, we have compared five simulation engines in two different hardware platforms, analyzing both the execution time and memory footprint. To perform a fair comparison between simulation engines that allow and do not allow model flattening, we did not flattened the benchmark in any case.

These five DEVStone models are executed against five different DEVS simulators, implemented in different programming languages such as C++, JAVA and Python. 

Results show that all the simulators were able to run the HOmem model at least for a significant range of \textit{width} and \textit{depth} values. Between all the five simulators, aDEVS, which is based on C++, had the lowest memory footprint at least in LI, HI, and HO models. With respect to execution time, xDEVS was the fastest one, specially in the set of HI and HO more complex models.

As future work, we propose the extension of this complete analysis to study the performance of DEVStone parallel and distributed simulations.

\begin{acks}
The authors would like to thank Dr. Gabriel Wainer and Dr. Sixuan Wang for their assistance in the compilation of CD++ under GNU/Linux.

This work is supported by the Spanish Ministry of Economy and Competitivity under research grants TIN2014-54806-R and TIN2013-40968-P.
\end{acks}

\bibliographystyle{SageV}
\bibliography{Bibliography}

\end{document}